\documentclass[traditabstract]{aa}
\usepackage{graphicx}
\usepackage{epsfig}
\usepackage{color}
\usepackage{txfonts}
\usepackage{natbib}
\bibpunct{(}{)}{;}{a}{}{,} % to follow the A&A style
\begin{document}

\title{Signatures of radial migration in  barred galaxies:  \\ Azimuthal variations in the metallicity distribution of old stars}
\titlerunning{Migration and azimuthal variations}

\author{P. Di Matteo\inst{1},  M. Haywood\inst{1},   F. Combes\inst{2}, B. Semelin\inst{2}, O.~N. Snaith\inst{1}}%,  O. Snaith\inst{1}}

\authorrunning{Di Matteo et al.}

\institute{GEPI, Observatoire de Paris, CNRS, Universit\'e
  Paris Diderot, 5 place Jules Janssen, 92190 Meudon, France\\
\email{paola.dimatteo@obspm.fr}
\and
LERMA, Observatoire de Paris, CNRS, 61 Av. de l'Observatoire, 75014 Paris, France; Universit$\acute{\textrm{e}}$ Pierre et Marie Curie, 4 place Jussieu, 75005 Paris, France
}

\date{Accepted, Received}

\abstract{By means of N-body simulations, we show that radial migration in galaxy disks, induced by bar and spiral arms, leads to significant azimuthal variations in the metallicity distribution of old stars at a given distance from the galaxy center. Metals do not show an axisymmetric distribution during phases of strong migration. Azimuthal variations are visible during the whole phase of strong bar phase, and tend to disappear as the effect of radial migration diminishes, together with a reduction in the bar strength. These results suggest that the presence of inhomogeneities in the metallicity distribution of old stars in a galaxy disk can be a probe of ongoing strong migration.  Such signatures may be detected in the Milky Way  by Gaia (and complementary spectroscopic data), as well as in external galaxies, by  IFU surveys like CALIFA and ATLAS3D.
Mixing -- defined as the tendency toward  a homogeneous, azimuthally symmetric, stellar distribution in the disk -- and  migration turns out to be two distinct processes, the effects of mixing starting to be visible when strong migration is over.}

\keywords{Galaxies: abundances; Galaxies: evolution; Galaxies: structure; Galaxies: kinematics and dynamics; Methods: numerical}

\maketitle

\section{Introduction}

Stellar asymmetries in galaxy disks, such as bars and spiral patterns, have been known for a long time to drive galaxy evolution.
Stellar bars, in particular, constitute an efficient way to redistribute angular momentum among the different galaxy components -- gas, stars and dark matter \citep{bour02, ber07, atha02}.
Angular momentum redistribution is particularly efficient at resonances, and a coupling between the action of a bar and that of spiral arms can cause a rapid migration of stars through the disk  \citep{min10, min11}, with characteristic time scales significantly lower than those predicted when the action of spiral arms alone is taken into account \citep{sel02, ros08}. Mergers and perturbations from satellite galaxies can also induce radial migration, as shown by \citet{min09}, \citet{bekki11} and \citet{bird12}.\\
If radial migration has received considerable attention in the last few years, driven by the need to interpret a number of galactic and extragalactic observations \citep{hay08, bov12, bak11, yoa12,  rad12}, it is still  not clear how important this process is in galaxy evolution, whether all disk galaxies go through a phase of strong migration, and how the Hubble type and bar strength affect it. 
\citet{bru11}, for example, showed that not all barred galaxies experience strong diffusion, this depends on the bar strength,  and thus ultimately on the stability of the disk. If more work is needed to quantify systematically the importance migration has in galaxy evolution, it is also essential to understand the signatures this process leaves in a galaxy.
Truncated stellar density profiles, upturns in the stellar age profiles, flattening of stellar metallicity gradients, thick disks, etc. can be the result of migration processes \citep{deb06, ros08b, loe11, min11, min12}. Unfortunately, however, they are not uniquely caused by migration \citep{elm06, san09, qu11, bour09}.
Is there any other way to quantify the strength of the process, and to understand if a galaxy is currently experiencing a strong redistribution of stars in its disk?\\
In this work we report on the results of a numerical study which shows that in the phase of strong migration induced by a bar galaxy disks exhibit significant azimuthal variations in the metal distribution of their old stellar component. These azimuthal variations remain strong during the phase of high bar strength ($\sim 1$Gyr in our simulations), when radial migration is maximum, and decrease later on. At this point, the inner stellar disk becomes homogeneous, on kpc scales, within few rotational periods after the end of the migration phase.
We suggest  that the presence of azimuthal inhomogeneities in the metal distribution of old stars in a barred galaxy can be used as evidence that its disk is going through a phase of significant stellar migration, related to a strong bar.\\
The paper is organized as follows: in Sect.\ref{method} a description of the initial conditions and numerical method adopted for the simulation are given; in Sect.\ref{results} the results are presented, firstly discussing the mode and timing of radial migration in barred galaxies, and then presenting the effect of this migration on the redistribution of metals in a stellar disk, together with a discussion on its dependency on the initial metallicity gradient;  in Sect.\ref{conclusions} the main conclusions of this work are described.

\section{Initial conditions and numerical method}\label{method}

We report on the results of a collisionless N-body  simulation of an isolated galaxy, with an initial bulge-to-disk ratio equal to 0.1, and no gas in its disk. This simulation is one of a set of three dissipationless high-resolution simulations,  with varying bulge-to-disk ratio.  We present the analysis only for one of the three simulations, since the results and conclusions found are common to all the cases analyzed. Note that hydrodynamics is not included in this work, since we are interested in  quantifying the effect of radial migration alone. Studying the complex interplay between gas, induced star formation, and stellar dynamics is beyond the scope of this paper. However, for a Milky Way-like galaxy, with a quiescent star formation of few $\rm{M}_{\odot}$/yr, we do not expect the inclusion of star formation to  drastically change the results, since the amount of newly formed stars per rotation constitutes only a small percentage of the total stellar mass in the disk.

The halo and the bulge of the galaxy are modeled as a Plummer sphere  \citep{BT87}, with characteristic masses $M_{\rm B}=9\times10^9M_{\odot}$ and $M_{\rm H}=1.02\times10^{11}M_{\odot}$\footnote{This value corresponds to the mass inside a sphere of radius R= 35 kpc, and guarantees realistic rotation curves \citep[see][for a similar set of galaxy models]{chili10}} and characteristic radii $r_{\rm B}$=1.3~kpc and $r_{\rm H}$=10~kpc. The stellar disk follows a Miyamoto-Nagai density profile  \citep{BT87}, with mass $M_*=9\times10^{10}M_{\odot}$  and vertical and radial scale lengths given, respectively, by $h_*$=0.5~kpc and $a_*$=4~kpc. The initial disk size is 13 kpc and the Toomre parameter is set equal to Q=1.8.
The galaxy is represented by $N_{\rm tot}=30 720 000$ particles redistributed among dark matter ($N_{\rm H}=10 240 000$) and stars ($N_{\rm stars}=20 480 000$). 
To initialize particle velocities, we adopted the method described in  \citet{hern93}. 

To model the galaxy evolution, we employed a Tree-SPH code, %in which gravitational forces are calculated using a hierarchical tree method \citep{bar86} and gas evolution is followed by means of smoothed particle hydrodynamics \citep{lucy77, gin82}. The code has been 
presented in \citet{sem02} and we refer the reader to this paper for a full description.
For the dissipationless simulations analyzed in this paper, the SPH part of the code has been switched off, % and the code deals only with the evaluation of gravitational forces and integration of particle orbits with time.
  and gravitational forces are calculated using a tolerance parameter $\theta=0.7$, including terms up to the quadrupole order in the multiple expansion. A Plummer potential is used to soften gravitational forces, with a constant smoothing 
length $\epsilon=50~{\rm pc}$ for each of the different species of particles. With this spatial resolution, it is possible to resolve the vertical structure of thin disks, and follow small scale inhomogeneities in the disk. The equations of motion are integrated using a leapfrog algorithm with a fixed time step of $\Delta t=0.25$~Myr. 

Stars in the disk are assigned a metallicity $z_m$ which depends on their initial distance $R$ from the galaxy center, according to the formula $
%\begin{equation}
z_m=z_010^{-0.07R}$, 
%\end{equation}
$z_0=3z_{\odot}$ being the central ($R=0$) metallicity, and $z_{\odot}$ the solar value. We thus assume initially no azimuthal variation at any given radius. For more details on this choice, see  Appendix~\ref{appmet}.The initial metallicity gradient has been also varied from its  value $\Delta_{[Fe/H]}=-0.07~\textrm{dex/kpc}$ \citep{maciel94, vanzee98} to study the dependence of the results on this initial choice, as discussed in Sect.\ref{results2}.

%%%%%%%%%%%%%%%%%%%%%%%%%%%%%%%%%%%%%%%%%%%%%

\section{Results}\label{results}

%\subsection{Stellar bars and radial migration}\label{results0}

\begin{figure}
\centering
\includegraphics[width=5.cm,angle=270]{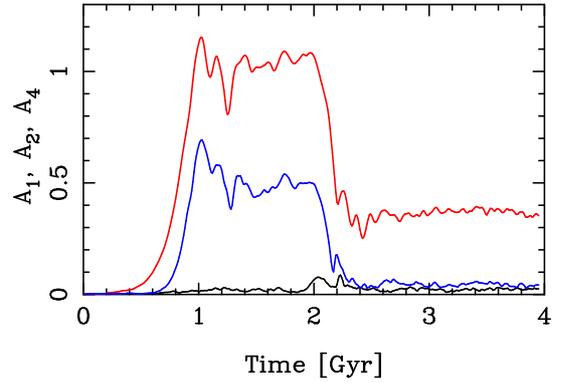}
\caption{Strength of the $A_1$ (black line), $A_2$ (red line) and $A_4$  (blue line) stellar asymmetries as a function of time.}
\label{asym}
\end{figure}

\begin{figure}
\centering
\includegraphics[width=3cm,angle=270]{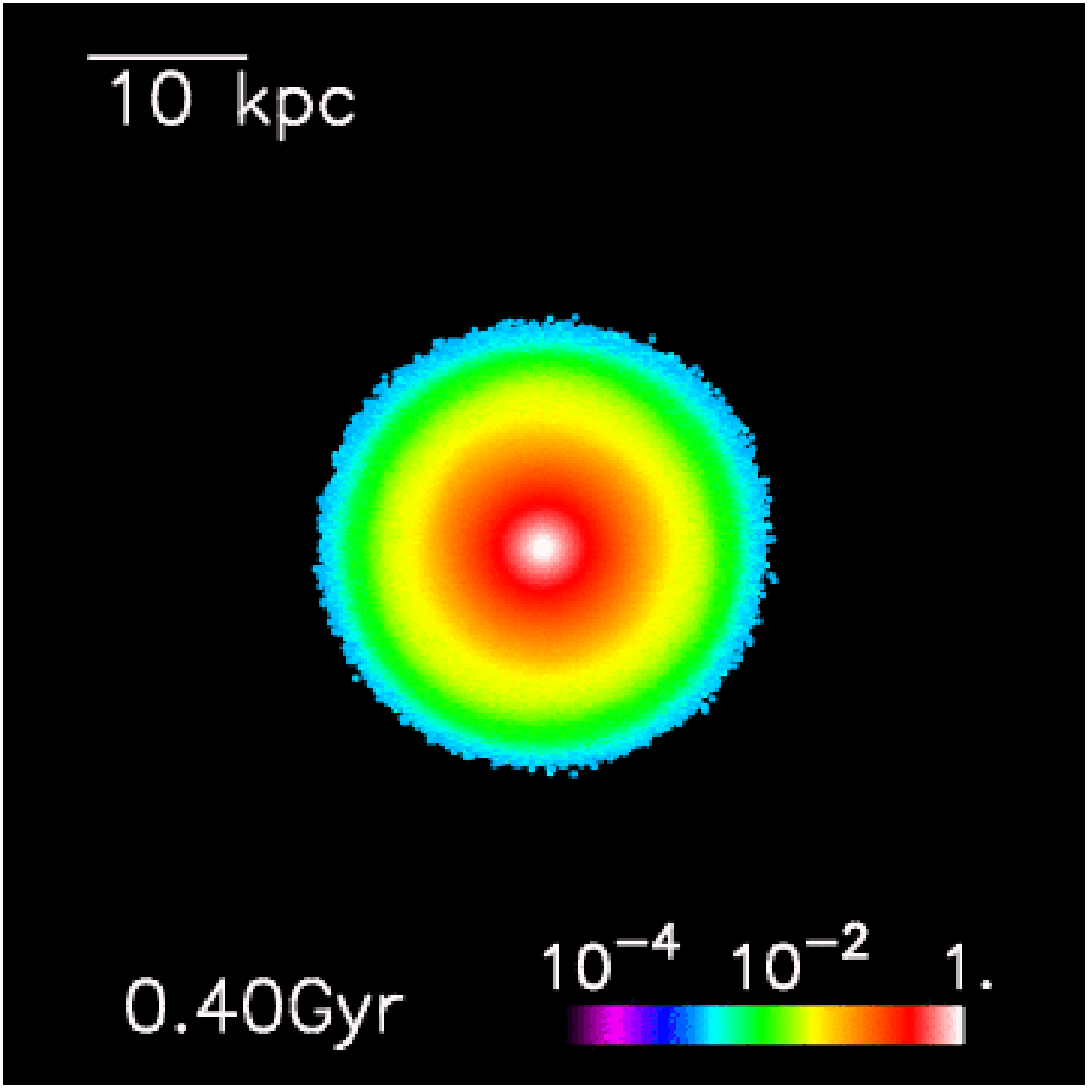}
\hspace{-0.2cm}
\includegraphics[width=3cm,angle=270]{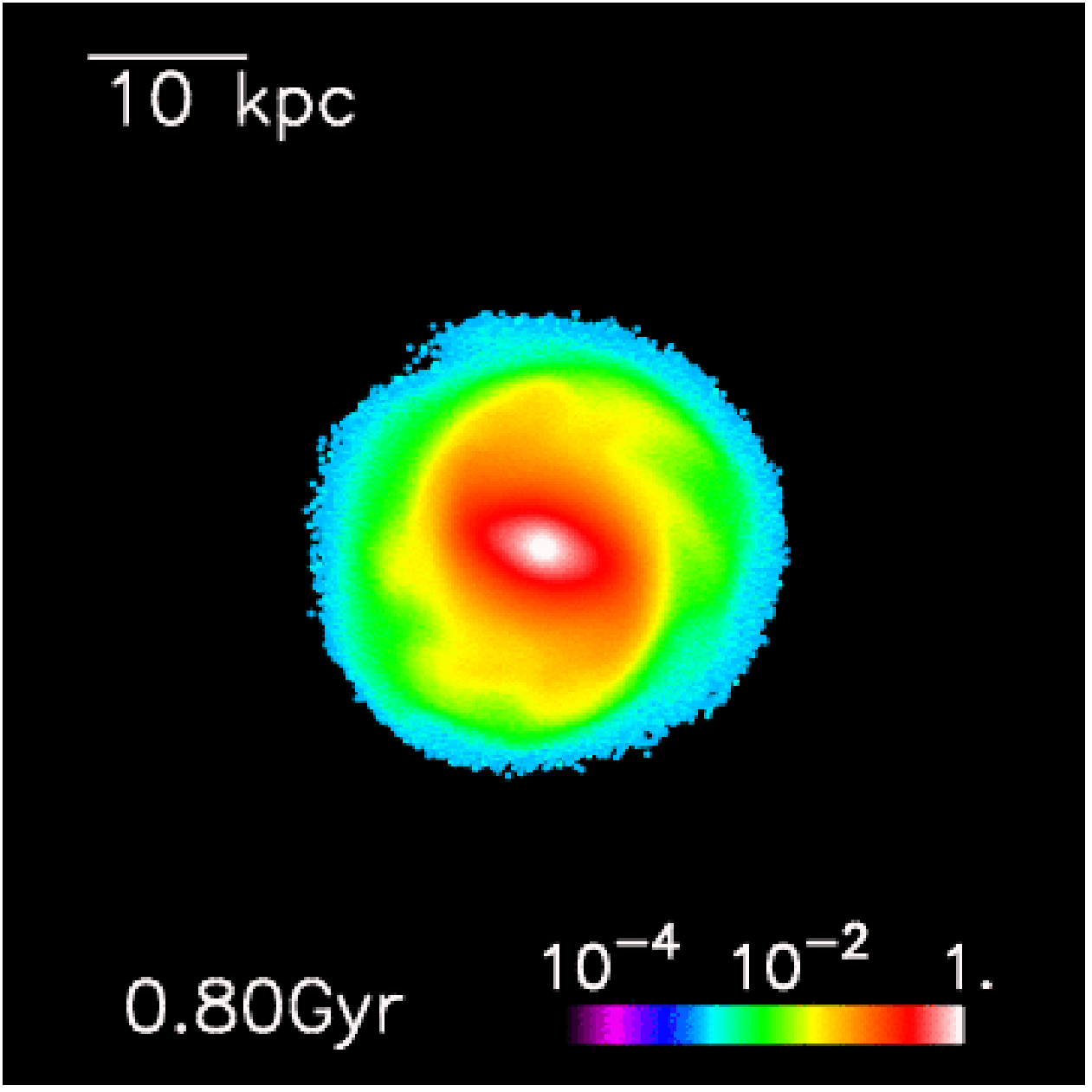}
\hspace{-0.2cm}
\includegraphics[width=3cm,angle=270]{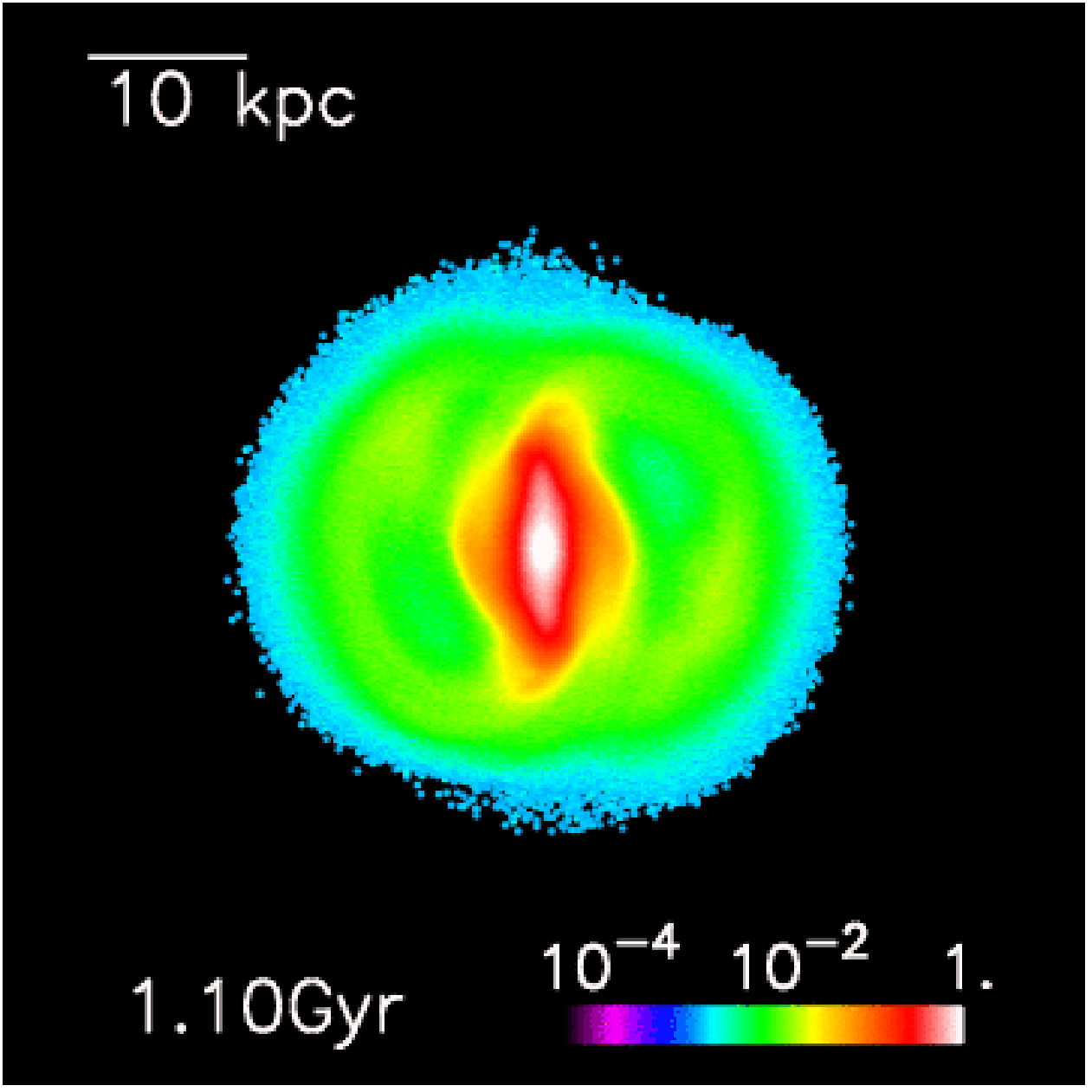}

\includegraphics[width=3cm,angle=270]{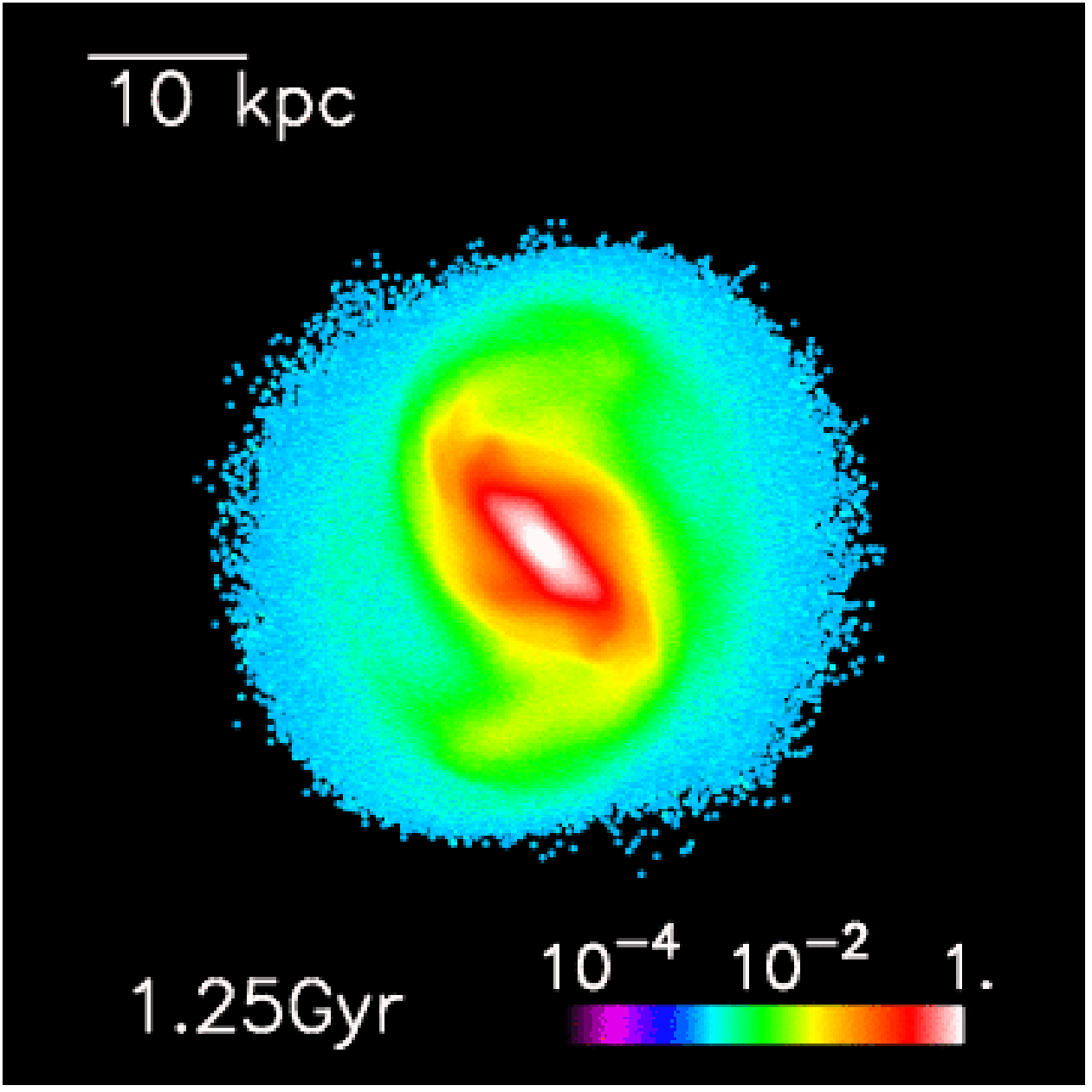}
\hspace{-0.2cm}
\includegraphics[width=3cm,angle=270]{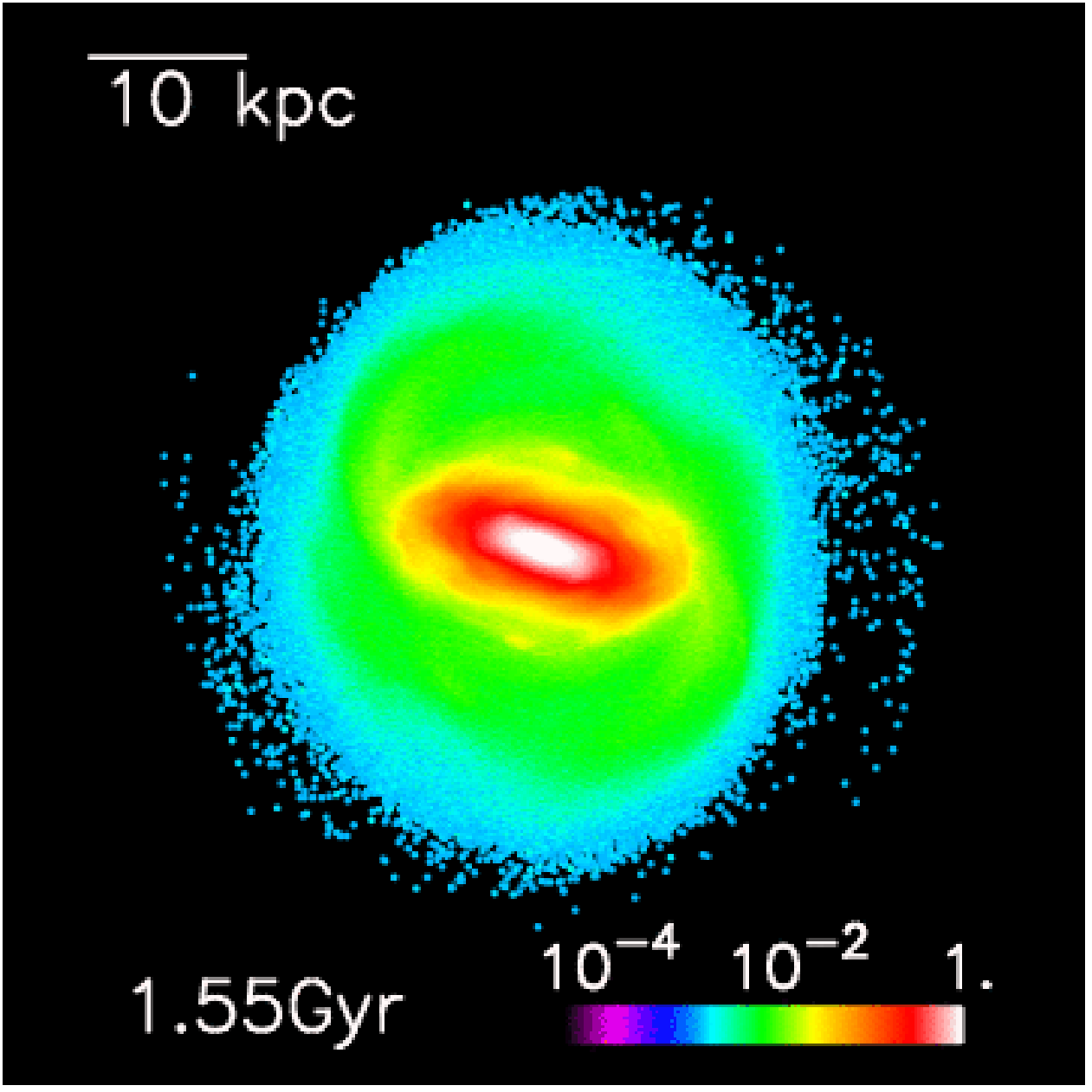}
\hspace{-0.2cm}
\includegraphics[width=3cm,angle=270]{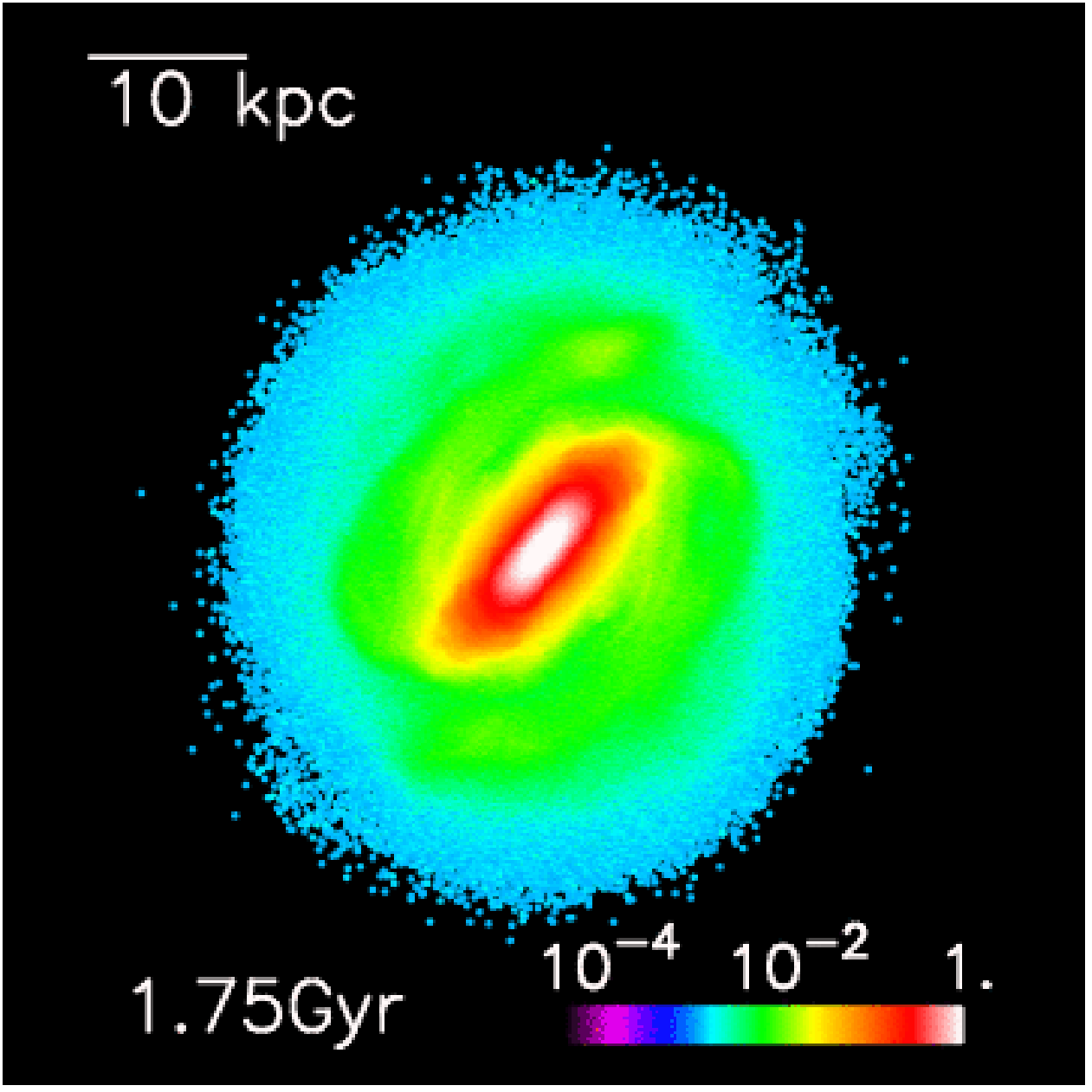}

\includegraphics[width=3cm,angle=270]{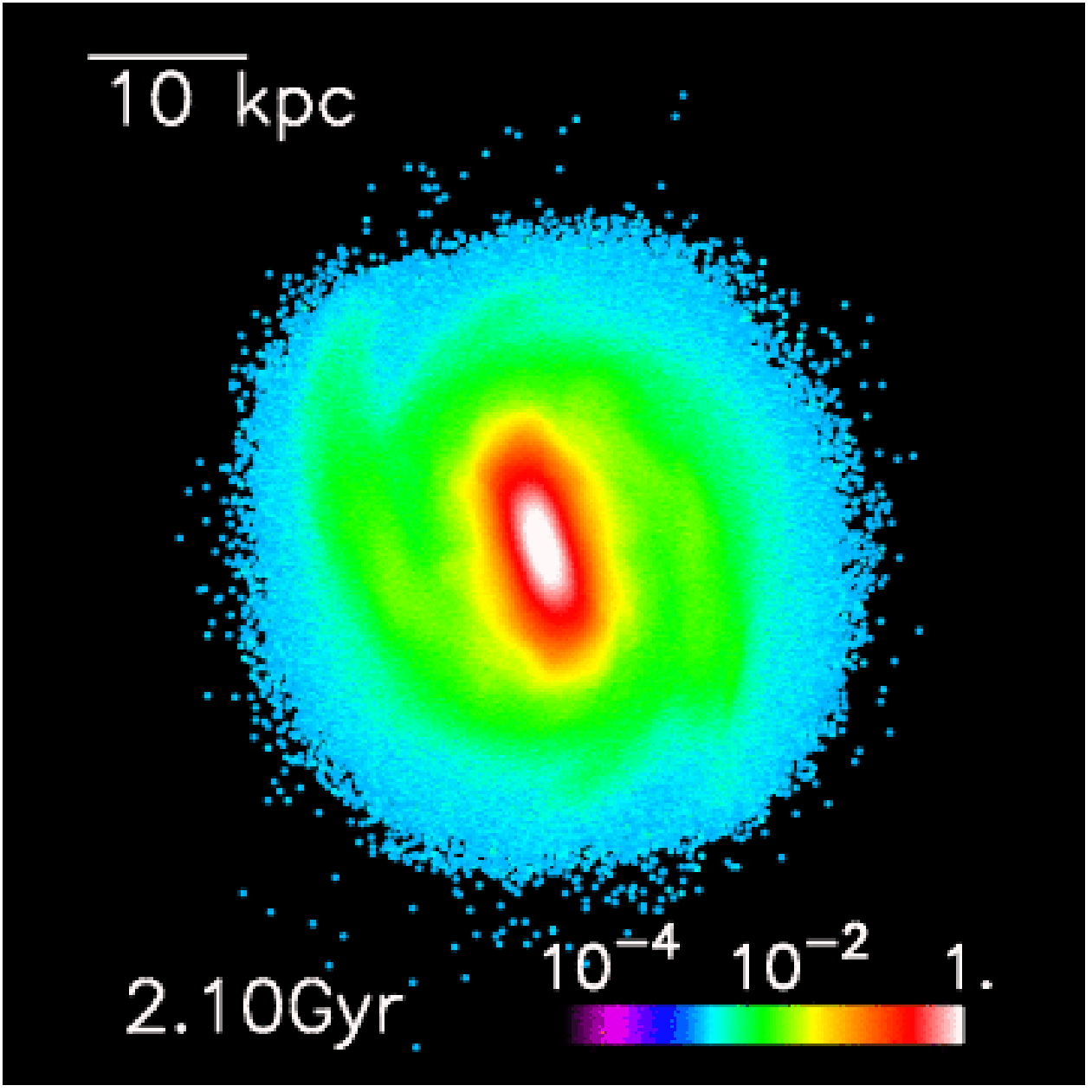}
\hspace{-0.2cm}
\includegraphics[width=3cm,angle=270]{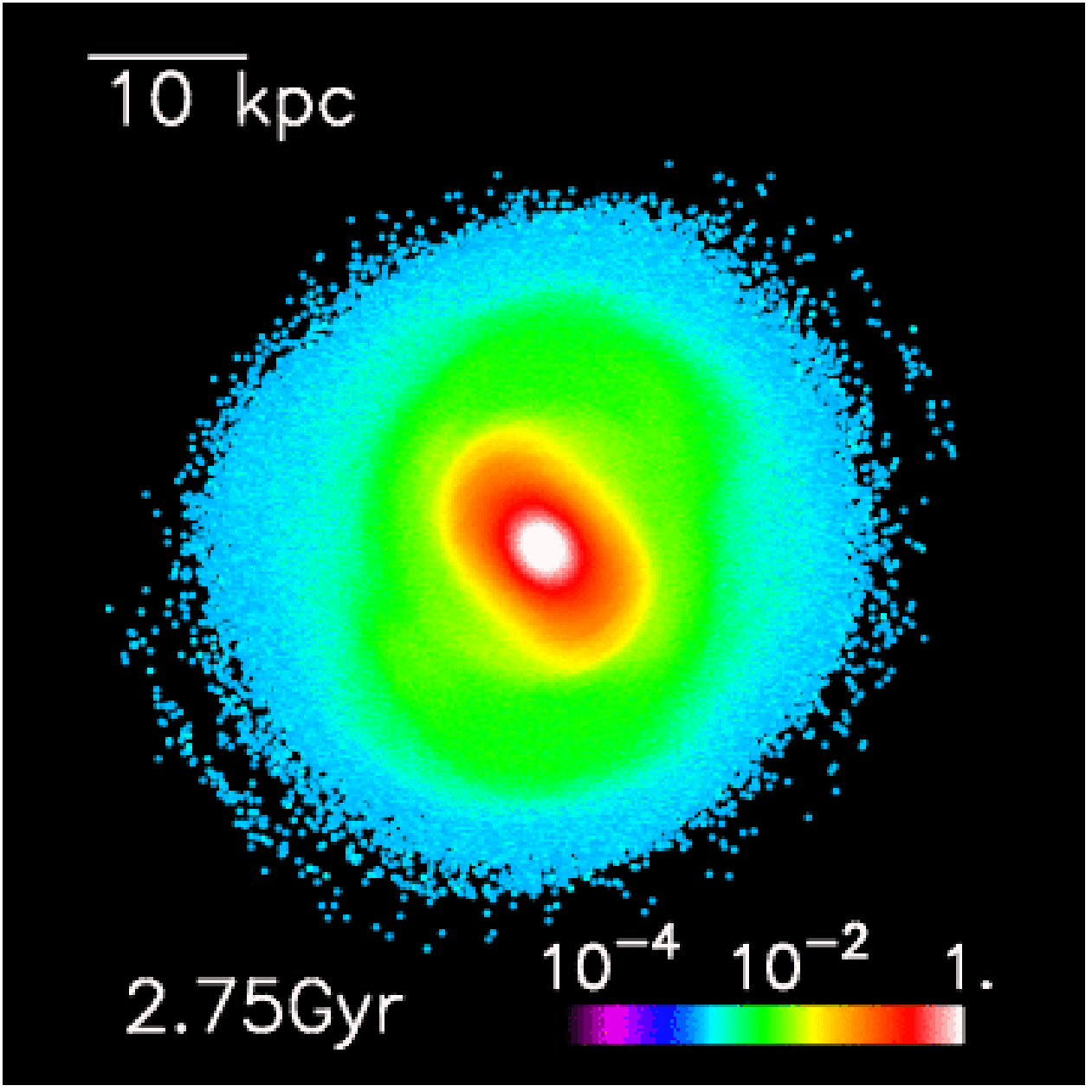}
\hspace{-0.2cm}
\includegraphics[width=3cm,angle=270]{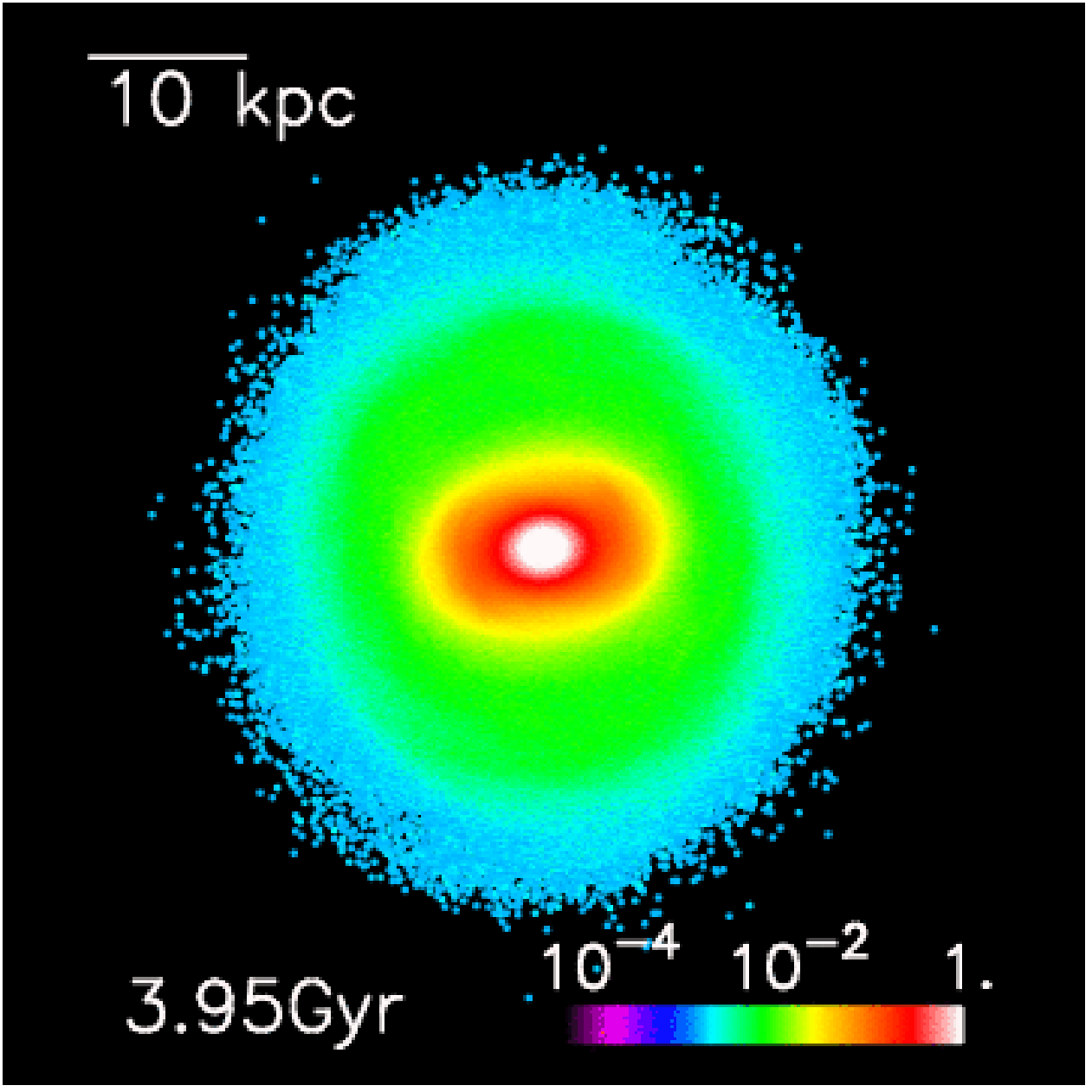}

\caption{Density maps of the stellar disk, seen face-on, at different times.  Each panel is 70 kpc$\times$70 kpc in size.%, corresponding to $t=$~0.4, 0.8, 1.1, 1.25, 1.55, 1.75, 2.01, 2.75, 3.95~Gyr.
}
\label{smaps}
\end{figure}

\begin{figure}
\centering
\includegraphics[width=3cm,angle=270]{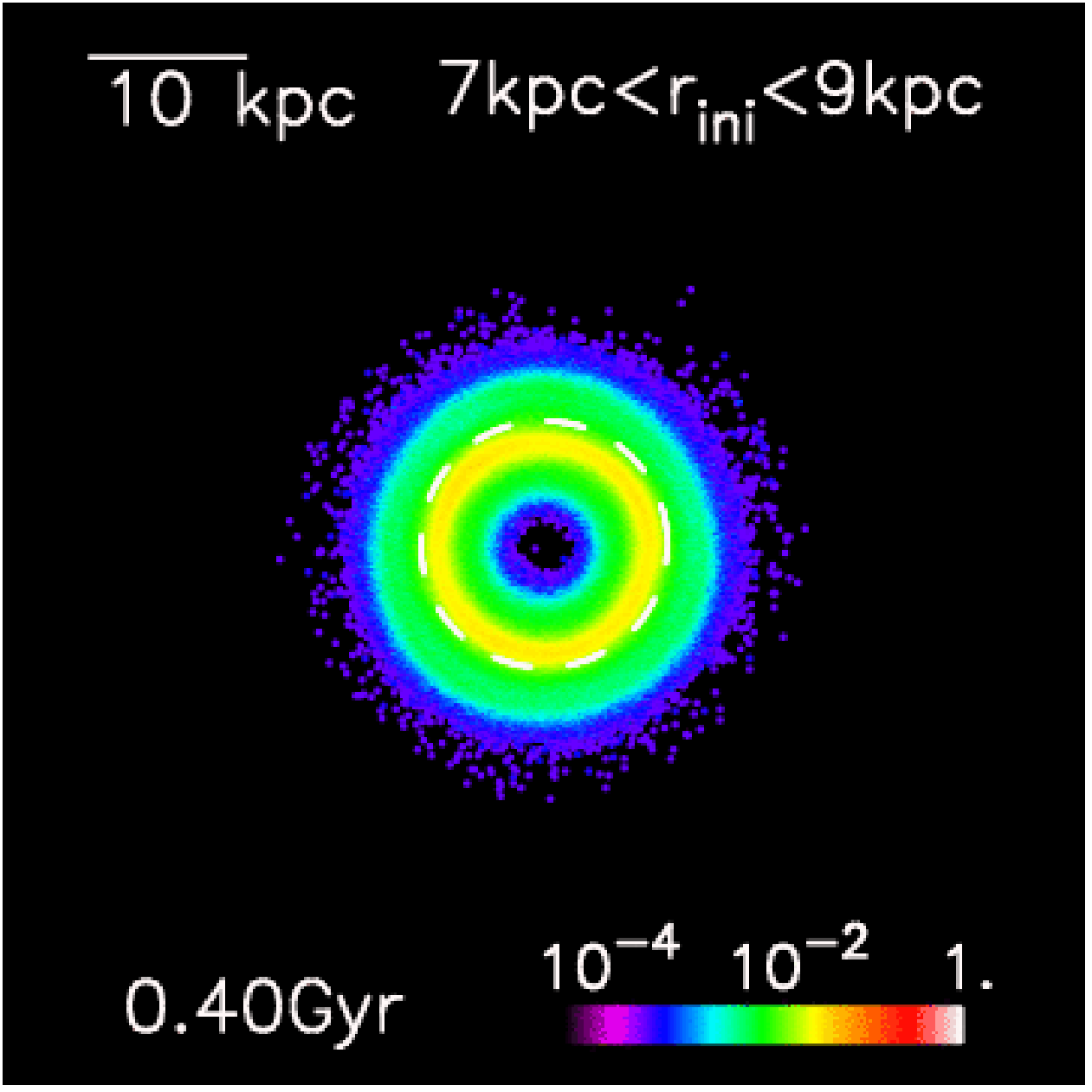}
\hspace{-0.2cm}
\includegraphics[width=3cm,angle=270]{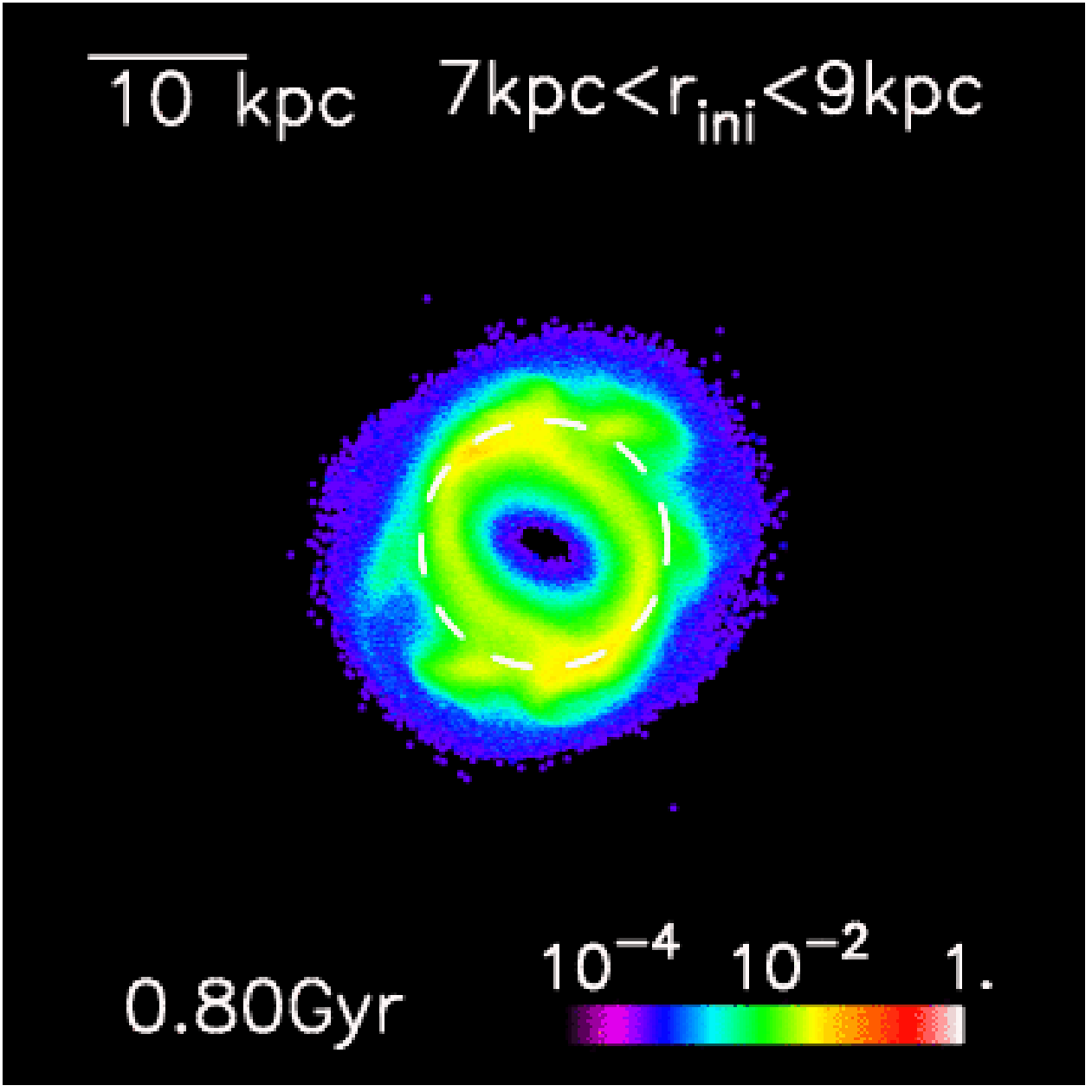}
\hspace{-0.2cm}
\includegraphics[width=3cm,angle=270]{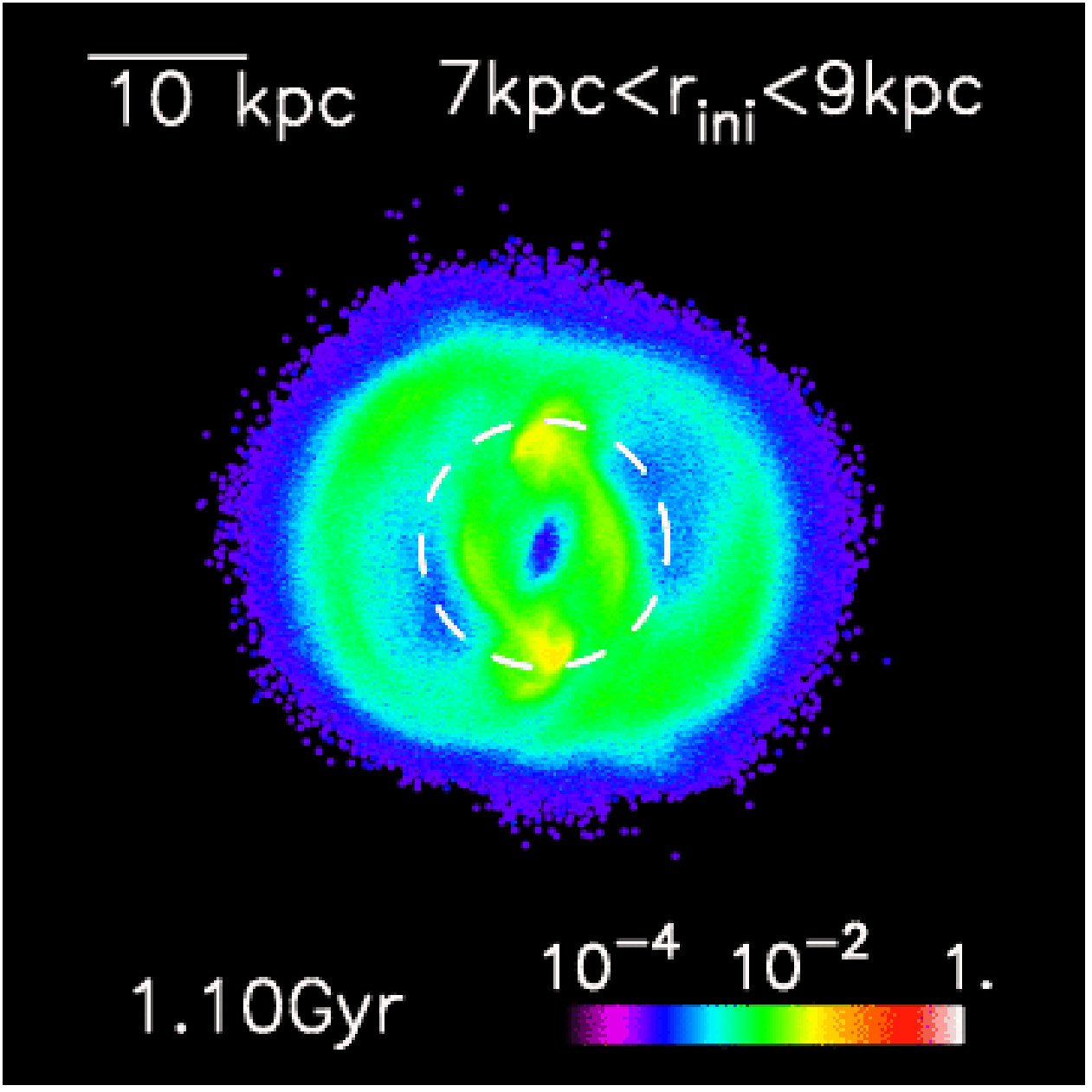}

\includegraphics[width=3cm,angle=270]{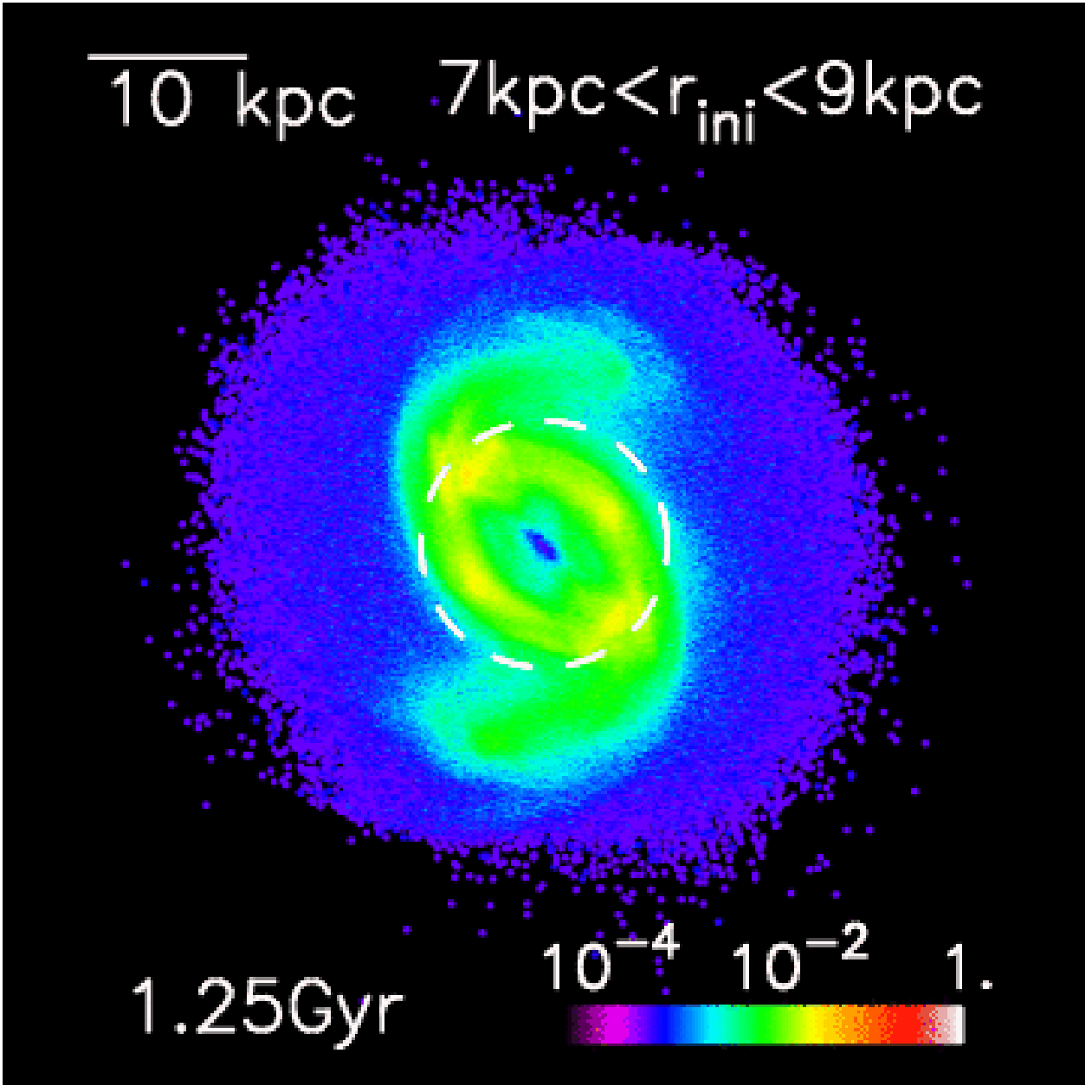}
\hspace{-0.2cm}
\includegraphics[width=3cm,angle=270]{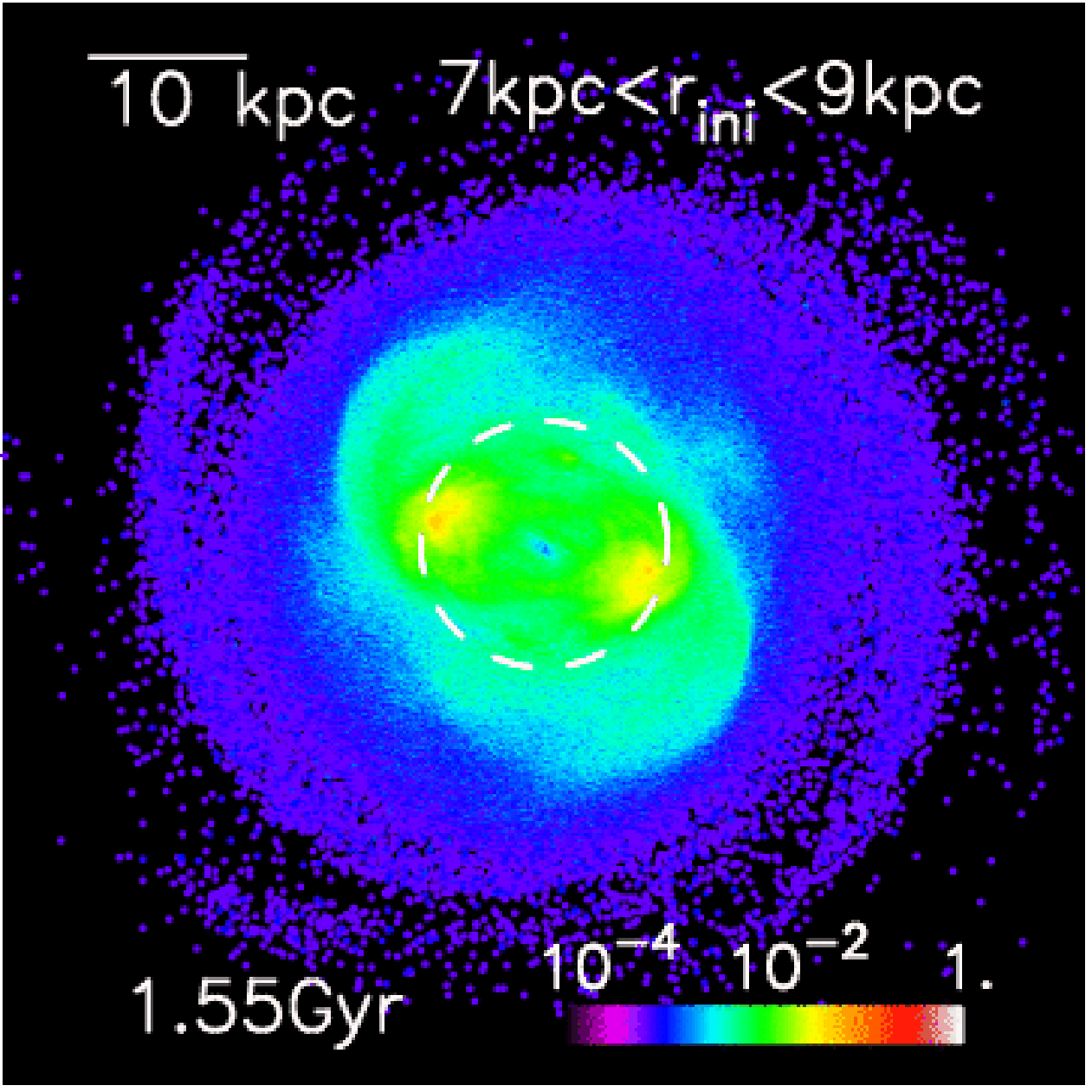}
\hspace{-0.2cm}
\includegraphics[width=3cm,angle=270]{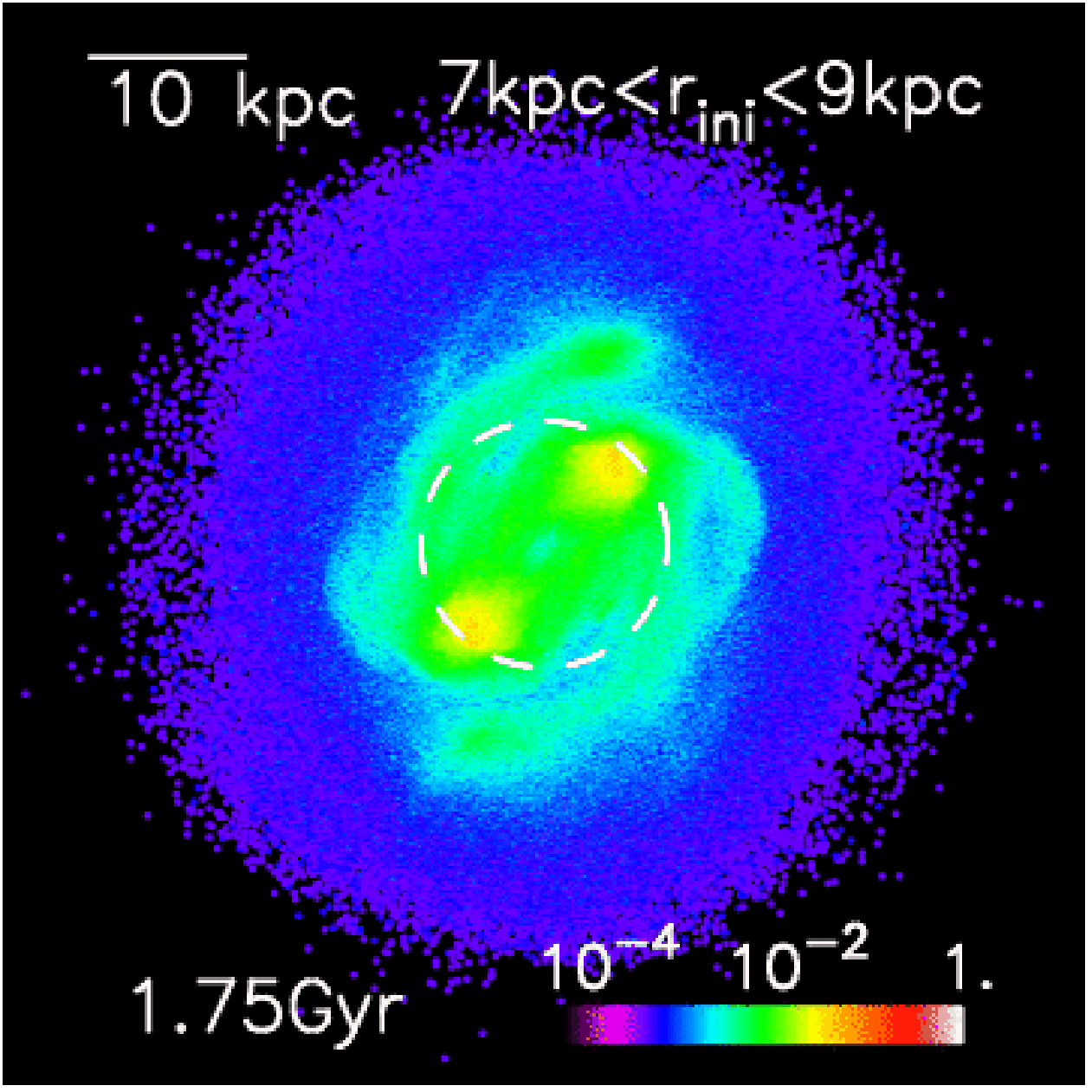}

\includegraphics[width=3cm,angle=270]{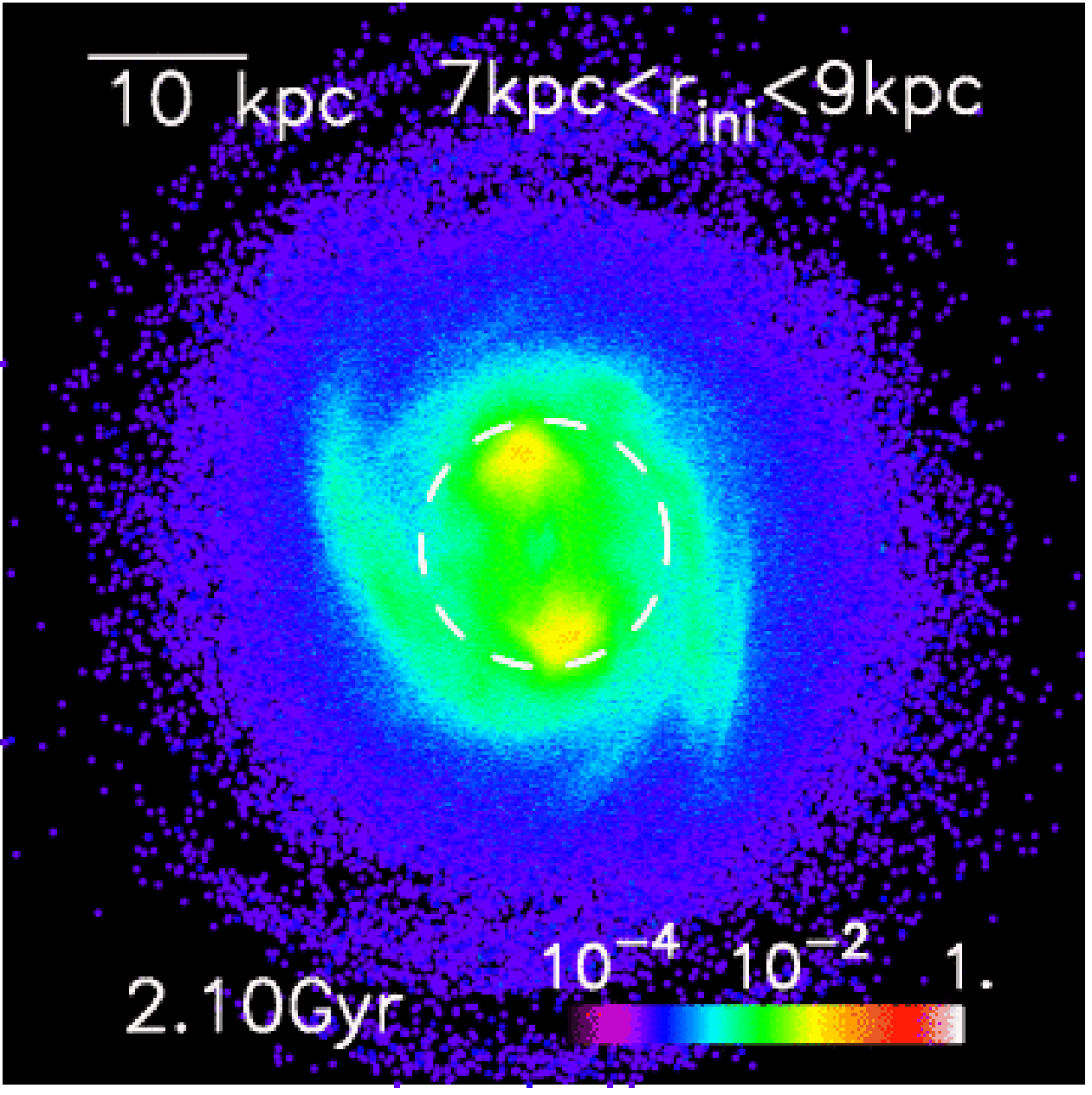}
\hspace{-0.2cm}
\includegraphics[width=3cm,angle=270]{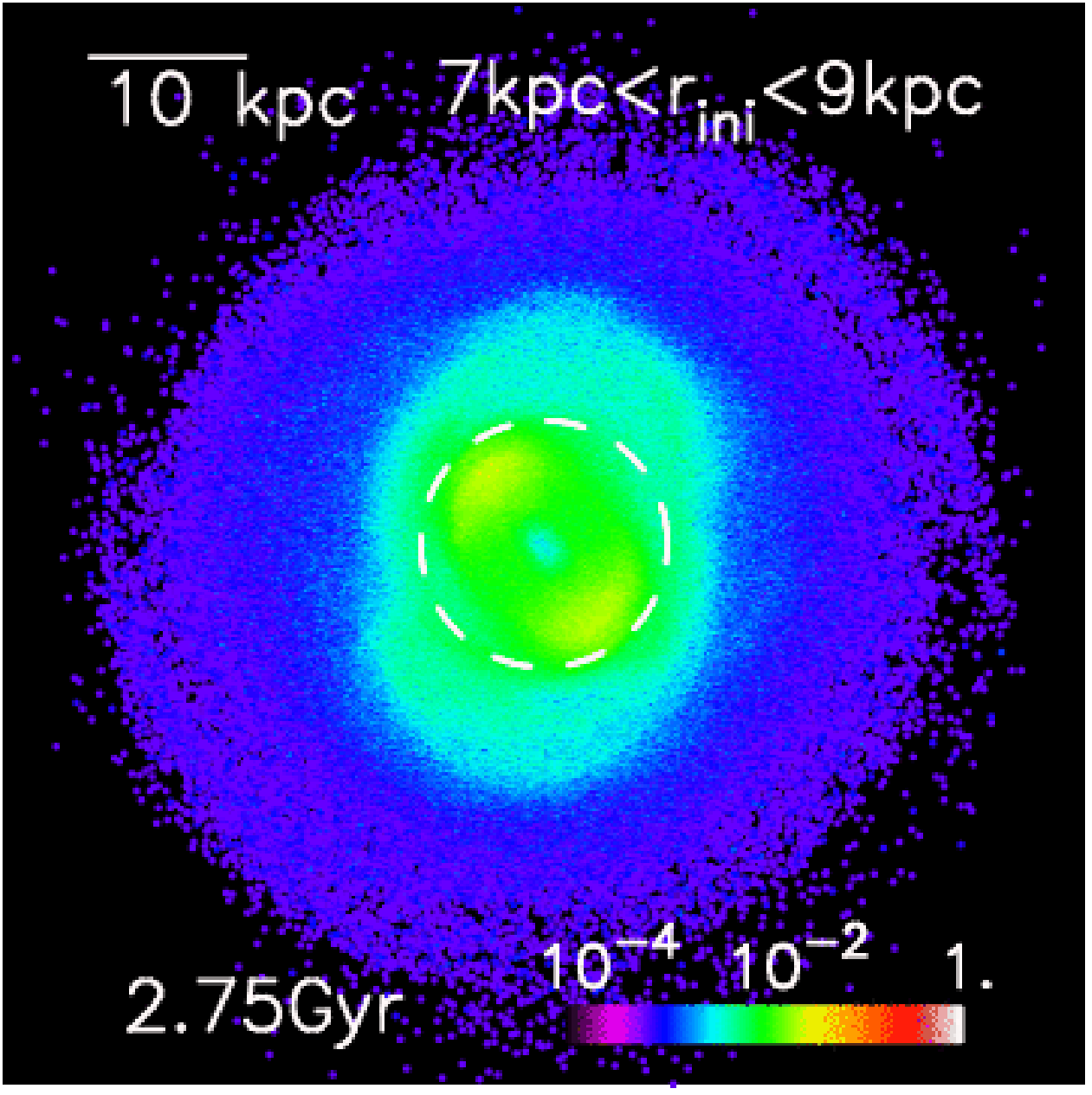}
\hspace{-0.2cm}
\includegraphics[width=3cm,angle=270]{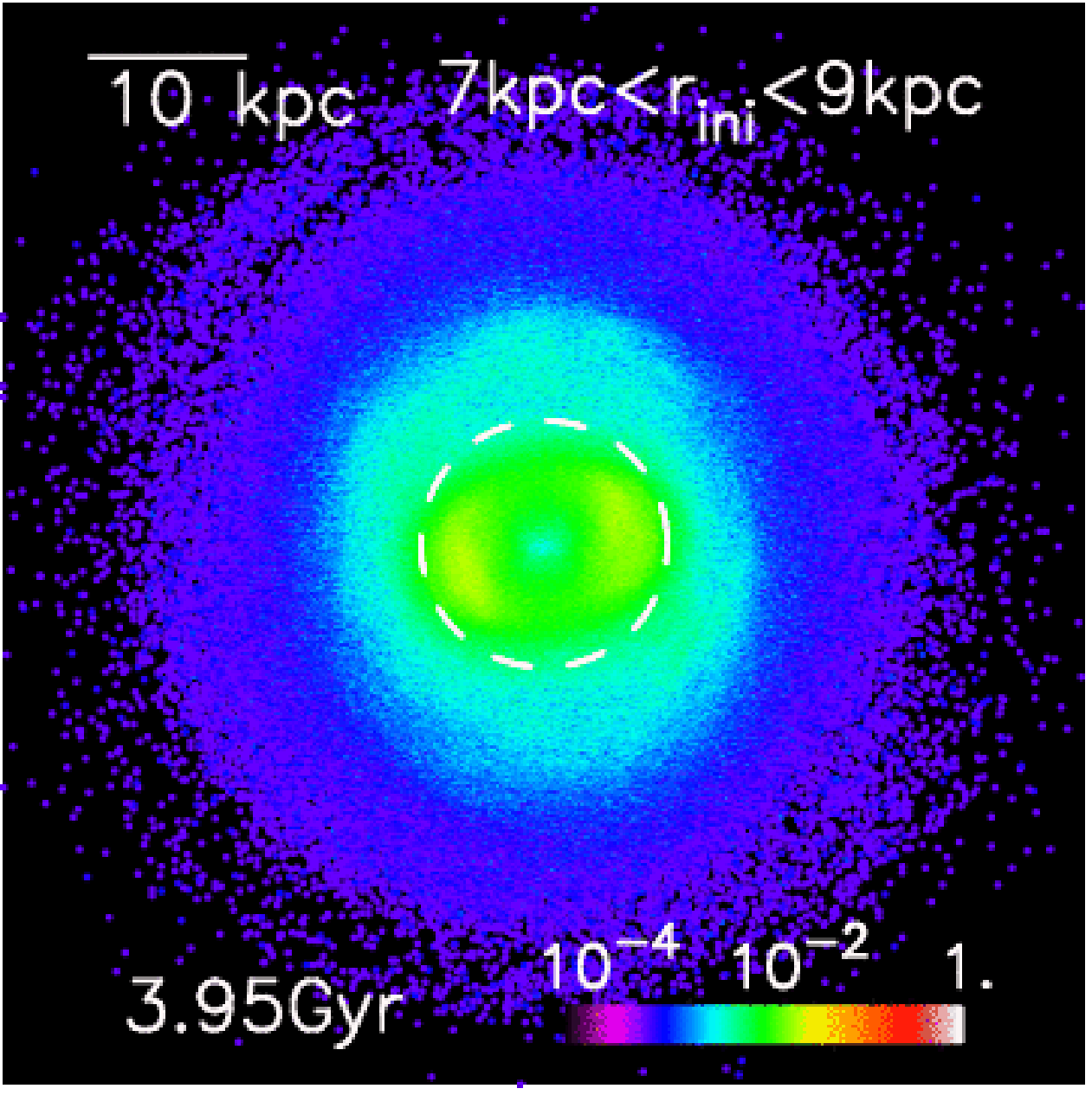}

\caption{Face-on density distribution of stars initially in a radial annulus 7~kpc$\le R \le$9~kpc ( in each panel, the average value, $R=8$~kpc, is indicated, by a dashed white circle).}
\label{mix}
\end{figure}

\begin{figure}
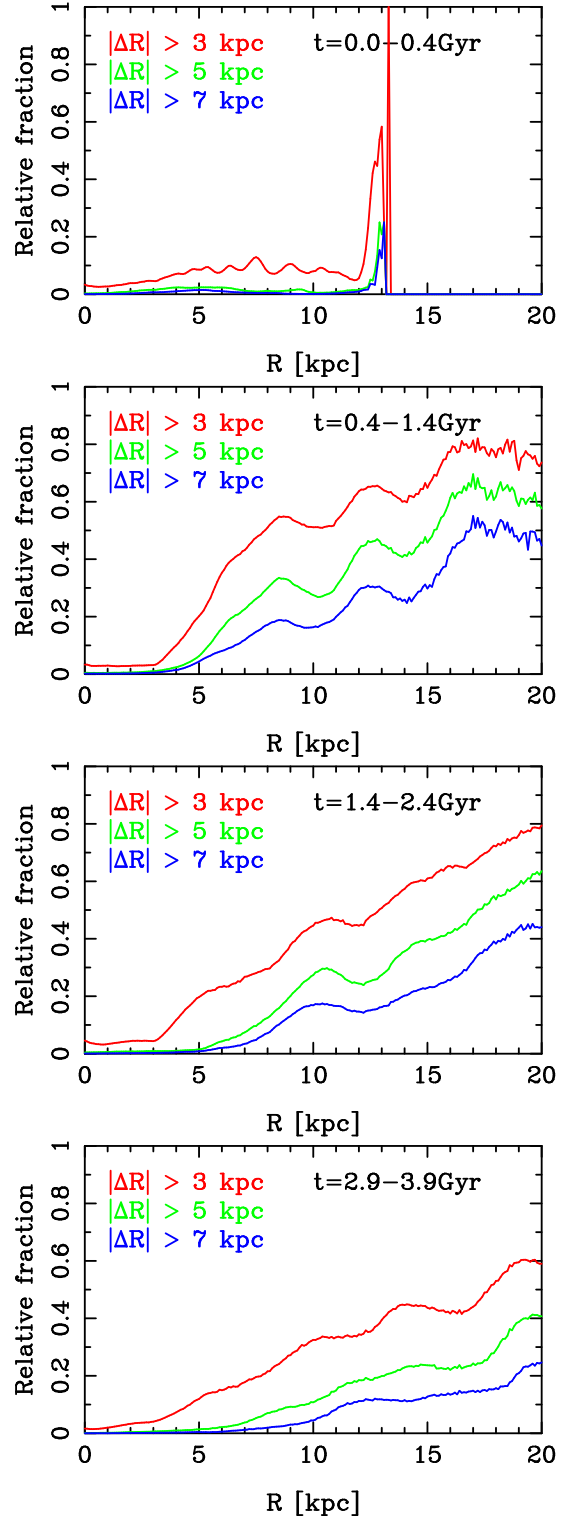

\centering
\includegraphics[width=5.cm,angle=270]{pmixbl_tini=001G_gS0_q1p8_BD0p10.ini_040.ps}
%\vspace{0.05cm}
\includegraphics[width=5.cm,angle=270]{pmixbl_tini=040G_gS0_q1p8_BD0p10.ini_140.ps}
%\vspace{0.1cm}
\includegraphics[width=5.cm,angle=270]{pmixbl_tini=140G_gS0_q1p8_BD0p10.ini_240.ps}
%\vspace{0.1cm}
\includegraphics[width=5.cm,angle=270]{pmixbl_tini=290G_gS0_q1p8_BD0p10.ini_390.ps}
\caption{Fraction of stars that migrate of $\mid \Delta R \mid > 3, 5, 7$~kpc from their location $R$ at the beginning of the time window, at four different time intervals (\emph{from top to bottom}): $t \in [0.-0.4]$~Gyr; $t \in [0.4-1.4]$~Gyr; $t \in [1.4-2.4]$~Gyr; $t \in [2.9-3.9]$~Gyr. Bins are 0.1~kpc in width. }
\label{migrprob}
\end{figure}

%\begin{figure}
%\centering
%\includegraphics[width=5.cm,angle=270]{pmixvstime_DT=1Gyr_10kpc.ps}
%\caption{Fraction of stars that have migrated by more than $\Delta R=7$~kpc (blue histogram and white dashed line) from the radius $R=10$~kpc in the time interval [$t-1~$Gyr, $t$]. For comparison, the fraction of stars that have migrated of more than $\Delta R=5$~kpc in the same time interval is also shown (green histogram and dashed grey line).} 
%\label{migrvstime}
%\end{figure}

\subsection{Radial migration in barred galaxies: how and when?}\label{when}
Due to the high number of particles employed and the reduced noise in the gravitational force field, it takes about 0.8~Gyr to see the appearance of asymmetric structures in the stellar disk. Indeed, only at this epoch strong $m=2,4$ asymmetries, associated with a bar and spiral structure, develop, maintaining a nearly constant strength for about 1 Gyr (see Fig.\ref{asym} and \ref{smaps}). The decline that follows in the bar and spiral arms strength at about t=2 Gyr is associated to a conspicuous vertical buckling of the bar, and the subsequent formation of a boxy/peanut shaped bulge, as observed many times in N-body simulations \citep{cosan81, com90, mar06, atha08}. It is worth noting that at later times the galaxy is still barred, but with isophotes rounder than those characterizing the phase of high bar strength (Fig.~\ref{smaps}).\\
Motivated by the results of Fig.~\ref{asym}-\ref{smaps}, in the following, when discussing the disk evolution, we will refer to some distinct temporal phases: early phase ($t <$ 0.8~Gyr); bar growth (very rapid phase, from $t=0.8$~Gyr to $t=1$~Gyr); strong bar activity (corresponding to the time of high, nearly constant bar strength, i.e. 1~Gyr$\le t \le$ 2 Gyr); weak bar ($t \ge$ 2 Gyr).  \\

At the epoch of bar formation, and during the whole phase of strong bar activity, stars in the disk experience a significant spatial redistribution, with a  probability of migration  maximal at the bar resonances, in agreement with the findings of \citet{min10}. Fig.~\ref{mix}, for example, shows the spatial distribution over time of stars that at the beginning of the simulation were in the radial annulus  7~kpc~$\le R <9$~kpc, $R$ being the distance from the galaxy center. This region includes the corotation radius at early times (the bar corotation is initially located at $R\sim8$~kpc, but it moves rapidly outwards to $R\sim10$~kpc). Before asymmetric structures appear in the disk, stars initially belonging to this annulus maintain a radially symmetric distribution in the plane. As soon as the bar starts to develop, the spatial distribution of stars becomes asymmetric: elongated along the bar axis in the inner regions, and showing signs of spiral structures in the outer regions.  Stars initially confined in this annulus can migrate toward the inner and outer disk. In particular, stars  moving outwards do so via spiral arms, which form together with the bar. These arms are visible during the phase of strong bar activity, fading away only 2 Gyr after the beginning of the simulation, when the bar strength declines (note that the length of the bar is always $\le$ 9 kpc, as it can be deduced by the location of the two density maxima --yellow clumps in Fig. ~\ref{mix} for $t > 1.1$~Gyr -- which identify its end).\\
As we will discuss in the next sections, the way radial migration occurs, through these asymmetric structures,  has an important impact on the spatial redistribution of metals and its variation with time.\\

Fig.~\ref{mix} clearly shows that in these simulations radial migration is initiated by the growth of the bar, at $t\sim$~0.8 Gyr. But for how long does it last? Is radial migration only a transient phenomenon, related to the formation of the bar, or does it last longer than few hundreds of Myrs? To answer this question, we have evaluated the fraction of stars that migrate of more than $\mid \Delta R\mid=\mid R(t_f)-R(t_i)\mid=3, 5, 7$~kpc from their location at the beginning of the time window $[t_i,t_f]$, with $[t_i,t_f]=[0.,0.4], [0.4, 1.4],  [1.4, 2.4]~\rm{and}  ~[2.9, 3.9]~\rm{Gyr}$, respectively (Fig.~\ref{migrprob}). Before asymmetries appear in the stellar disk, the fraction of strong migrators ($ \mid{\Delta R\mid} > 5$~kpc) is null all over the disk, except at the outer edge, where an initial relaxation from the initial conditions leads the disk to slightly expand with respect to the truncation radius initially at $R=$13~kpc. The fraction of migrators drastically changes in the following Gyr, from $t=0.4$ to $t=1.4$~Gyr: the whole outer disk is affected by a significant redistribution of stars, with up to 30$\%$ of stars in this phase moving by more than 7~kpc. There are two radii where the relative fraction of migrators is higher: at $R=8.5$~kpc and $R=12.5$~kpc, corresponding respectively, to the corotation and outer Lindblad resonance at that time. If migration is strong at the epoch of bar growth, it is still significant at later times: in the time interval $t=1.4-2.4$~Gyr, mostly corresponding to the phase of strong activity of the bar, 20$\%$ of stars migrate by more than $\mid\Delta R\mid > 7$~kpc from the corotation (located at $R=10$~kpc at this epoch). A weaker, but still not negligible migration, is present also at later times ($t=2.9-3.9$~Gyr), when the bar strength has diminished, after the vertical buckling.\\
We note that the distribution of the fraction of migrators versus radius is significantly different from that predicted by \citet{bird12} (see Fig.~4 in that paper). While in their isolated galaxy models, this fraction follows the surface density profile of the disk, decreasing with radius, our models predict the opposite trend, with a very low probability of strong migration  in the inner disk -- where the bar traps a large fraction of stars -- and  local maxima in the outer disk associated to the location of the bar resonances. This discrepancy is probably related to the fact that their isolated disks are stable against bar formation, and, as a consequence, the effect of scattering at the bar resonances is not taken into account.

\subsubsection{Comparison with Minchev et al results}\label{minres}

The results discussed in Sect.~\ref{when} agree with those already presented in a number of previous works, concerning the study of radial migration in barred galaxies \citep{bru11, min10, min11, min12}. In particular, as discussed in several works by Minchev and collaborators, we can confirm that radial migration in these simulations is not only associated to the epoch of bar growth:  migration is indeed initiated at that time, but it lasts for all the phase of strong bar activity, and also, at lower levels, later on (for $t\ge$~2Gyr).
Moreover, the stability of the simulated disks over several hundreds of Myrs (see Fig.~\ref{asym} and \ref{smaps}) guarantees that radial migration observed in these simulations is not a consequence of initial violent instabilities in the disk, as recently suggested by \citet{rok12} to be the case for the simulations analyzed by \citet{min11,min12}.

\begin{figure}
\centering
\includegraphics[width=5.cm,angle=270]{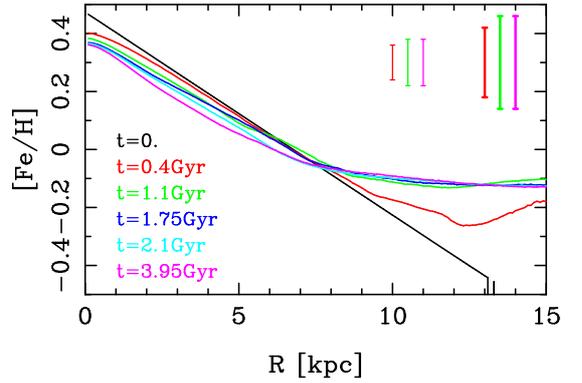}
\caption{Azimuthally averaged metallicity profile of the stellar disk at different times, as shown in the legend. The initial metallicity gradient is $\Delta_{\rm{[Fe/H]}}=-0.07$~dex/kpc. Metallicity dispersions at three different times (see corresponding colors), and at two different radii ($R=0$, thin bars; $R=8~$kpc, thick bars) are shown in the top-right part of the plot.}
\label{metprof}
\end{figure}

\begin{figure}
\centering
\includegraphics[width=3cm,angle=270]{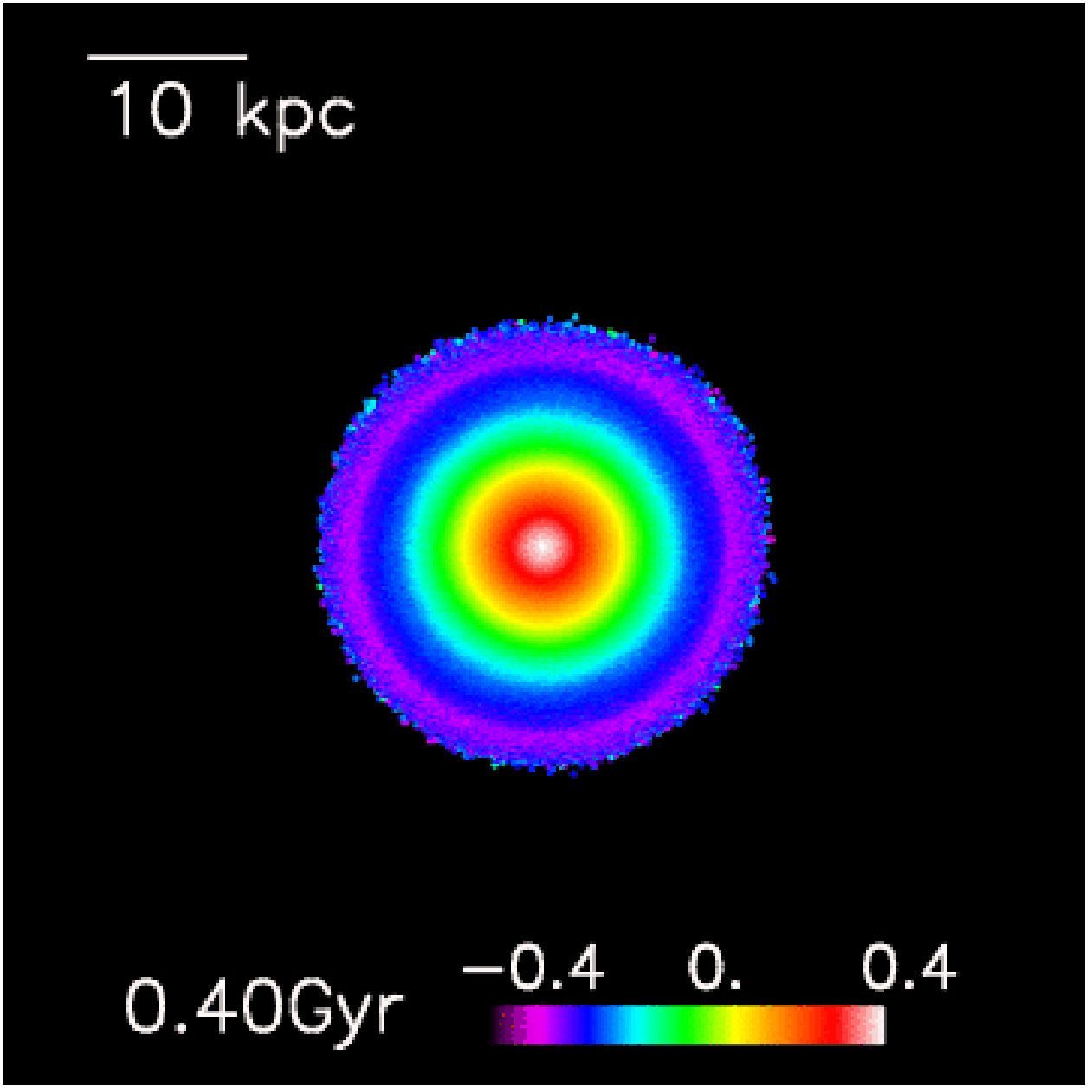}
\hspace{-0.2cm}
\includegraphics[width=3cm,angle=270]{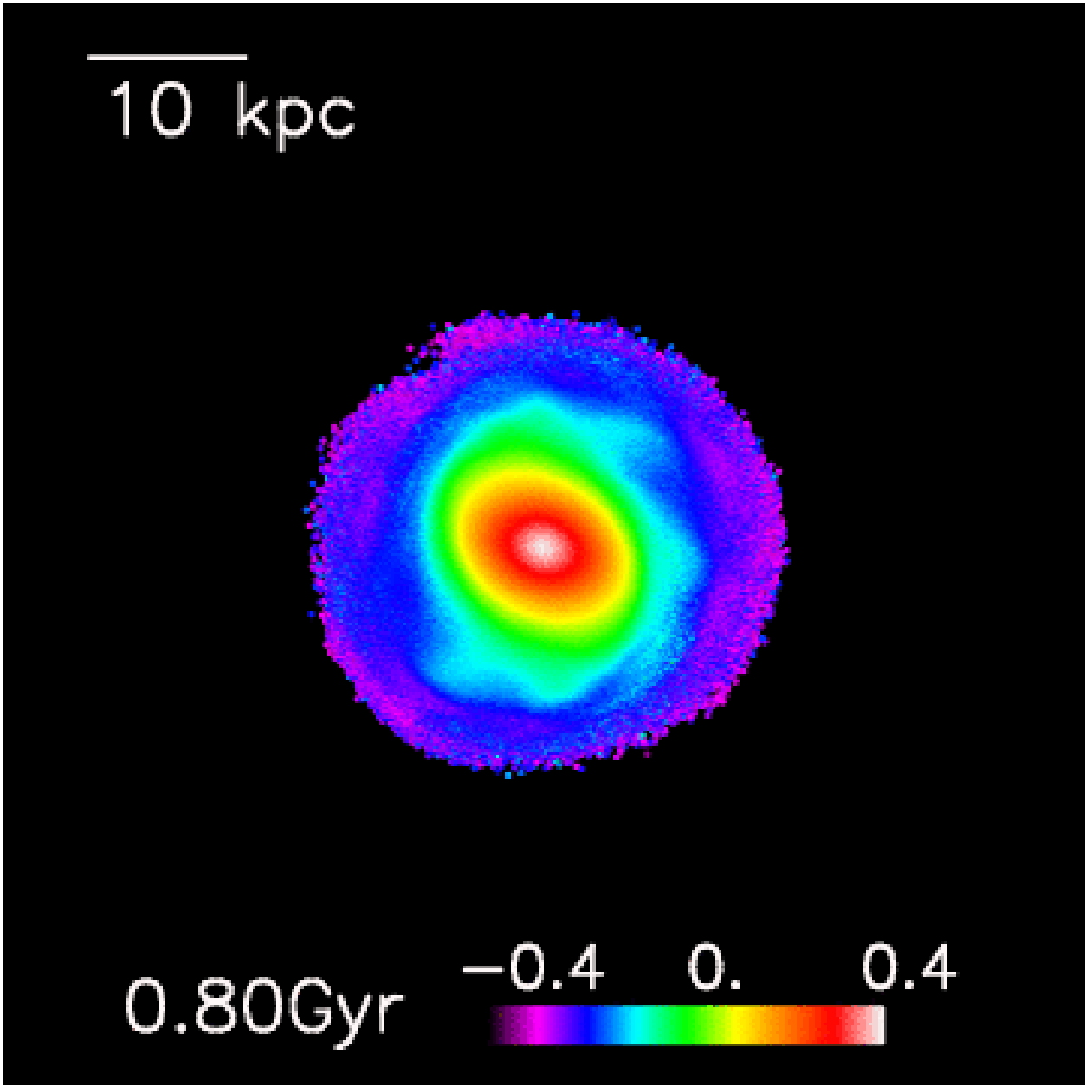}
\hspace{-0.2cm}
\includegraphics[width=3cm,angle=270]{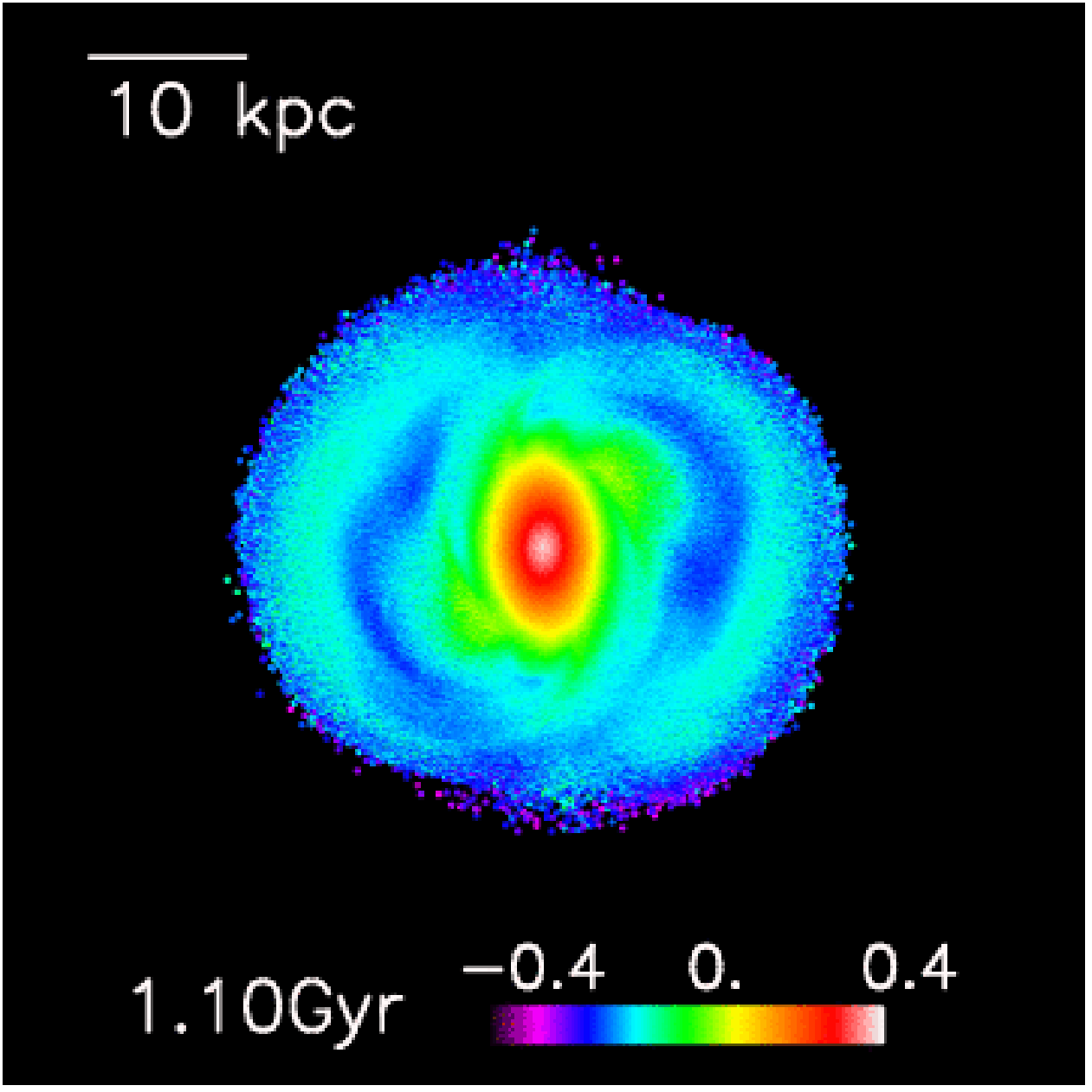}

\includegraphics[width=3cm,angle=270]{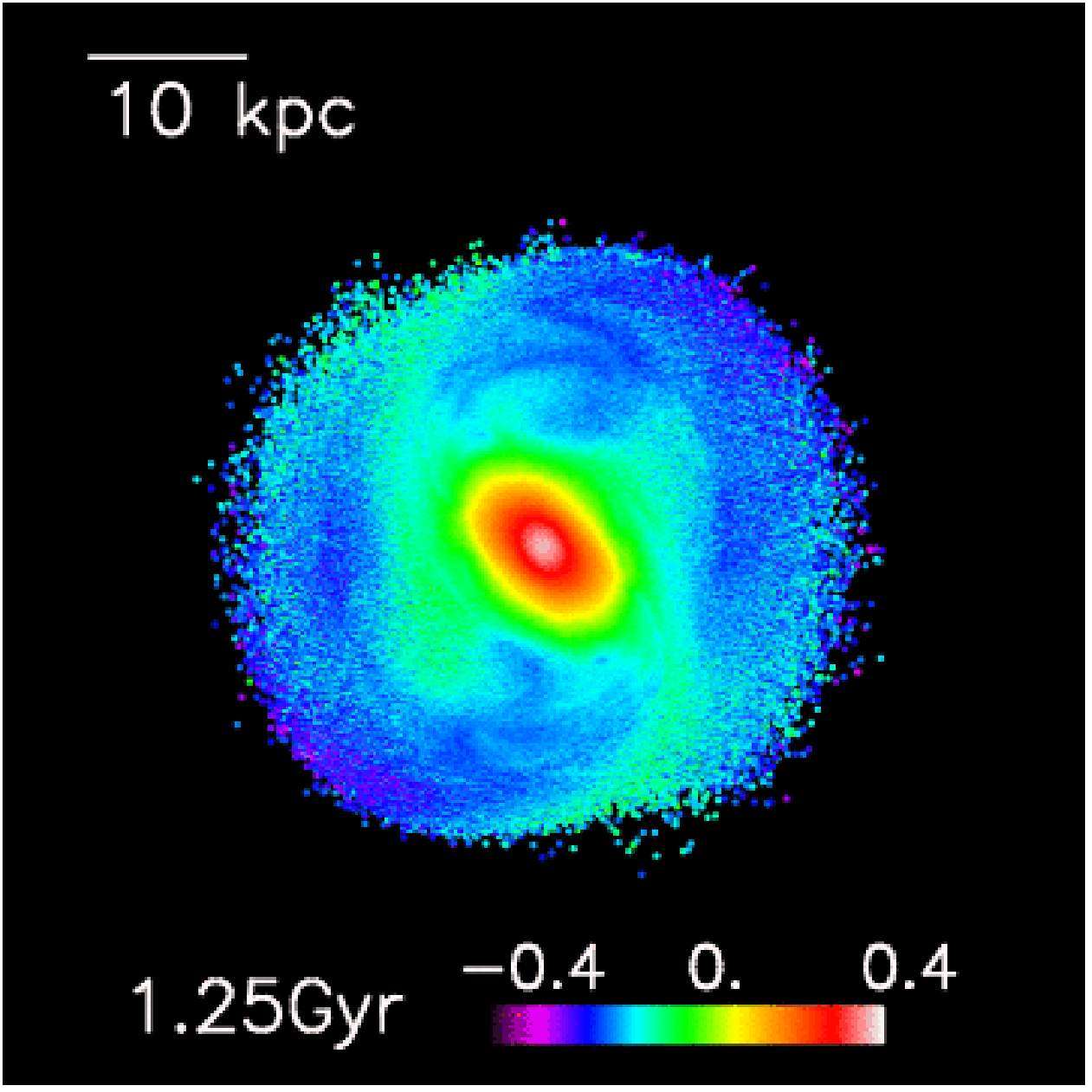}
\hspace{-0.2cm}
\includegraphics[width=3cm,angle=270]{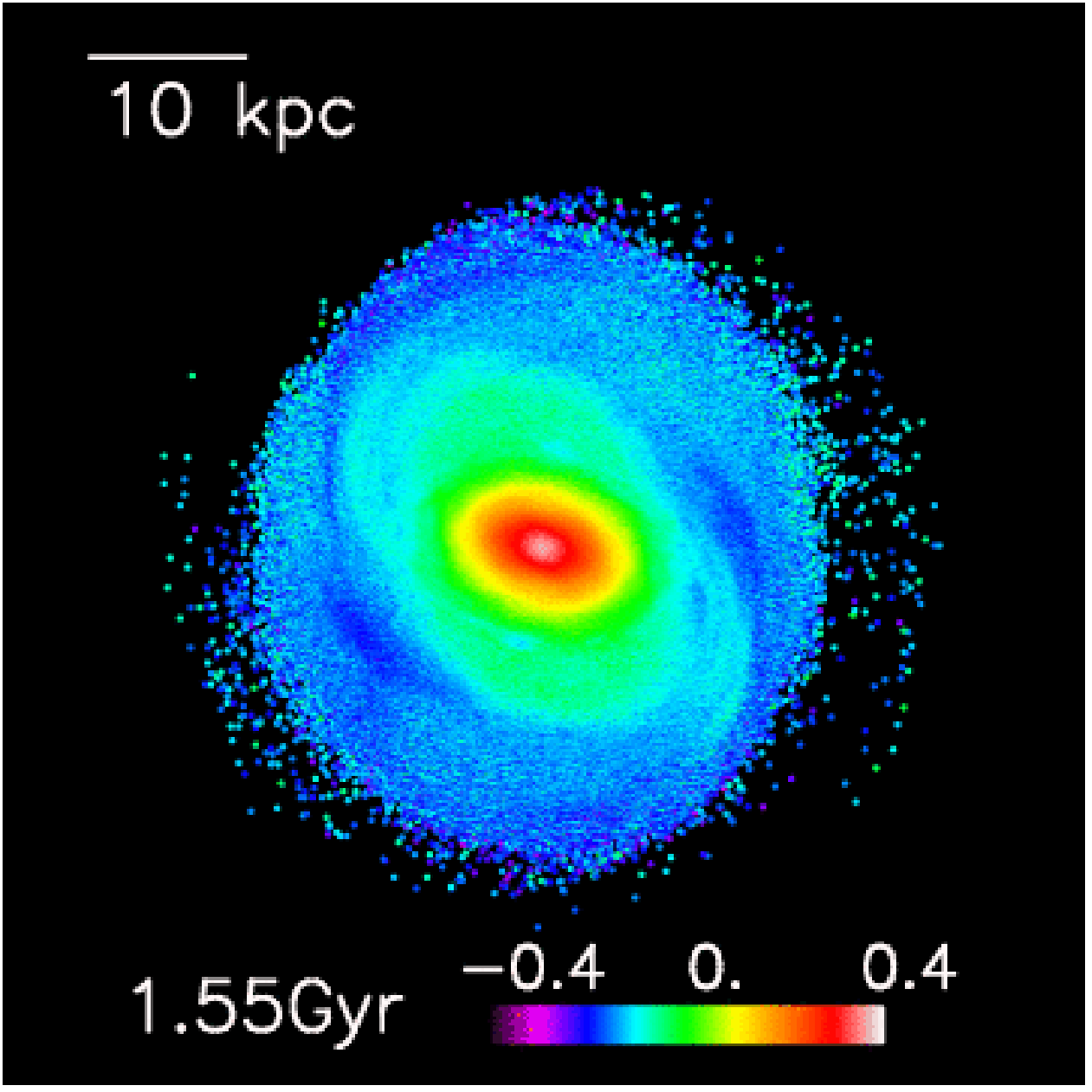}
\hspace{-0.2cm}
\includegraphics[width=3cm,angle=270]{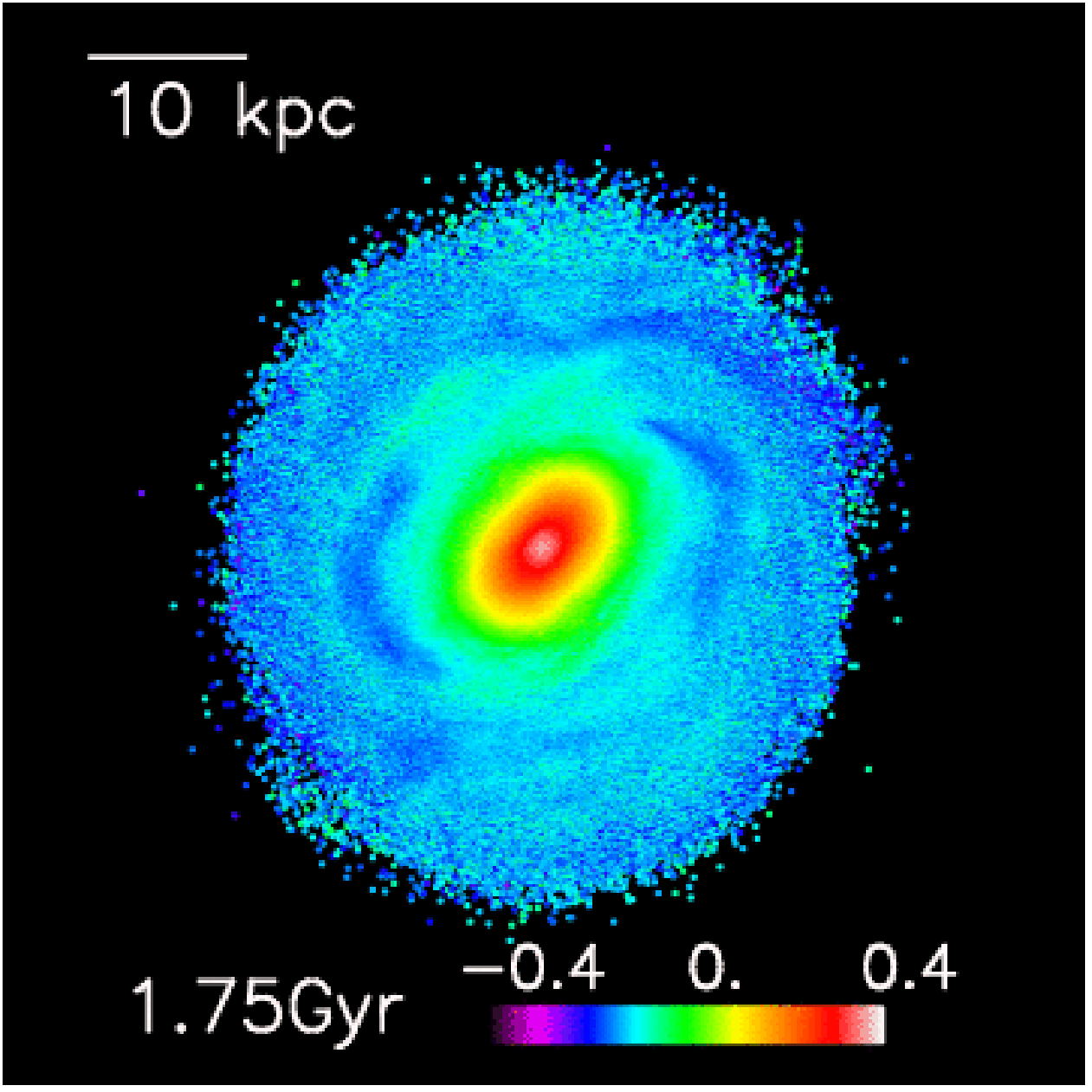}

\includegraphics[width=3cm,angle=270]{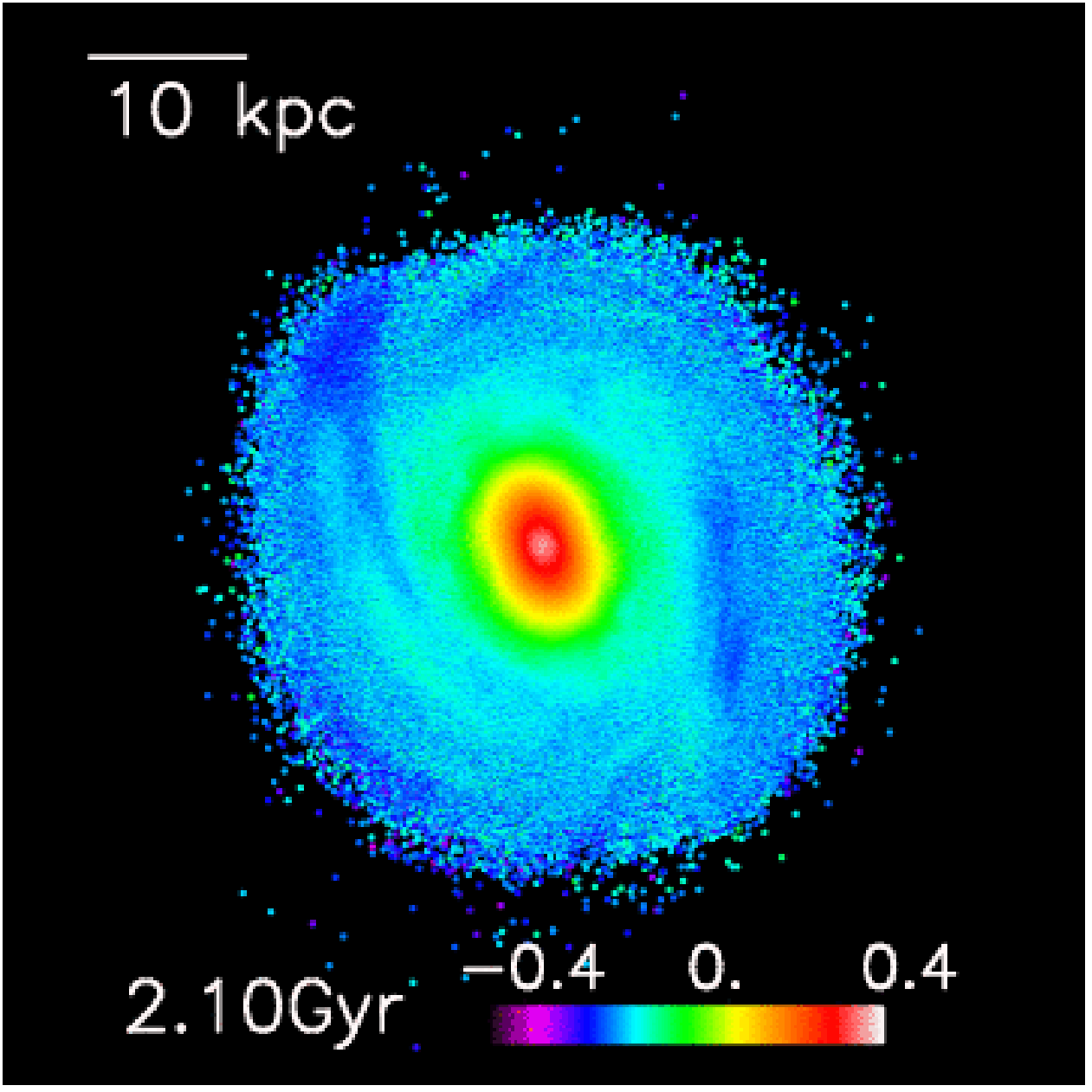}
\hspace{-0.2cm}
\includegraphics[width=3cm,angle=270]{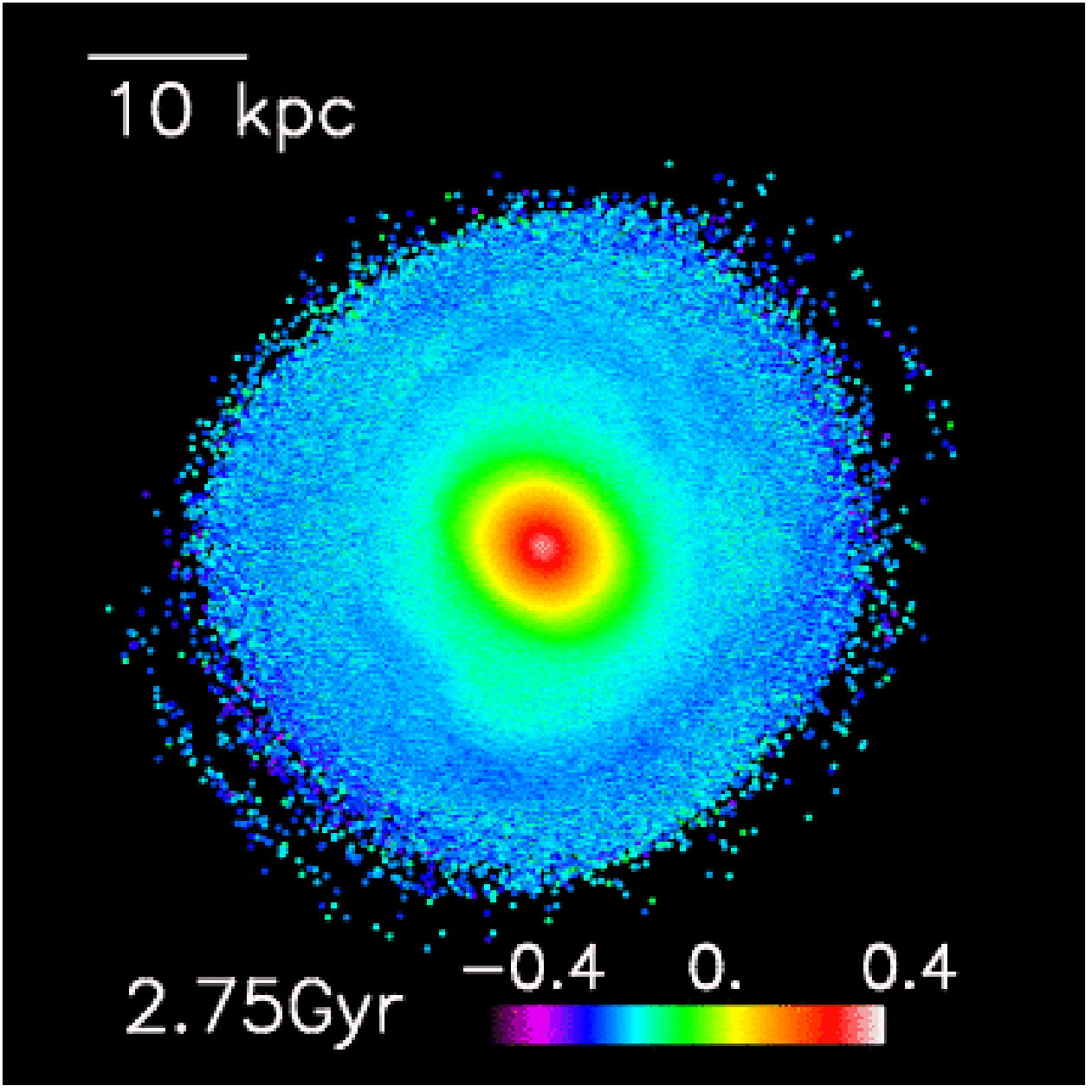}
\hspace{-0.2cm}
\includegraphics[width=3cm,angle=270]{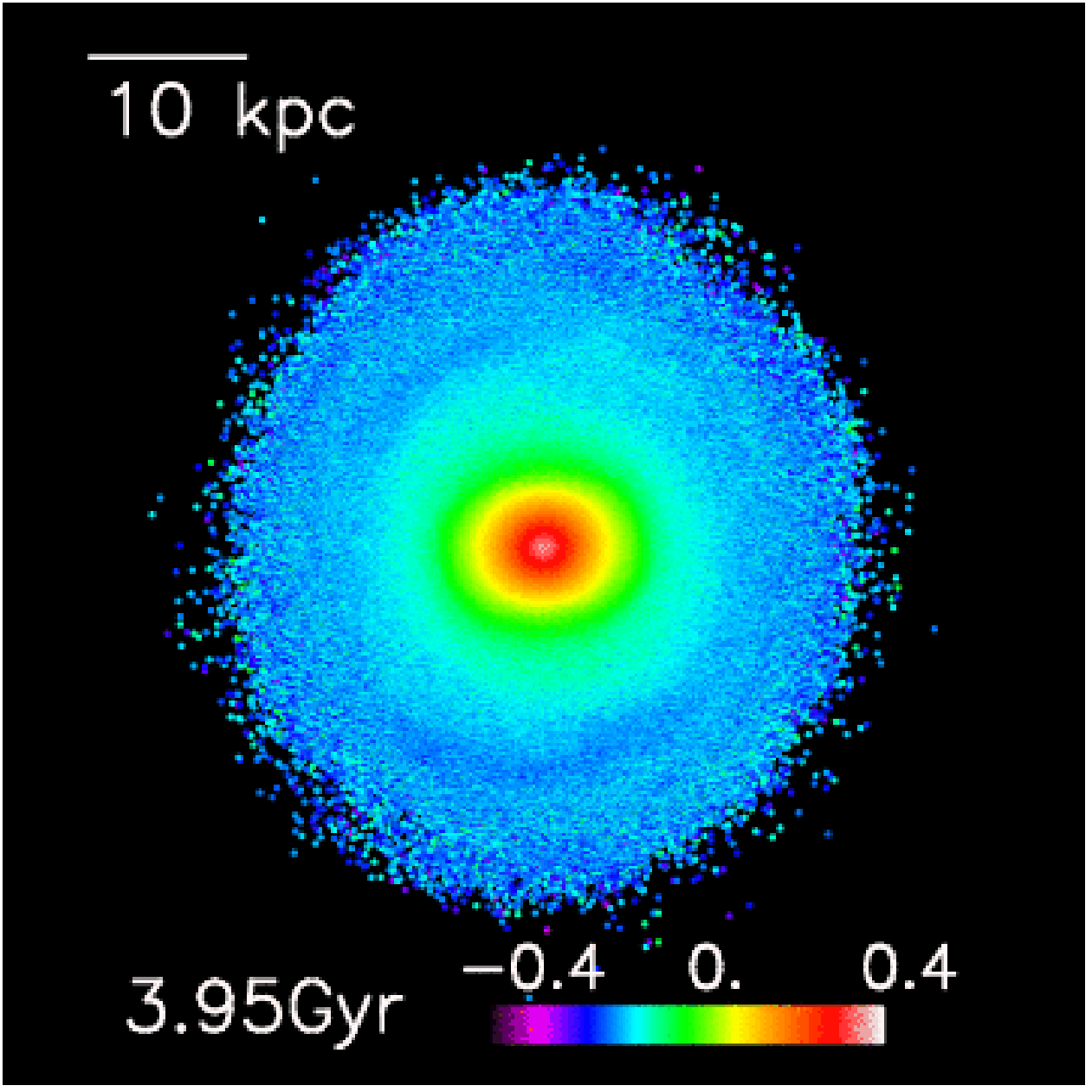}

\caption{2D [Fe/H] maps of the stellar disk, seen face-on, at different times.  Each panel is 70 kpc$\times$70 kpc in size.%, corresponding to $t=$~0.4, 0.8, 1.1, 1.25, 1.55, 1.75, 2.01, 2.75, 3.95~Gyr.
}
\label{FeHmaps}
\end{figure}

\begin{figure}
\centering
\includegraphics[width=3.cm,angle=270]{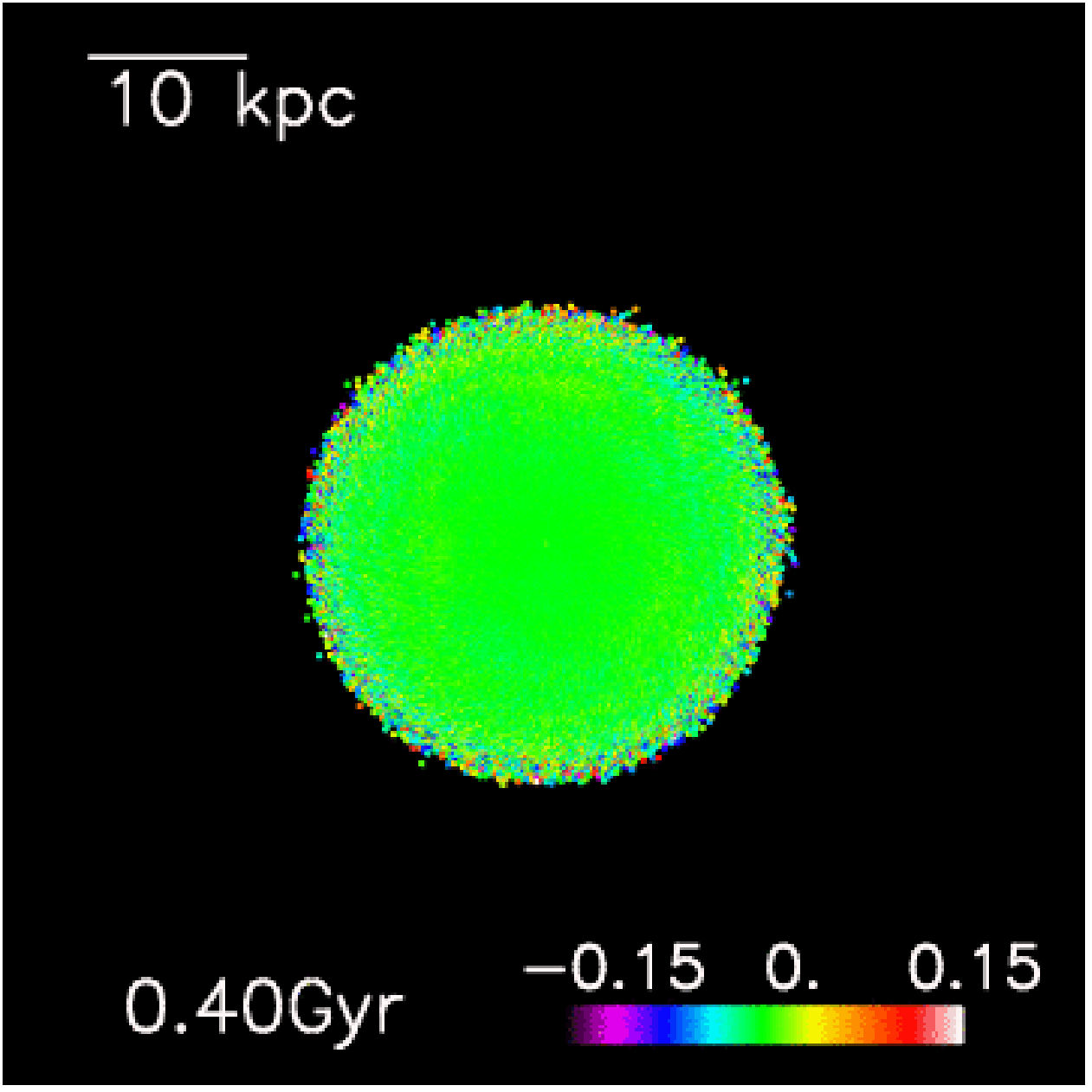}
\hspace{-0.2cm}
\includegraphics[width=3.cm,angle=270]{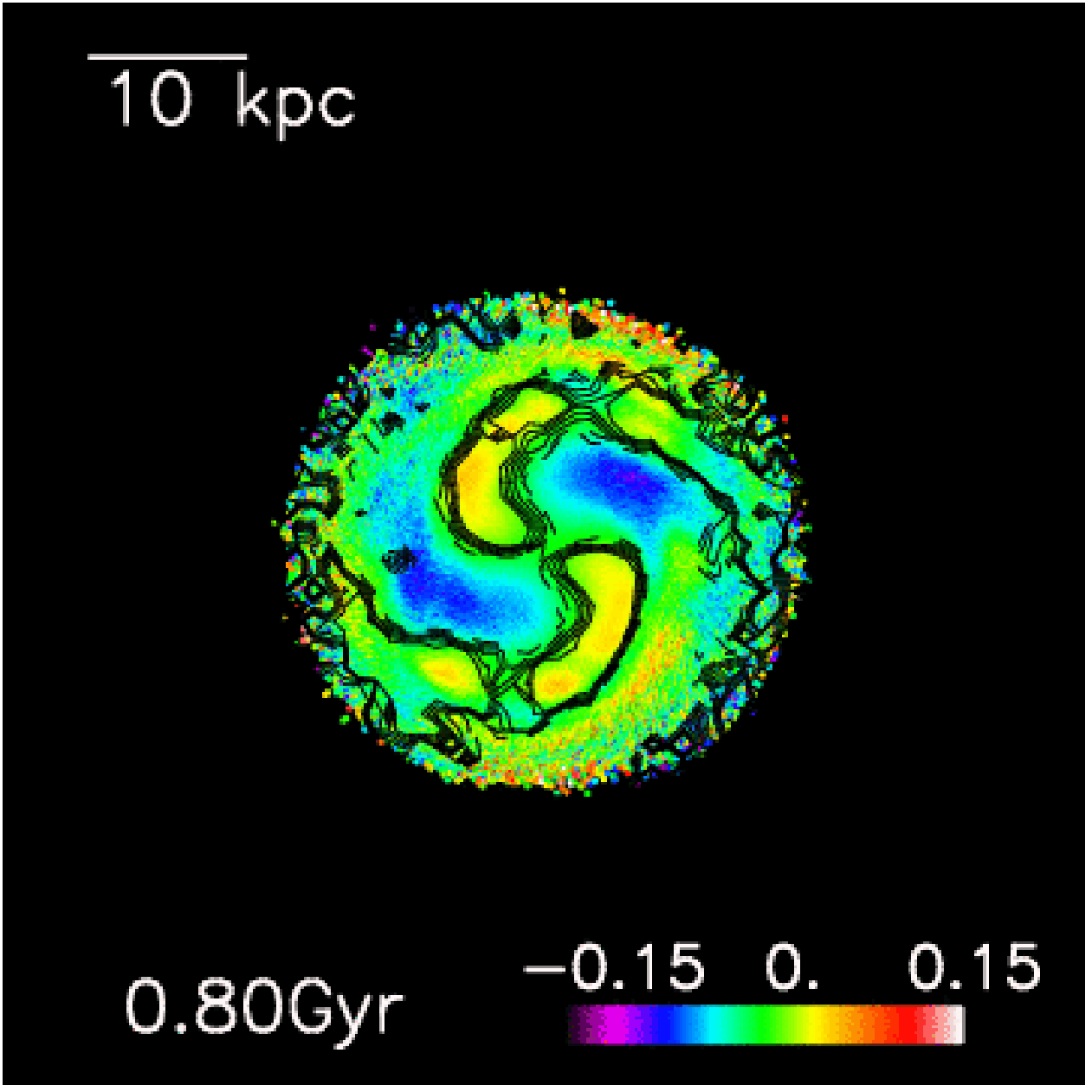}
\hspace{-0.2cm}
\includegraphics[width=3.cm,angle=270]{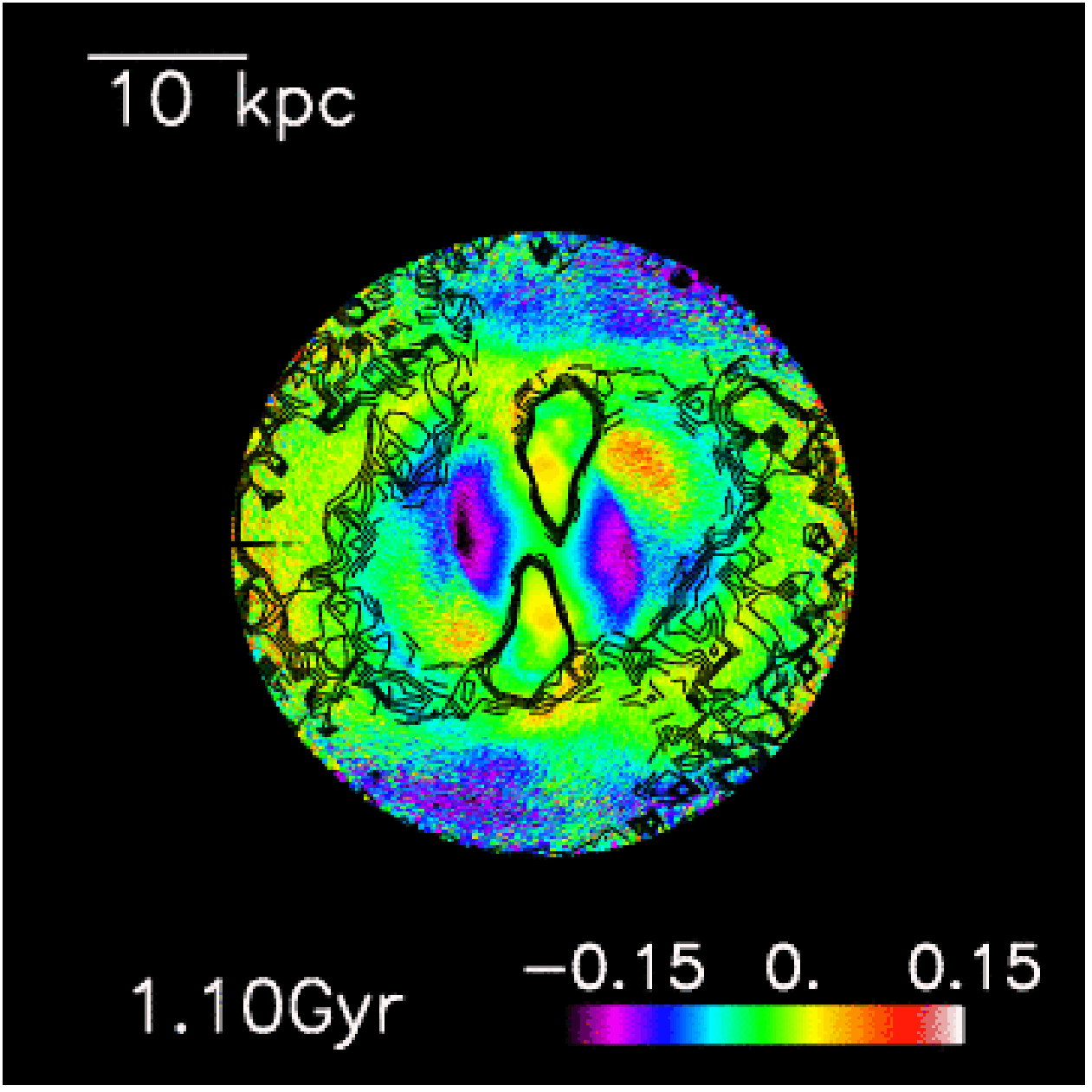}

\includegraphics[width=3.cm,angle=270]{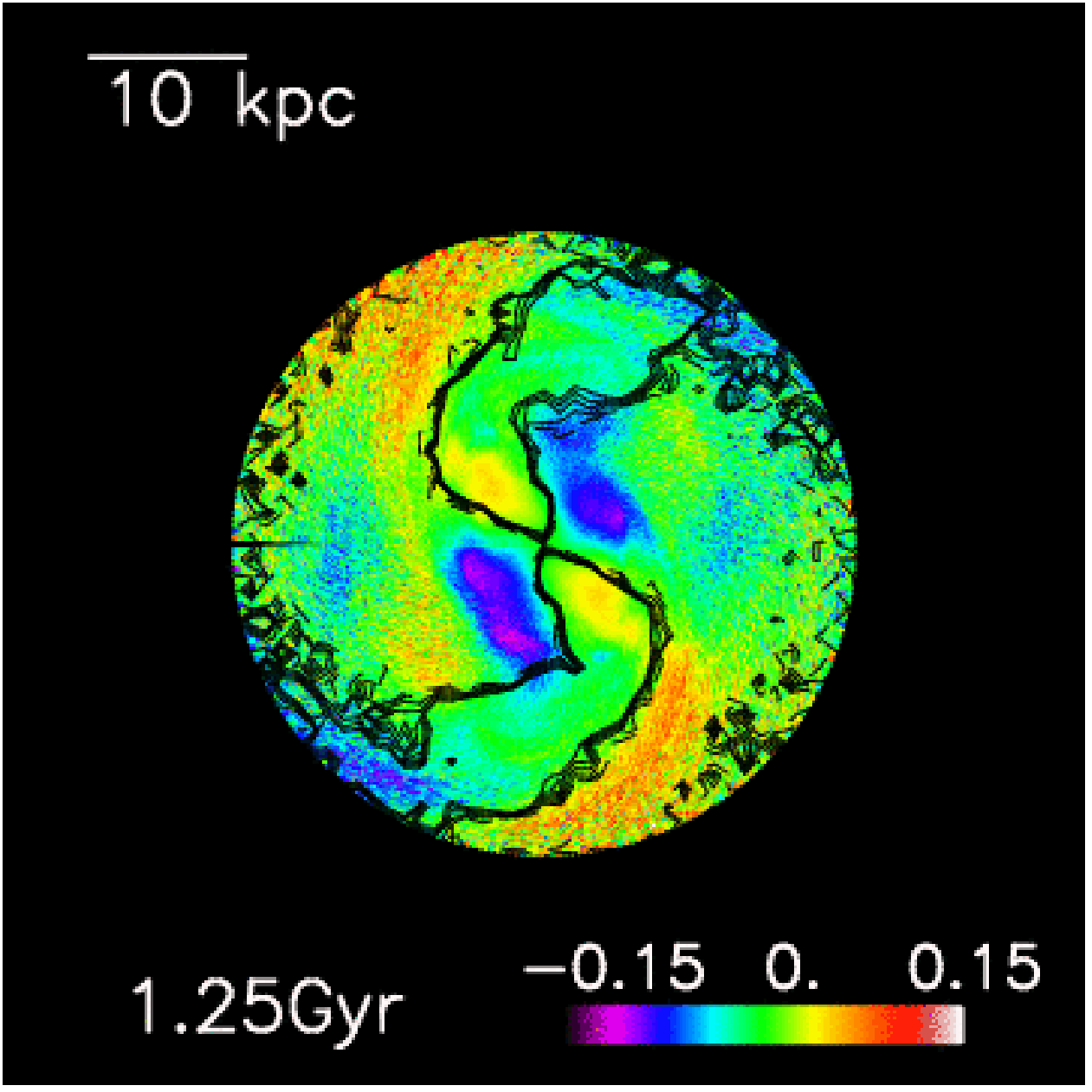}
\hspace{-0.2cm}
\includegraphics[width=3.cm,angle=270]{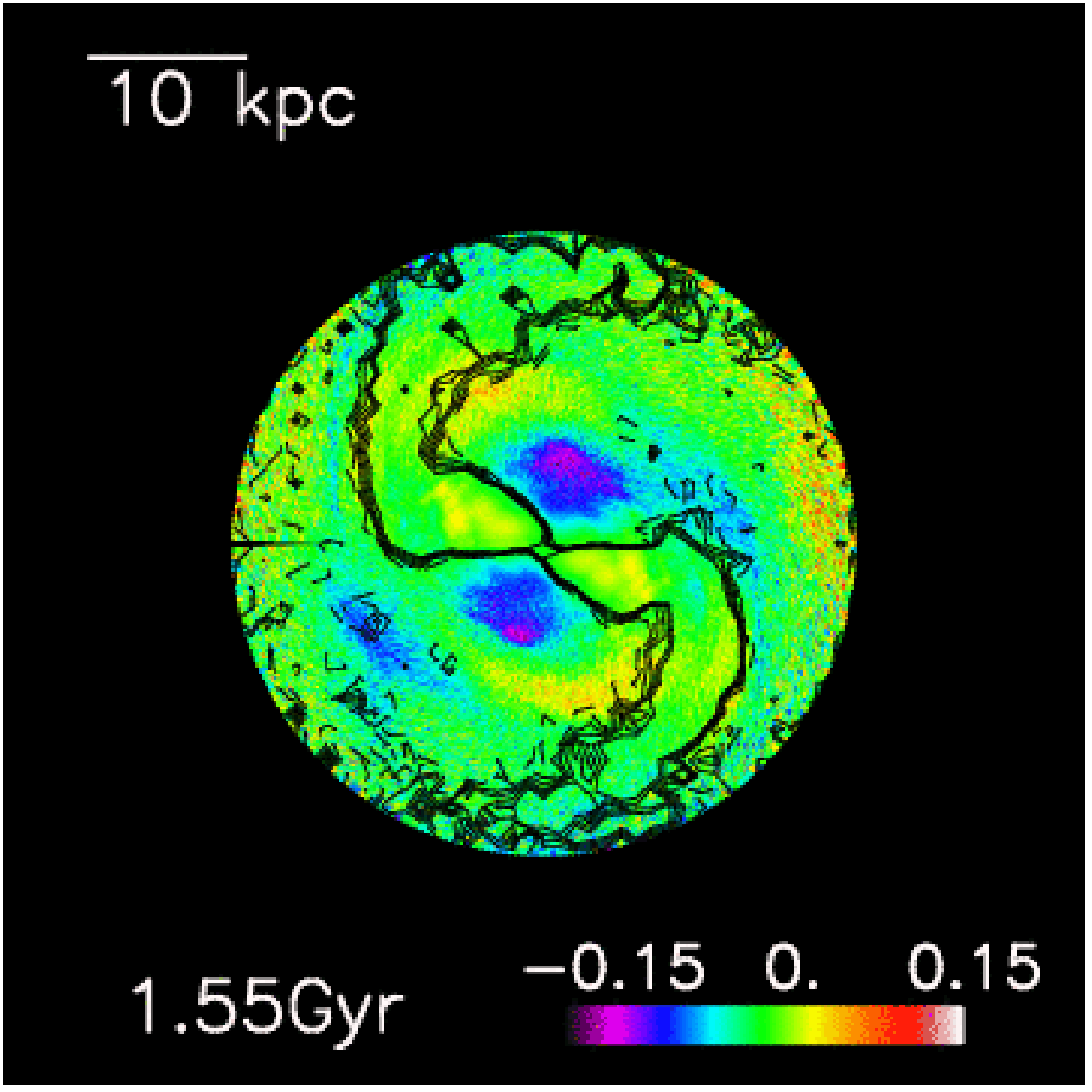}
\hspace{-0.2cm}
\includegraphics[width=3.cm,angle=270]{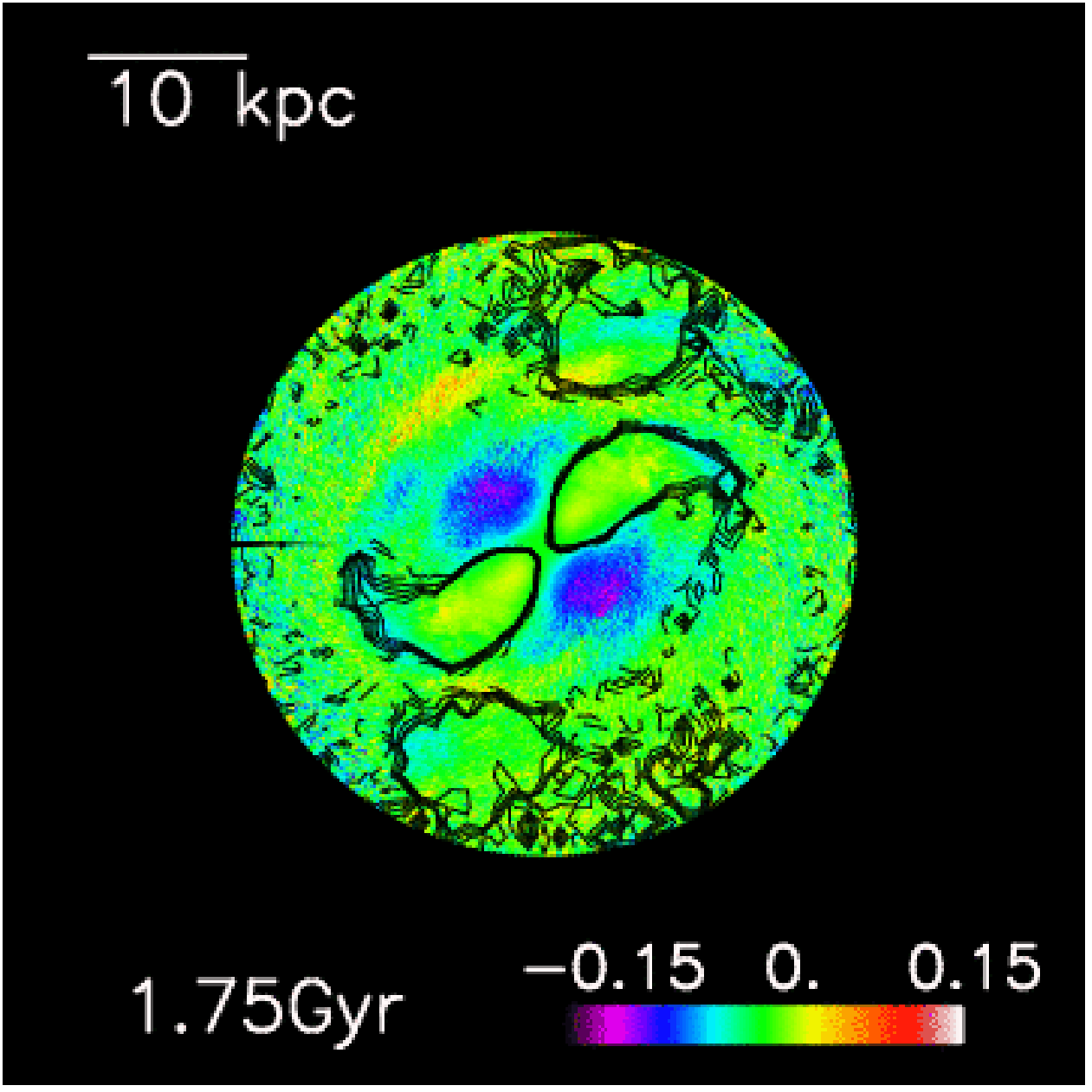}

\includegraphics[width=3.cm,angle=270]{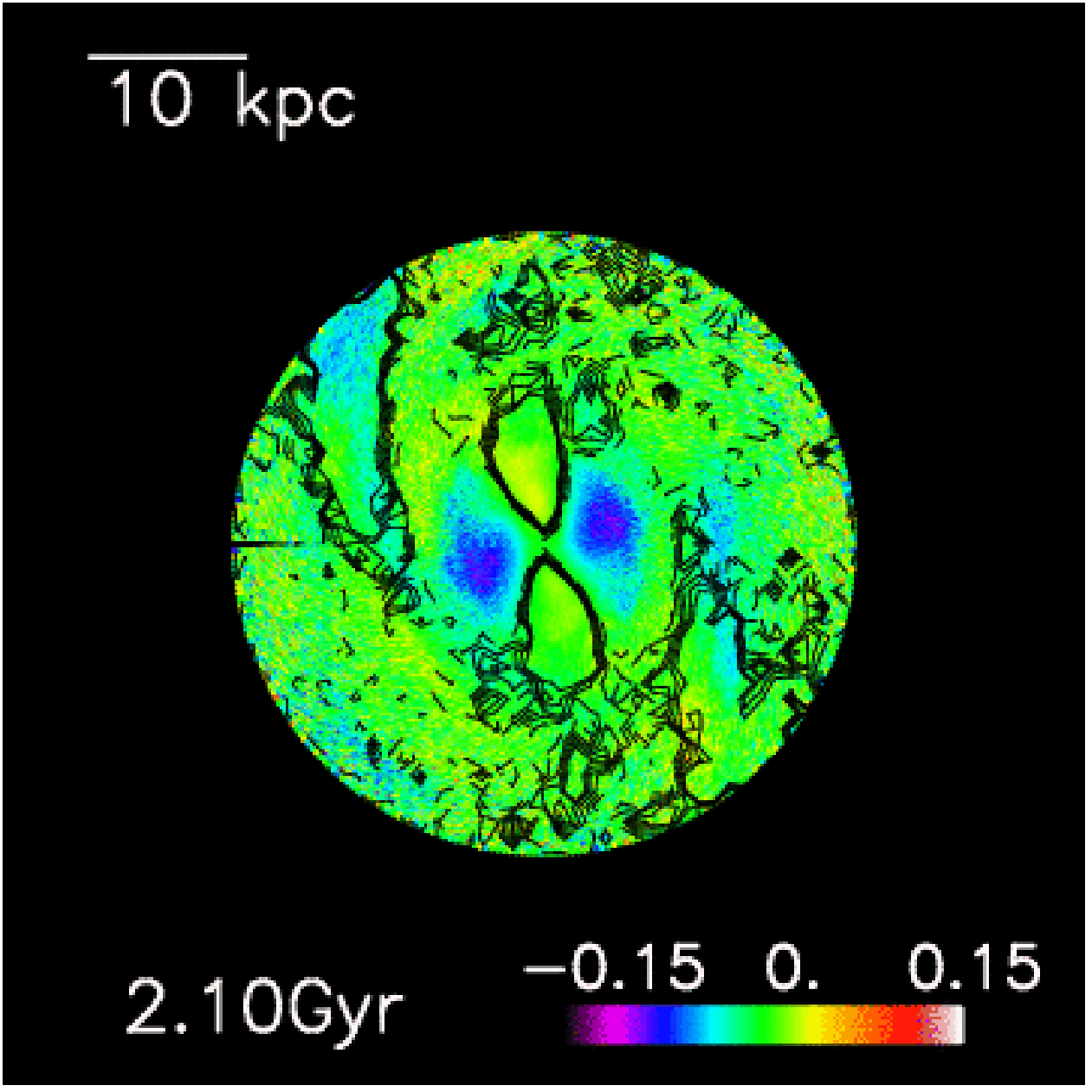}
\hspace{-0.2cm}
\includegraphics[width=3.cm,angle=270]{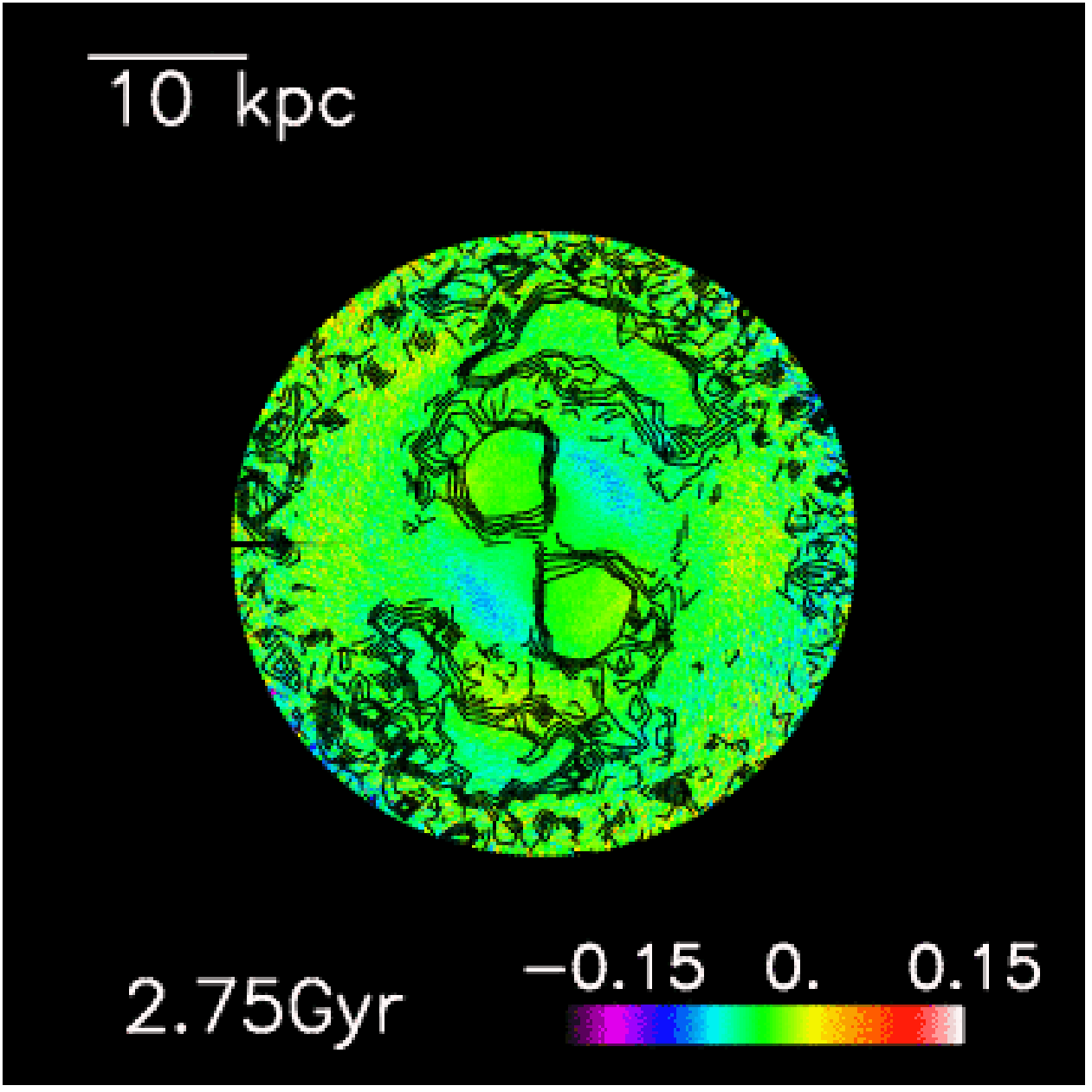}
\hspace{-0.2cm}
\includegraphics[width=3.cm,angle=270]{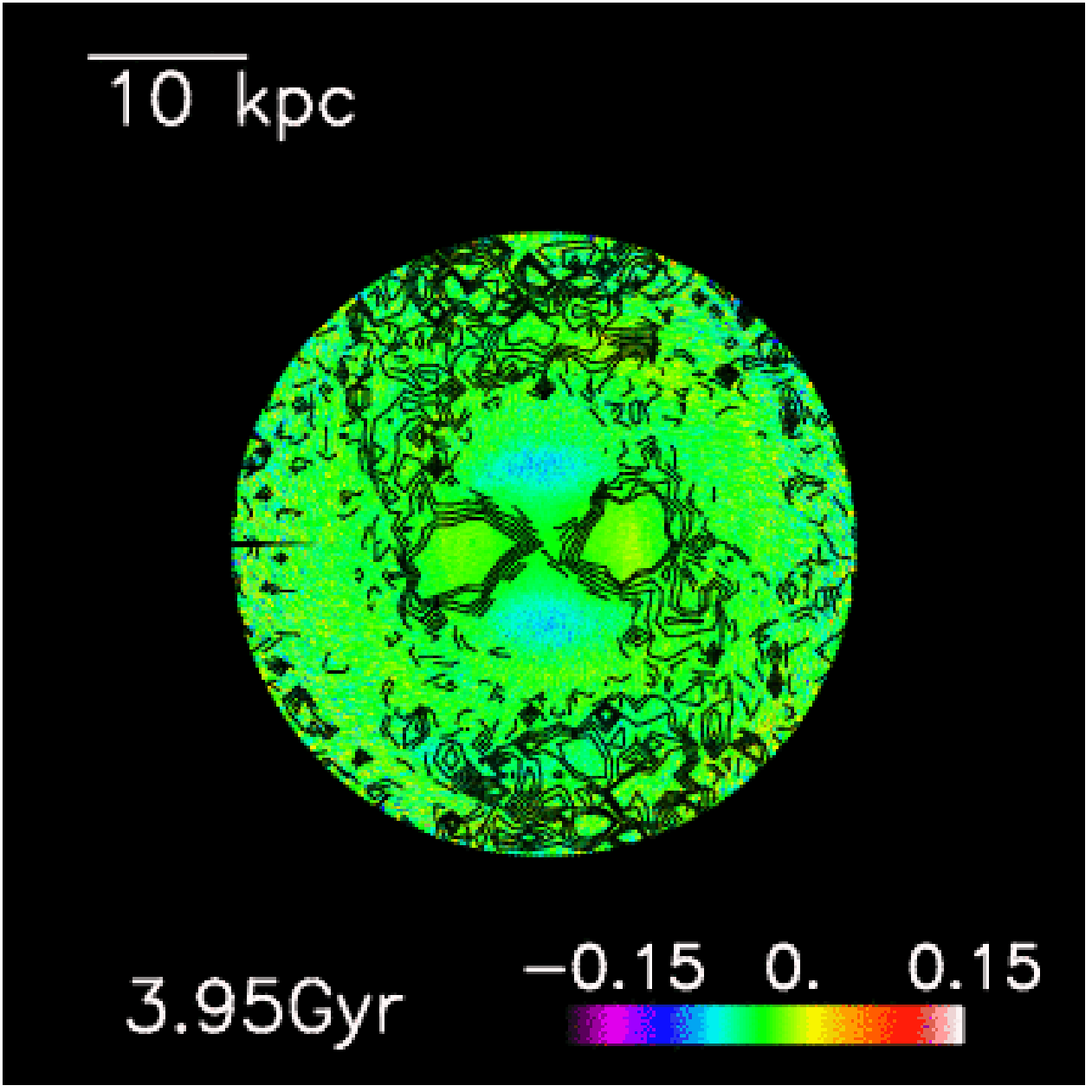}

\caption{Maps of azimuthal metallicity variations, $\delta_{\rm{[Fe/H]}}$ (dex), at the different times shown also in Fig.\ref{FeHmaps}. Black contours correspond to equally spaced isosurfaces of the differential stellar density, $\Sigma_{diff}$.}
\label{var1}
\end{figure}

\subsection{Azimuthal variations in the metallicity distribution}\label{results1}
%Due to the high number of particles employed and the reduced noise in the gravitational force field, it takes about 0.8~Gyr to see the appearance of asymmetric structures in the stellar disk. Indeed, only at this epoch strong $m=2,4$ asymmetries, associated to a bar and spiral structure, develop, maintaining a nearly constant strength for about 1 Gyr (see Fig.\ref{metprof}, left panel). The decline that follows in the bar and spiral arms strength at about t=2 Gyr is associated to a conspicuous vertical buckling of the bar, and the subsequent formation of a boxy/peanut shaped bulge, as observed many times in N-body simulations \citep{cosan81, com90, mar06, atha08}.\\
In the previous subsection we have seen that the presence of asymmetries in the stellar disks causes a significant radial redistribution of stars. This redistribution lasts all the time the bar is strong, and declines to lower levels after the bar buckling. Since migration is efficient mostly in the phase of strong bar activity (Fig.~\ref{migrprob}), is there any observational signature that can help in quantifying the strength of the bar? In other words, is there a way to quantify if the galaxy is experiencing a strong phase of bar activity and thus of strong migration?\\
In this section and in the following, we will show that the presence of azimuthal variations in the metallicity distribution of old stars in a disk is a powerful probe of the state of bar activity and of its impact on the stellar disk. 

We will start this analysis by discussing the evolution with time of the stellar metallicity profile (see Fig.~\ref{metprof}).
In the phase of strong migration, the initial metallicity gradient flattens out significantly in the outer disk regions, varying from its initial value $\Delta_{\rm{[Fe/H]}}=-0.07~\rm{dex/kpc}$  to $\Delta_{\rm{[Fe/H]}}=-0.01~\rm{dex/kpc}$ (see also \citet{friedli94, min11}), while changing only marginally in the inner regions (from $\Delta_{\rm{[Fe/H]}}=-0.07~\rm{dex/kpc}$  to $\Delta_{\rm{[Fe/H]}}=-0.06~\rm{dex/kpc}$). This leads to a characteristic change of slope in the metallicity profile, as often observed in disk galaxies \citep[for a recent work, see][]{lep12}. At the same time, the metallicity dispersion increases. At t=0, by construction, the dispersion is null; at t=0.4 Gyr, well before bar formation, the dispersion around the mean metallicity caused by epicycle "blurring"  is about 0.12 dex\footnote{The value is measured at $R=8$~kpc, but note that it depends on the distance from the galaxy center. At $R=0$, for example, the metallicity dispersion in equal to 0.06~dex at t=0.4 Gyr and changes only slightly at the epoch of strong bar activity, increasing up to 0.08~dex}  and as soon as the bar forms, the value raises up to about 0.16 dex, and stays constant till the end of the simulation. Radial migration processes related to the phase of strong activity of the bar thus add an extra dispersion with respect to the pre-migration value, of about 0.1~dex\footnote{This value has been obtained by quadratic sum, since the two dispersions, before and after migration, are independent.}.\\ 
 In Fig.\ref{metprof} we have evaluated the azimuthally averaged metallicity profile. But how far from axisymmetry is the real metallicity distribution? In other words, what is the variation around the mean value at any given distance from the galaxy center?
The face-on maps of the stellar [Fe/H] distribution at different times are shown in Fig.\ref{FeHmaps}.
Once formed (between 0.8 and 1 Gyr) the bar is clearly recognizable in the inner regions, where the metallicity, initially axisymmetric, becomes elongated along the bar major axis. %: more metal rich stars are elongated parallel to the bar, while less metal rich stars are found in regions orthogonal to the bar major axis.
Also, the distribution in the outer regions is far from axisymmetry, showing signs of spiral structure and metallicity inhomogeneities up to the edge of the disk.

These inhomogeneities are related to the way radial migration occurs in galaxies, through spiral patterns \citep[see  Sect.~\ref{when} and ][]{min12, grand12}: metal-rich stars which move to the outer disk are mostly from the region outside the corotation \citep{bru11}, and migrate through spiral patterns  to the outer parts of the disk. In other words, migration is not axisymmetric. To  show that  the variations observed in the metallicity maps are related to the appearance of the bar and spiral arms, and that they fade away with the stellar asymmetries, we present maps of the  azimuthal metallicity variations, $\delta_{\rm{[Fe/H]}}$, evaluated at different times during the galaxy evolution (Fig.\ref{var1}).
Here $\delta_{\rm{[Fe/H]}}$ = [Fe/H]-[Fe/H]$_{axy}$, with [Fe/H] the raw value of the metallicity and [Fe/H]$_{axy}$ the azimuthal average, evaluated, at each time, in radial bin of 0.1 kpc.
These variations are compared with the stellar density distribution at the same epoch, quantified through the differential stellar density $\Sigma_{diff}$=$(\Sigma-\Sigma_{axy})/\Sigma_{axy}$, where $\Sigma$ is the raw value of the stellar density and $\Sigma_{axy}$ is the azimuthally averaged density at each time \citep[see also][]{min12}.
These plots show that azimuthal variations in the metallicity of the old stellar component appear as soon as stellar asymmetries start to develop.  As the bar develops, $\delta_{\rm{[Fe/H]}}$ maps become quite complex, with maxima (and minima) inside the bar radius aligned with (mostly perpendicular to)  the bar major axis. Also, the outer disk -- which has expanded in size since the epoch of bar formation, in agreement with previous works \citep[among others,][]{min11, min12} -- is characterized by metallicity variations, the absolute magnitude of which can reach $0.1\rm{dex}$ for an initial gradient $\Delta_{\rm{[Fe/H]}}=-0.07~\rm{dex/kpc}$. Strong metallicity variations are present in the inner disk for all the time that stellar asymmetries are strong, and they fade away as the bar and spiral arms strength diminishes. 
However, while in the bar region metallicity variations are always aligned with the bar, in the outer disk maxima and minima do not necessarily coincide with maxima and minima in the differential stellar density. %, showing a more complex behavior: at t=0.8 Gyr, they describe a spiral pattern, which closely matches the spiral density distribution found in the $\Sigma_{diff}$ contour levels. At following times, maxima in the $\delta_{\rm{[Fe/H]}}$ not necessarily coincide  with  the spiral arms, but they can rather lead ($t=1.25$~Gyr) or trail ($t=1.55$~Gyr) the spiral pattern. 
This behavior is due to the fact that once they have migrated into the outer disk stars have an angular rotation velocity different from the pattern speed of these asymmetric structures, and  can thus  intercept the arms at different location during their motion around the galaxy.\\
%\textbf{It is worth noting that  the stellar density maps in Fig.~\ref{smaps} show that the galaxy is still barred at the end of the simulation, while the metallicity maps show only weak signs of azimuthal variations at that time. This suggests that the metallicity distribution of stars in the disk can be used to probe the state of the bar, and the strength of the associated radial migration.}

\begin{figure}
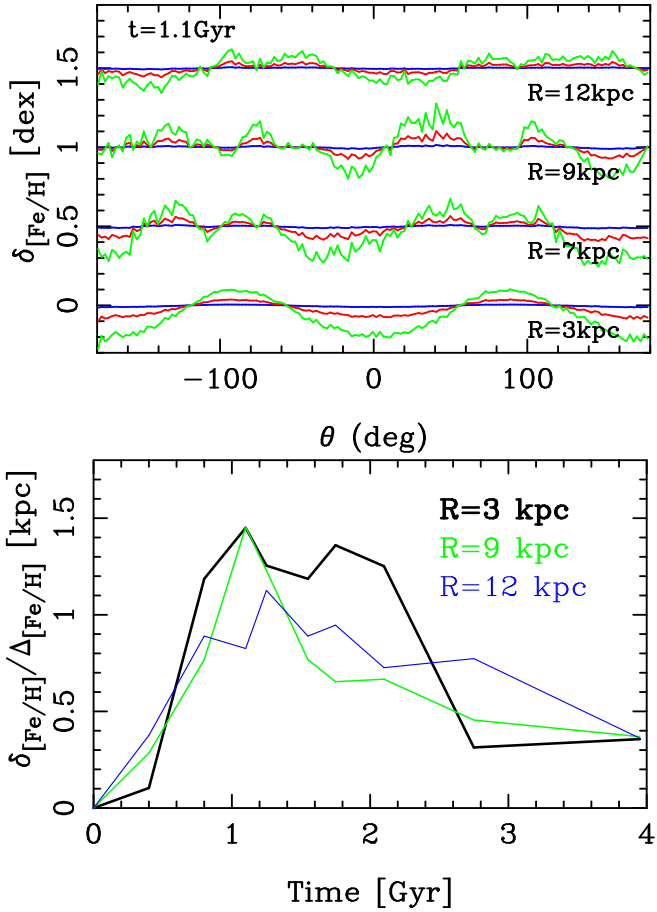

\centering
\includegraphics[width=6.cm,angle=270]{pvarALL_110_gS0_q1p8_BD0p10.ini.ps}
\includegraphics[width=6.cm,angle=270]{pvarALLvstimenewreg_gS0_q1p8_BD0p10.ini.ps}
\caption{\emph{Top panel: }Metallicity variations, $\delta_{\rm{[Fe/H]}}$, as a function of the azimuth, for three different values of the initial metallicity gradient: $\Delta_{\rm{[Fe/H]}}=-0.01$~dex (blue line),   $\Delta_{\rm{[Fe/H]}}=-0.07$~dex (red line),  $\Delta_{\rm{[Fe/H]}}=-0.1$~dex (green line), and at four different radii in the galaxy disk. Variations are shown at the time of strong bar activity. Note that the values corresponding to $R=7$~kpc, $R=9$~kpc, $R=12$~kpc  have been arbitrarily shifted of 0.5~dex, 1~dex, 1.5~dex with respect to their real values. \emph{Bottom panel: }$\delta_{\rm{[Fe/H]}}/\Delta_{\rm{[Fe/H]}}$ ratio as a function of time, for an initial metallicity gradient $\Delta_{\rm{[Fe/H]}}=-0.01$~dex/kpc.The ratio is shown at three different disk radii: $R=3$~kpc (black line), $R=9$~kpc (green  line), and $R=12$~kpc (blue line). Note that this ratio is independent of the gradient, as detailed in the text.}
\label{vargrad}
\end{figure}
%\begin{figure}
%\centering
%\includegraphics[width=4.5cm,angle=270]{pvarALLvstime2reg_gS0_q1p8_BD0p10.ini.ps}
%\caption{$\delta_{\rm{[Fe/H]}}/\Delta_{\rm{[Fe/H]}}$ ratio as a function of time, for three different values of the initial metallicity gradient: $\Delta_{\rm{[Fe/H]}}=-0.01$~dex/kpc (blue),   $\Delta_{\rm{[Fe/H]}}=-0.07$~dex/kpc (red),  $\Delta_{\rm{[Fe/H]}}=-0.1$~dex/kpc (green). The values are shown at two different disk radii: $R=3$~kpc (black thick line) and $R=7$~kpc (grey thin line).}
%\label{vargrad2}
%\end{figure}
%\vspace{-0.6cm}

\subsection{Dependence on the initial metallicity gradient}\label{results2}

The amplitude of the inhomogeneities, formerly discussed,  depends on the initial metallicity gradient of stars in the disk, as shown in  Fig.\ref{vargrad} (top panel), where the value of $\delta_{\rm{[Fe/H]}}$ is shown as a function of azimuth, for three different values of the initial gradient, steep ($\Delta_{\rm{[Fe/H]}}=-0.19~\rm{dex/kpc}$), moderately steep ($\Delta_{\rm{[Fe/H]}}=-0.07~\rm{dex/kpc}$) and nearly flat ($\Delta_{\rm{[Fe/H]}}=-0.01\rm{dex/kpc}$).  At t=1.1 Gyr, when the bar is strong, the maximum variations (of the order of 0.2~dex) are found for the case of steep gradient, while very small variations ($\sim 0.01$~dex) are produced if the initial gradient is nearly flat. Once the phase of strong activity of the bar is over (t=3.95 Gyr), the disk is chemically homogeneous, with the metallicity variations fading toward zero for any choice of the initial gradient.\\
It is interesting to note that a small $\delta_{\rm{[Fe/H]}}$ does not necessarily  imply a weak or absent radial migration. Small variations, due to an initial nearly flat metallicity gradient, may indeed hide a strong migration. 

To quantify the amount of spatial redistribution of stars in a disk it is indeed necessary to normalize to the value of the gradient, as shown in Fig.\ref{vargrad} (bottom panel). Here the ratio  $\delta_{[Fe/H]}/\Delta_{\rm{[Fe/H]}}$ is plotted as a function of time, for three different values of the initial metallicity gradient\footnote{In this plot, $\delta_{[Fe/H]}$ represents the maximum absolute value (over $\theta$) of the abundance variation at any given radius $R$}. In evaluating this ratio, we have used the initial value of the metallicity gradient, $\Delta_{\rm{[Fe/H]}}$, since it varies only marginally over time in the bar region\footnote{Note that this choice leads to underestimate the ratio  $\delta_{[Fe/H]}/\Delta_{\rm{[Fe/H]}}$ in the outer disk, where the gradient  flattens considerably after bar formation. } (cf Fig.\ref{metprof} and discussion in Sect.~\ref{results1}). \\%If the current value of the metallicity gradient at time $t$ is adopted, instead of the initial one, the ratio  $\delta_{[Fe/H]}/\Delta_{\rm{[Fe/H]}}$ would be a factor $\sim$1.2 higher than the values found in Fig.~\ref{vargrad}.
As expected, Fig.\ref{vargrad} shows that this ratio is independent on the initial metallicity gradient, and it is a tracer of the strength of radial migration, i.e. of the contamination, at any given radius, of stars coming from different disk radii.
As a first approximation, this value can be seen as the difference between $\tilde{R}$ and $R$, with $\tilde{R}$ being the radius where migrating stars come from, i.e.:
%\begin{eqnarray*}
 $\delta_{\rm{[Fe/H]}}/\Delta_{\rm{[Fe/H]}} = (\Delta_{\rm{[Fe/H]}} \tilde{R} -\Delta_{\rm[Fe/H]} R)/ \Delta_{\rm{[Fe/H]}}= \tilde{R}-R$. 
%\end{eqnarray*} 
In reality, migration contaminates a given radius $R$ with stars coming from a variety of different initial radii $\bar{R}$, however this measure can be interpreted as a way to quantify the intensity of radial migration for an initial (not null) gradient. If its value is greater than unity, for example (as it is the case in the inner disk in the phase of strong bar) then at any given radius, the metallicity variations in azimuth are similar to those expected from moving radially (in the center or anticenter direction) by more than 1 kpc.\\

The ratio  $\delta_{[Fe/H]}/\Delta_{\rm{[Fe/H]}}$ is also an indicator of the duration of the phase of strong  migration: in the bar region, its value is indeed above unity for all the duration of high bar strength (cf Fig. \ref{asym} and \ref{vargrad}), thus indicating that inhomogeneities at any given distance from the center persist, i.e. stars are not mixed yet. The effect of mixing indeed starts at the end of the strong bar activity and migration phase (corresponding to $t\sim$ 2 Gyr in our simulations): at this time, the ratio $\delta_{[Fe/H]}/\Delta_{\rm{[Fe/H]}}$  diminishes by a factor greater than 3, together with a decrease in bar and spiral arms strength. Once strong migration has ended, the timescale to disperse inhomogeneities of size  $\Delta R$ in the disk is $\propto 2\pi/[(d\Omega/dR)\Delta R]$, with $\Omega$ being the angular velocity of stars in the disk. For a fluctuation of size $\Delta R=1~$kpc, this corresponds to about  2 Gyr at $R=10$~kpc, and 0.3 Gyr at $R=4$~kpc.

\section{Conclusions}\label{conclusions}

In this work we study,  by means of dissipationless, N-body simulations, the signatures that radial migration induced by a bar imprints on the old stellar population of a disk galaxy.

%\textbf{The presence of non-axisymmetric structures alone in a face-on disk does not constrain the activity of the bar, or the strength of radial migration they induce in the stellar disk. By means of dissipationless N-body simulations, we show that  metallicity variations in the old stellar disk  can be used to constrain the phase of the bar, and its impact  on the whole disk.}
%By means of dissipationless N-body simulations we \textbf{study the impact  radial migration has on the spatial redistribution of metals in an old stellar disk.}%show that migration induced by bar and spiral arms  leads to significant azimuthal variations in the metallicity distribution of old stars at any given distance from the galaxy center.

Firstly, we show that migration is initiated at the time of bar formation, and that it remains significant over the whole  phase of high bar strength, in agreement with previous results \citep{min11, min12, bru11}. \\
Then we show that, as a consequence of this migration, significant azimuthal variations  are generated in the metallicity distribution of old disk stars. \\
At the peak of bar strength, the metallicity variations  above or below the 
azimuthally averaged value are of the order of the initial gradient. They are particularly strong in the bar region, with the higher metallicities observed along the bar major axis and the lower metallicities parallel to the bar minor axis. \\
While the strength of the azimuthal variations depends on the initial stellar metallicity gradient, the ratio $\delta_{\rm{[Fe/H]}}/\Delta_{\rm{[Fe/H]}}$ does not. This value is indeed independent on the initial metal distribution and, as we show, it is a measure of the strength of the bar and \emph{thus} of the ongoing migration. 
%The ratio $\delta_{\rm{[Fe/H]}}/\Delta_{\rm{[Fe/H]}}$ can be used also to quantify the timescale of the migration process: it takes indeed about 1 Gyr this ratio to decrease below unity, this time thus corresponding to the duration of the phase of inhomegeneous stellar distribution in the stellar disk. \\
Since migration, induced by bar and spiral arms, causes an inhomogeneous stellar distribution in the stellar disk -- the duration of this phase being quantified by a ratio  $\delta_{\rm{[Fe/H]}}/\Delta_{\rm{[Fe/H]}}$ greater than unity -- migration and mixing do not occur at the same time. Mixing, by definition, requires a homogeneous stellar distribution in the disk, and this is not compatible, as we show, with the phase of strong migration. Only when migration is over, mixing can be established.\\
The presence of non-axisymmetric structures in a (face-on) galaxy disk does not provide information about the level of activity of the bar. It is not possible, for example, to determine, by means of morphological information alone, if a bar is in an early phase of evolution or if it has evolved already through a phase of vertical instabilities. %\citep[see][for a kinematics-based diagnostic]{gad05}. 
Also, from morphology alone, is not possible to quantify the level of stellar redistribution a disk is experiencing. Our work shows that metallicity variations in old stellar populations of galaxy disks, possibly coupled with kinematic information \citep{gad05},  can be used as a diagnostic tool for quantifying these processes, since their amplitude is strongly related to the level of activity of the bar and subsequent migration. Hence, chemical inhomogeneities over the whole disk would testify an ongoing strong bar activity,  while inhomogeneities limited to the outer disk would suggest a fading bar activity.
Barred galaxies with axisymmetric distribution of metallicity could be the result of either a bar that has always 
been weak, or one that has now fade to weak activity, but note that the difference could still be detectable 
in the level of vertical dispersions \citep{gad05}.

%Asymmetric abundance patterns have been found also in the HII regions of the barred galaxy NGC 4736 \citep{davies09}. 
The predictions of our models can be detectable in barred galaxies with  IFU surveys like CALIFA and ATLAS3D. In this sense, some encouraging results have been already published by means of long-slit spectroscopy studies. Recently, \citet{psb11} have indeed compared  the metallicity properties of stars in the disk  with those along the bar  for two early-type galaxies, showing that in both cases bars show higher metallicities, and flatter gradients than disk stars in the same region.  Our models predict the same trend and thus can be an important piece of information to add to the ongoing interpretation of these findings. \\
Also in the Milky Way, azimuthal inhomogeneities have been reported, but currently only among young stars  (\citet{luck06}, but see \citet{luck11}; \citet{davies09}): the variations found may be thus the result of a patchy star formation. The detection (or not) of similar signatures in the old stellar component of the Galactic disk by future surveys like Gaia (and complementary spectroscopic data) may help in constraining the last epoch of strong activity of the bar and the amount of current radial migration.\\
Finally, since radial migration can be caused by several different physical mechanisms, as detailed in the introduction, it will be important to compare them to understand if all leave signatures similar to those discussed in this paper.

\section*{Acknowledgments}

All the simulations have been performed on the supercomputer CURIE at CCRT, CEA, in the framework of the "Projet Grand Challenge" and thanks to the GENCI grant "x2012040507". The authors acknowledge the support of the French Agence Nationale de la Recherche (ANR) under contract ANR-10-BLAN-0508 (Galhis project). PDM warmly thanks M.~D.~Lehnert for a careful reading of a first version of this manuscript and his precious comments, R.~Cid Fernandes and E.~Emsellem for stimulating discussions.  We thank the anonymous referee who helped improving the quality of this paper.

\Online

\begin{appendix}

\section{On the choice of initial conditions : are they realistic?}\label{appmet}
\begin{figure}
\centering
\includegraphics[width=3.cm,angle=270]{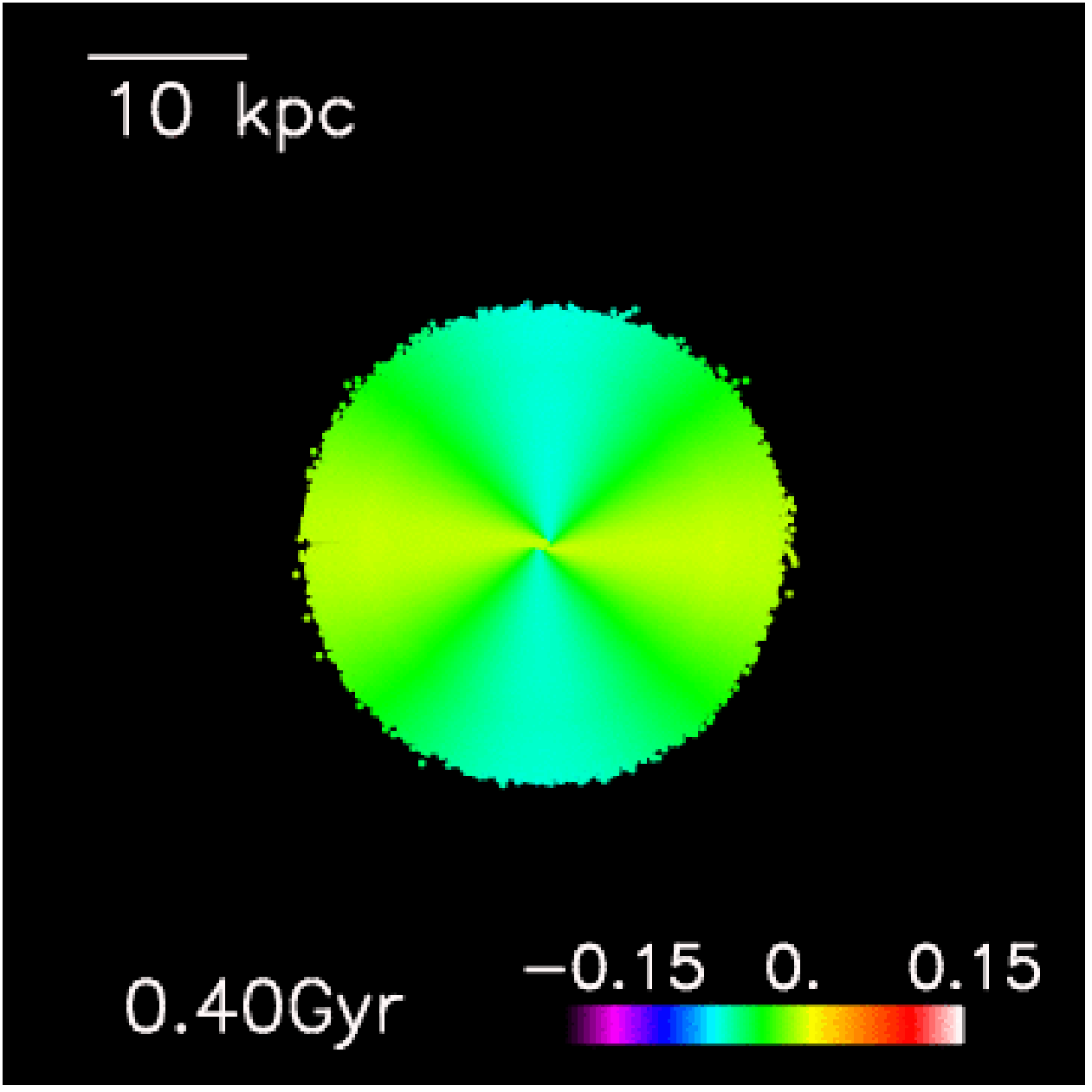}
\hspace{-0.2cm}
\includegraphics[width=3.cm,angle=270]{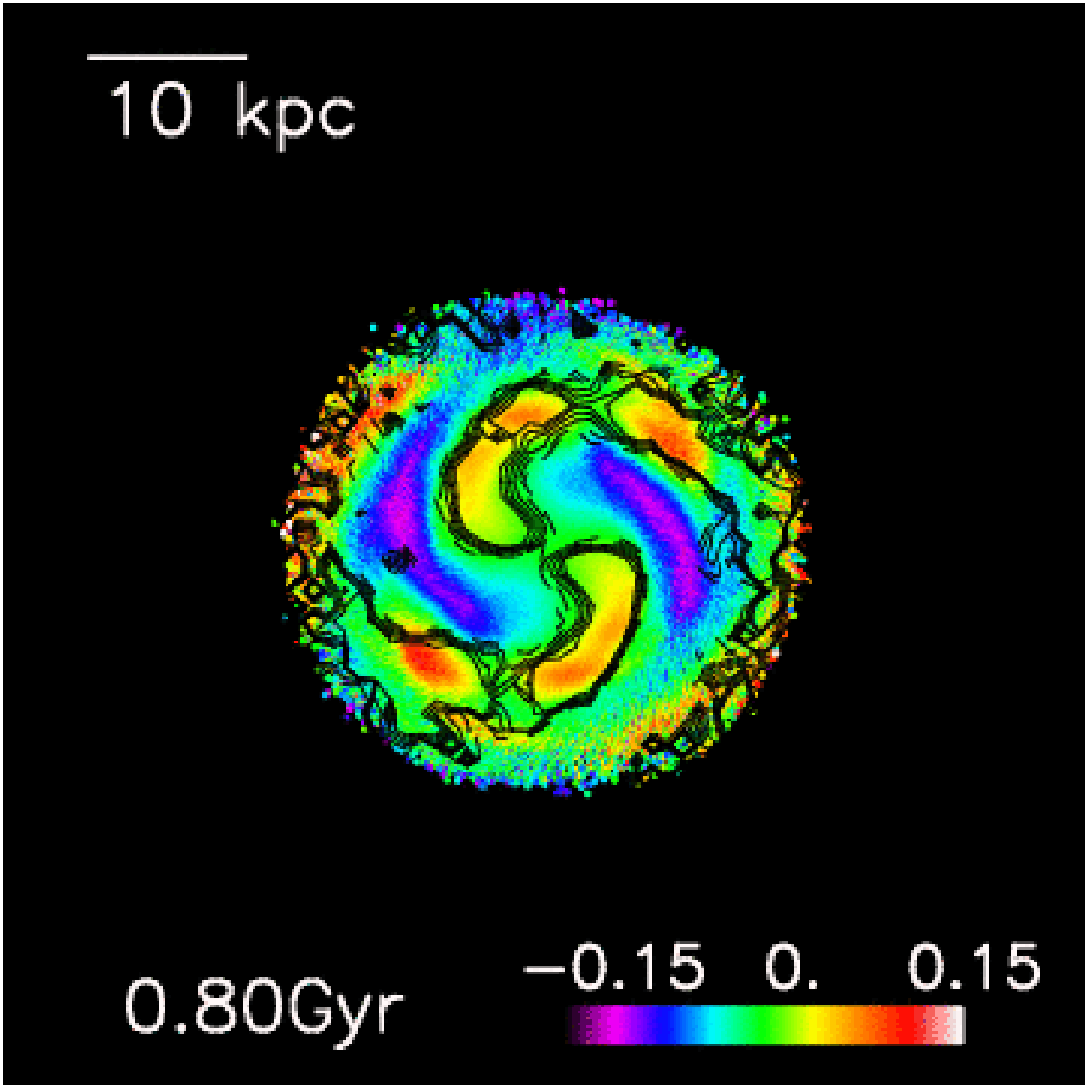}
\hspace{-0.2cm}
\includegraphics[width=3.cm,angle=270]{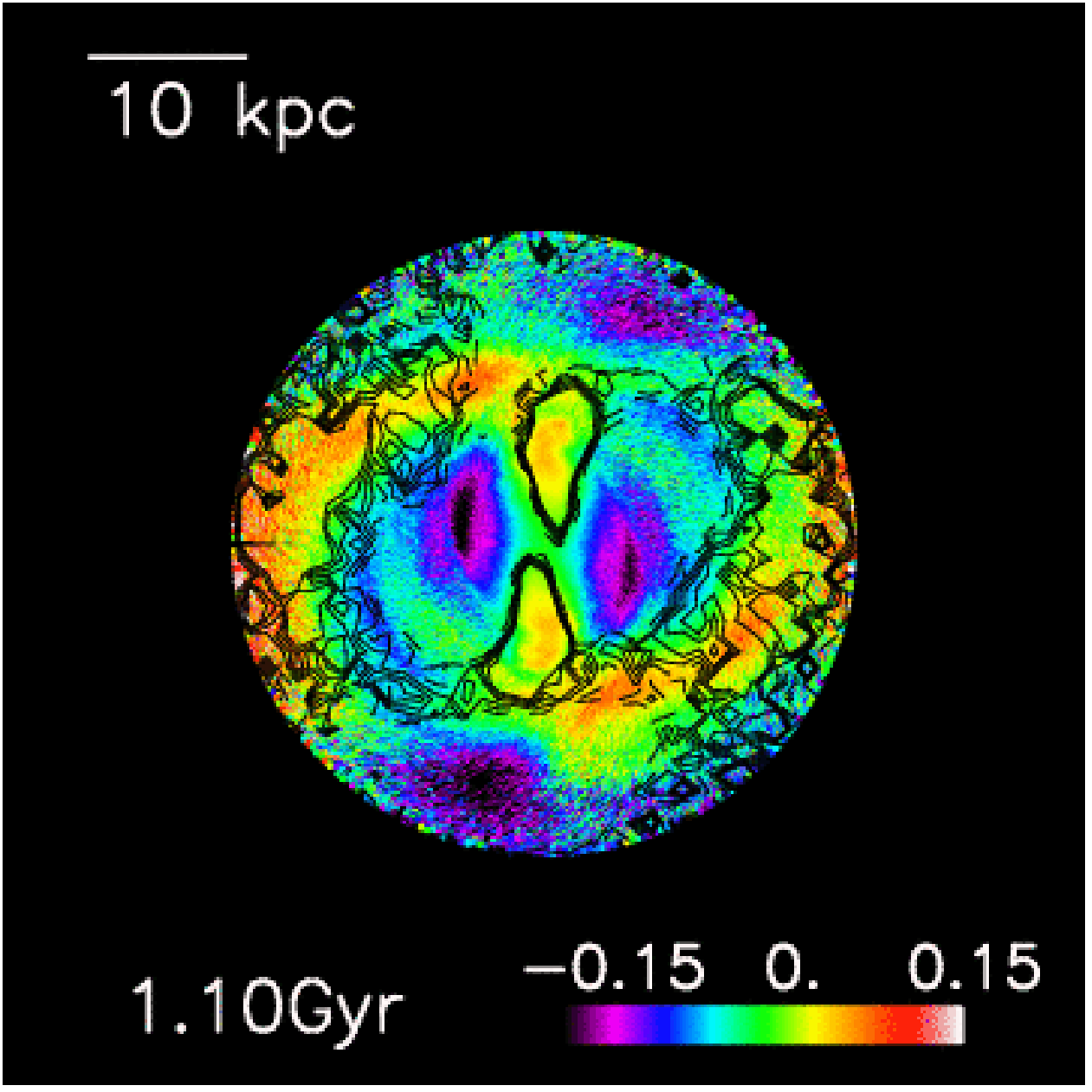}

\includegraphics[width=3.cm,angle=270]{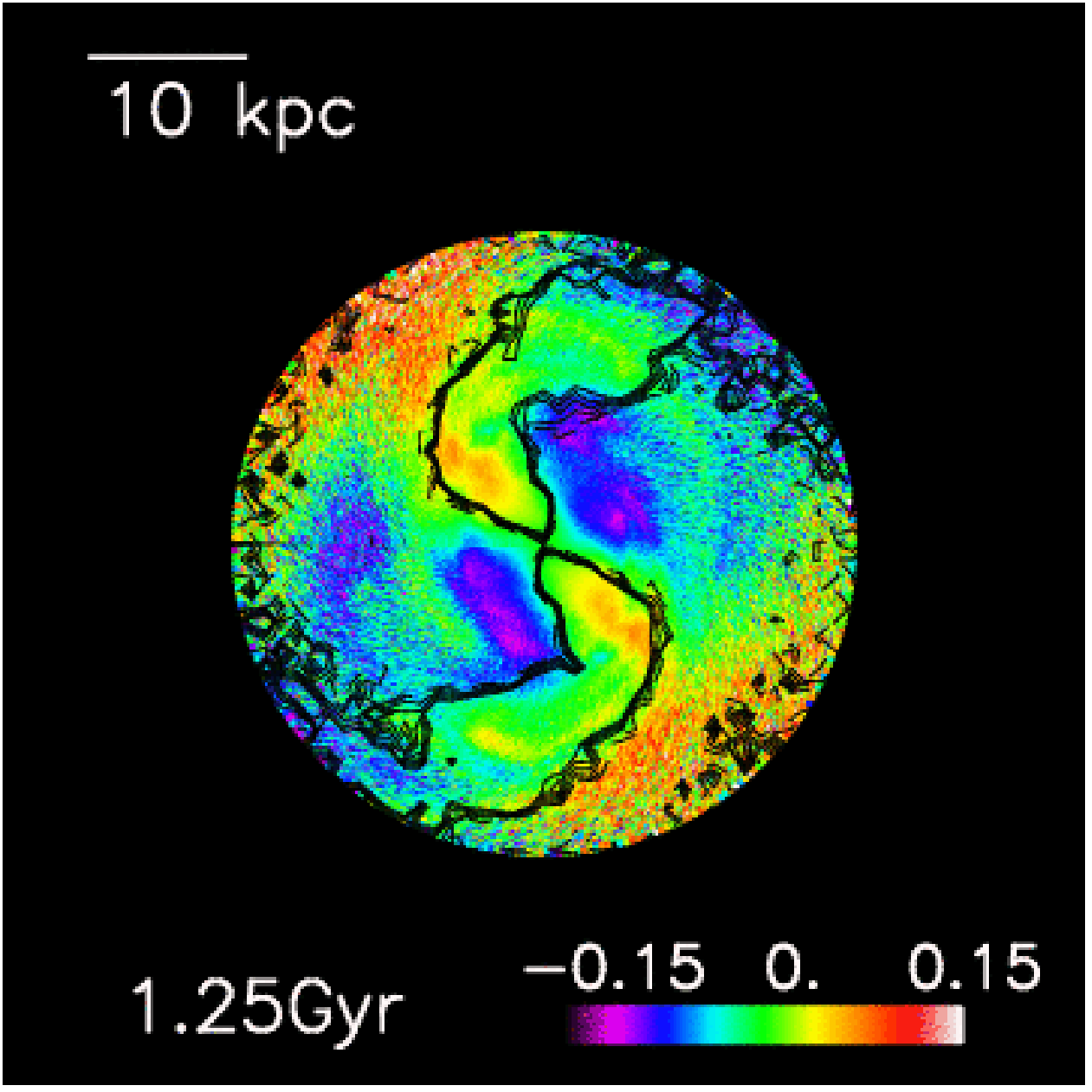}
\hspace{-0.2cm}
\includegraphics[width=3.cm,angle=270]{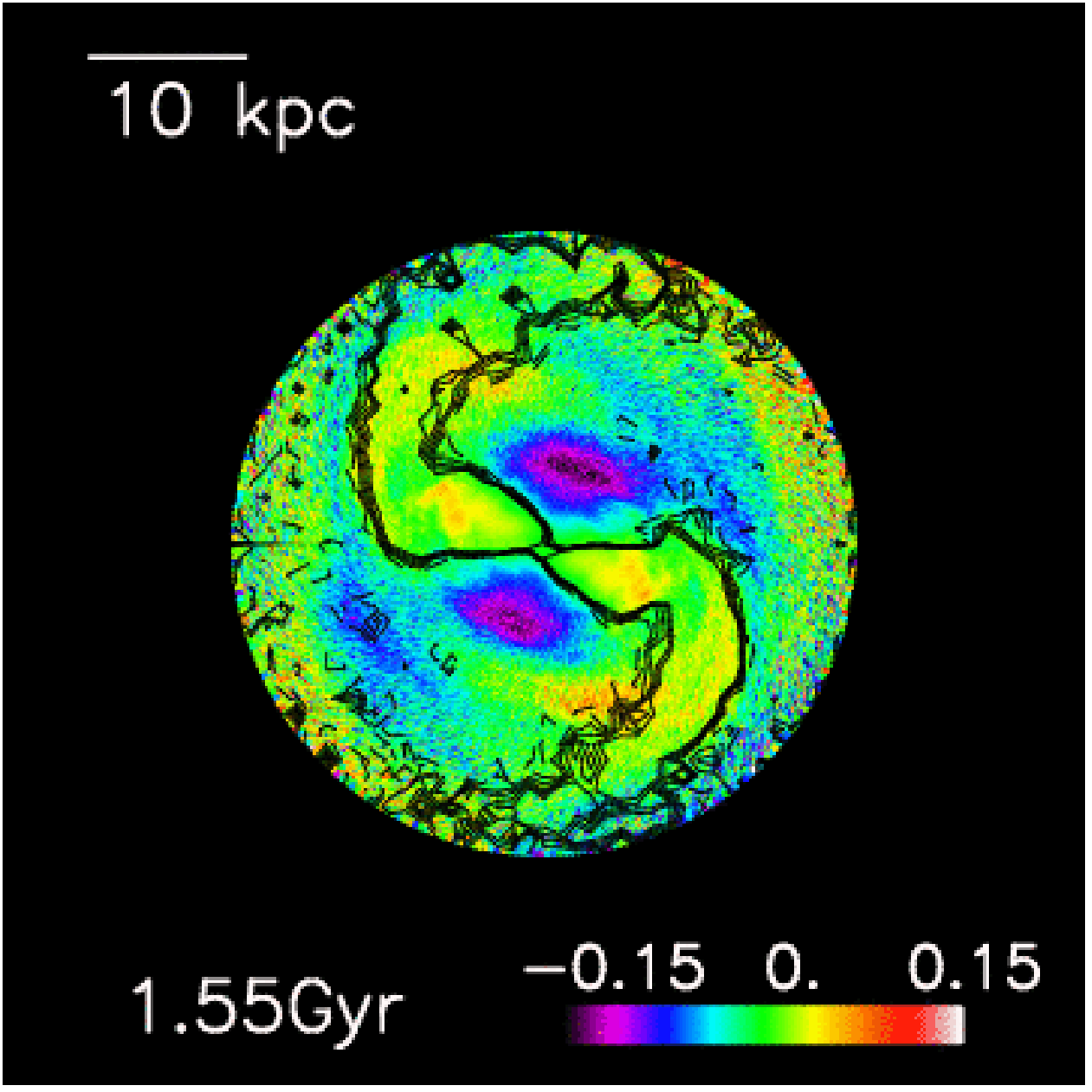}
\hspace{-0.2cm}
\includegraphics[width=3.cm,angle=270]{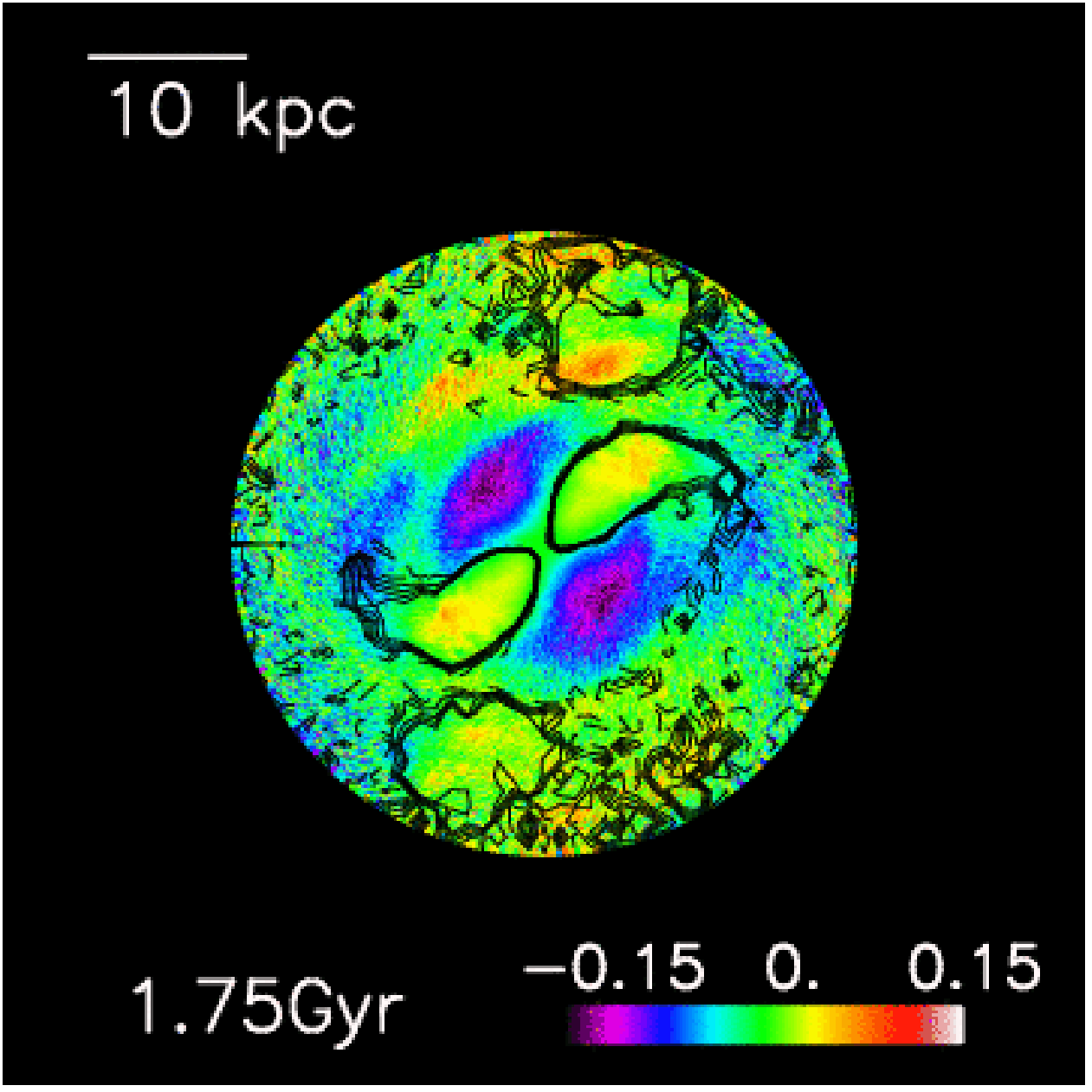}

\includegraphics[width=3.cm,angle=270]{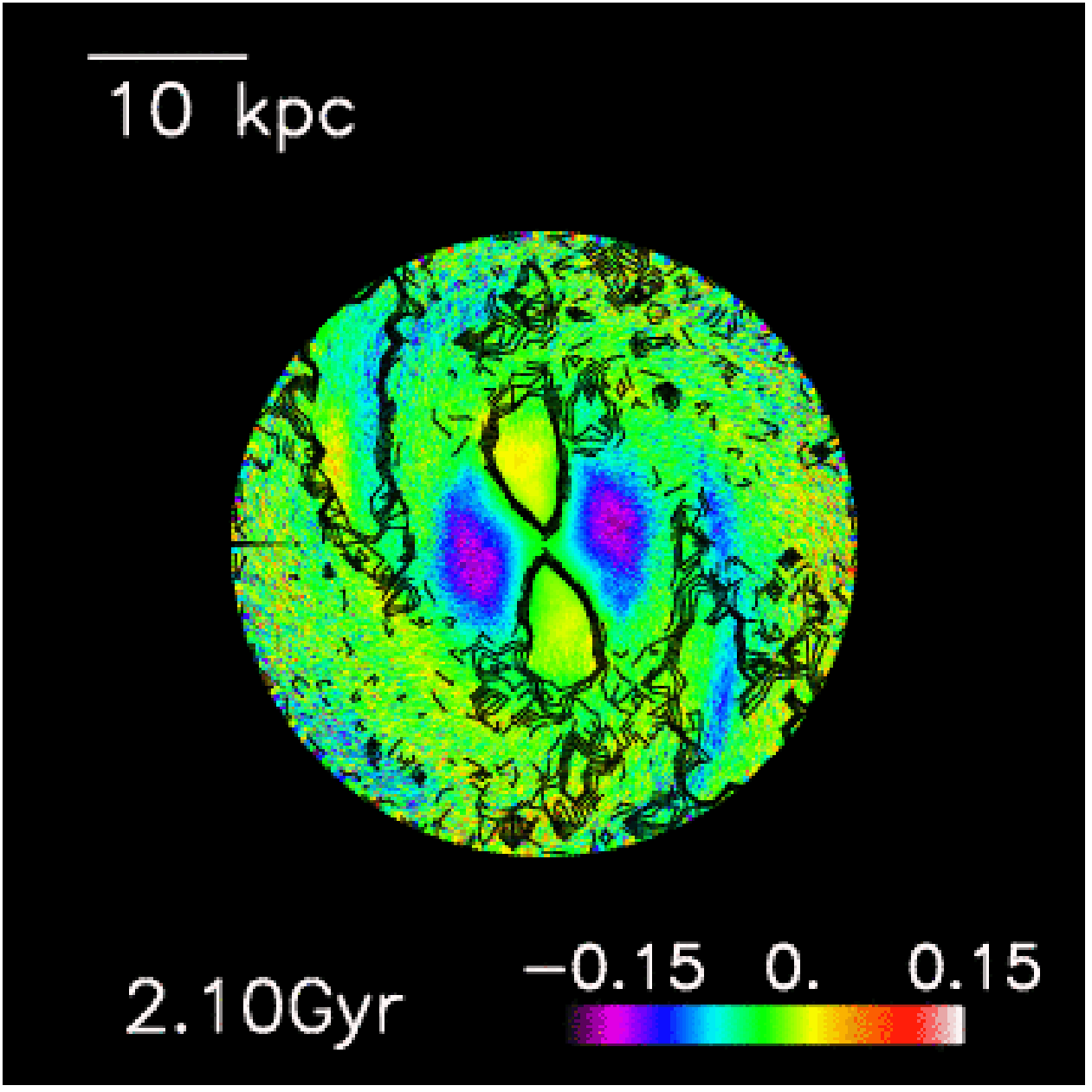}
\hspace{-0.2cm}
\includegraphics[width=3.cm,angle=270]{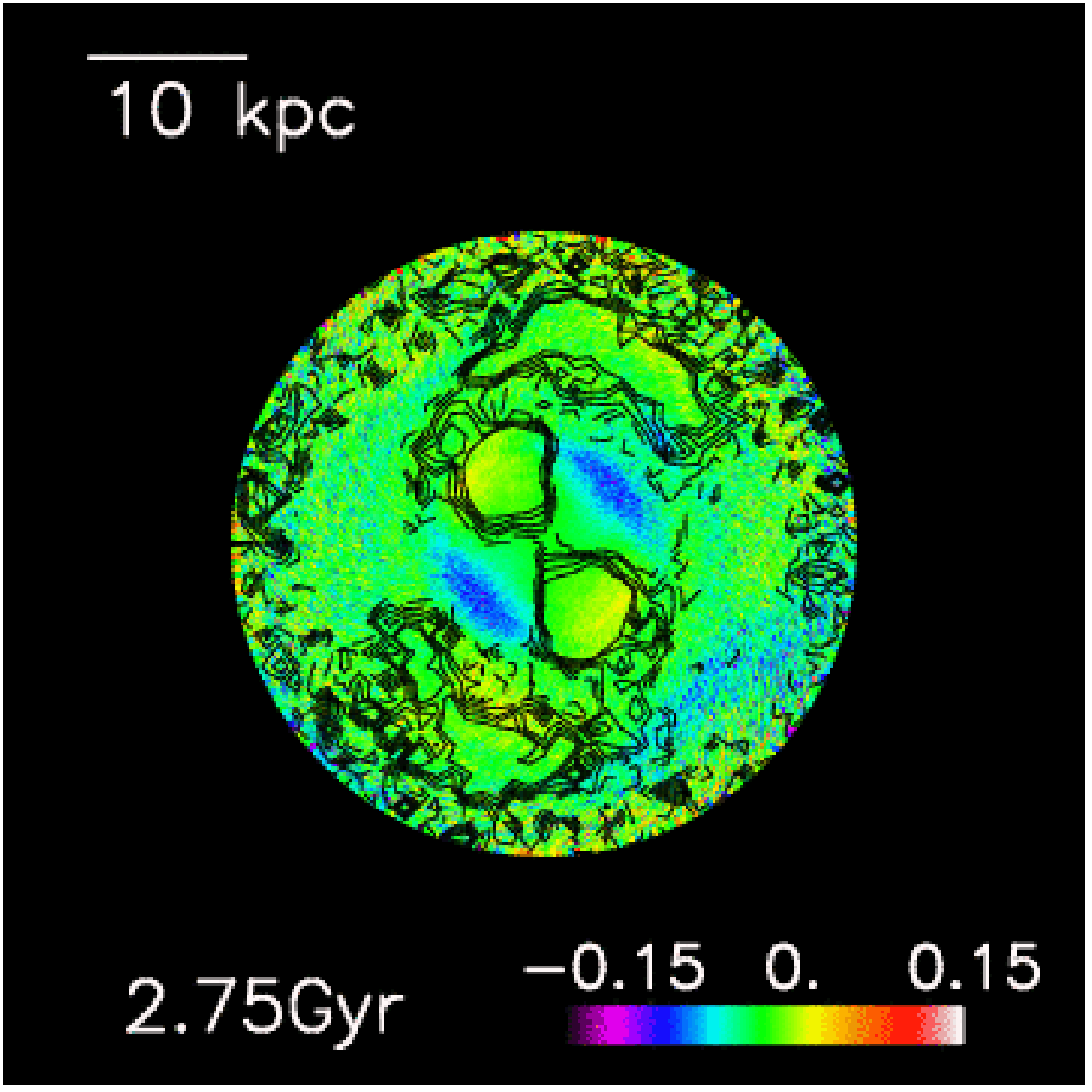}
\hspace{-0.2cm}
\includegraphics[width=3.cm,angle=270]{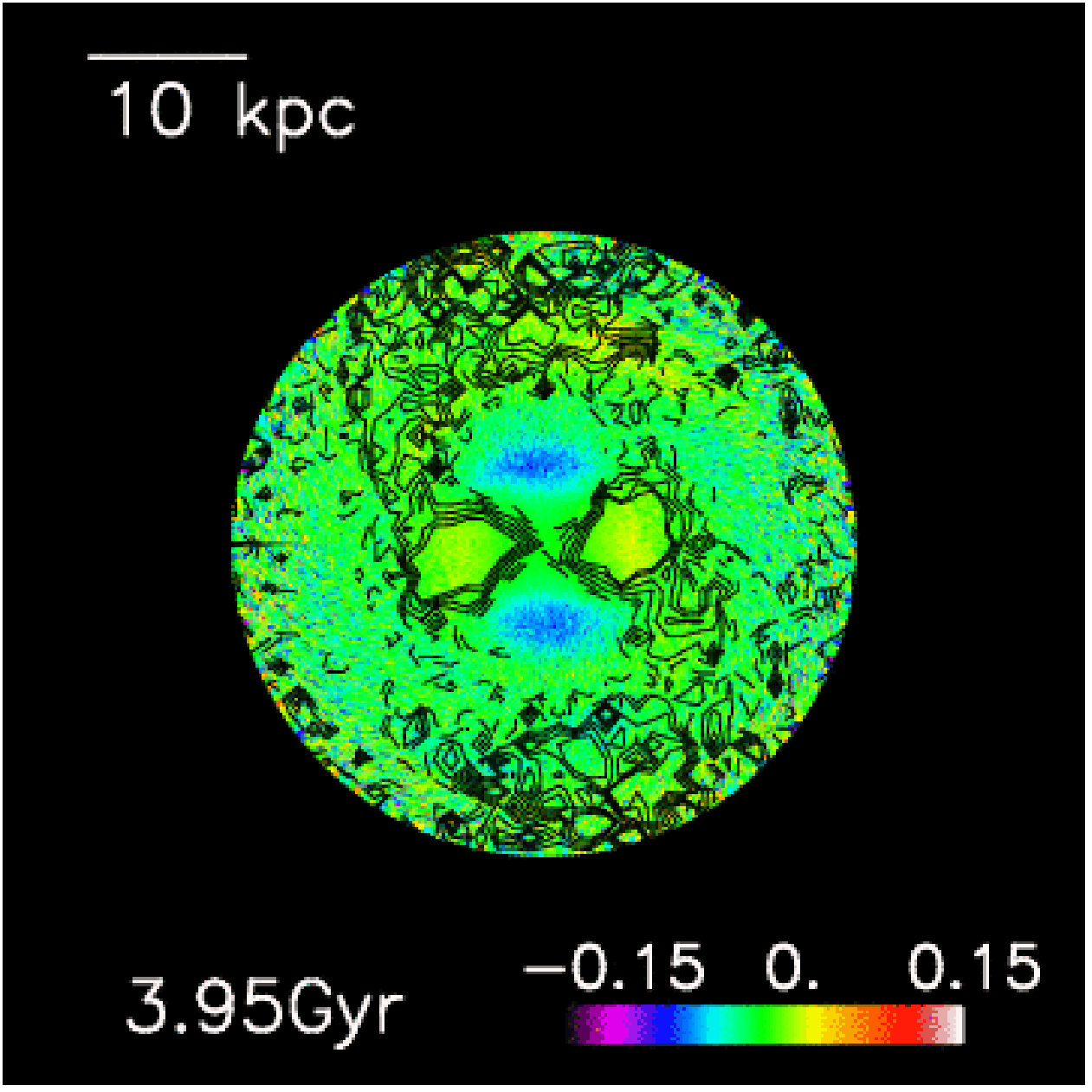}

\caption{Maps of azimuthal metallicity variations, $\delta_{\rm{[Fe/H]}}$ (dex), at different times for a stellar disk with azimuthal inhomogeneities $\delta\sim0.4\Delta$ at t=0.4~Gyr (see text). Black contours correspond to equally spaced isosurfaces of the differential stellar density, $\Sigma_{diff}$. See Fig.\ref{var1} for comparison. }
\label{initvar}
\end{figure}

\begin{figure}
\centering
\includegraphics[width=3.cm,angle=270]{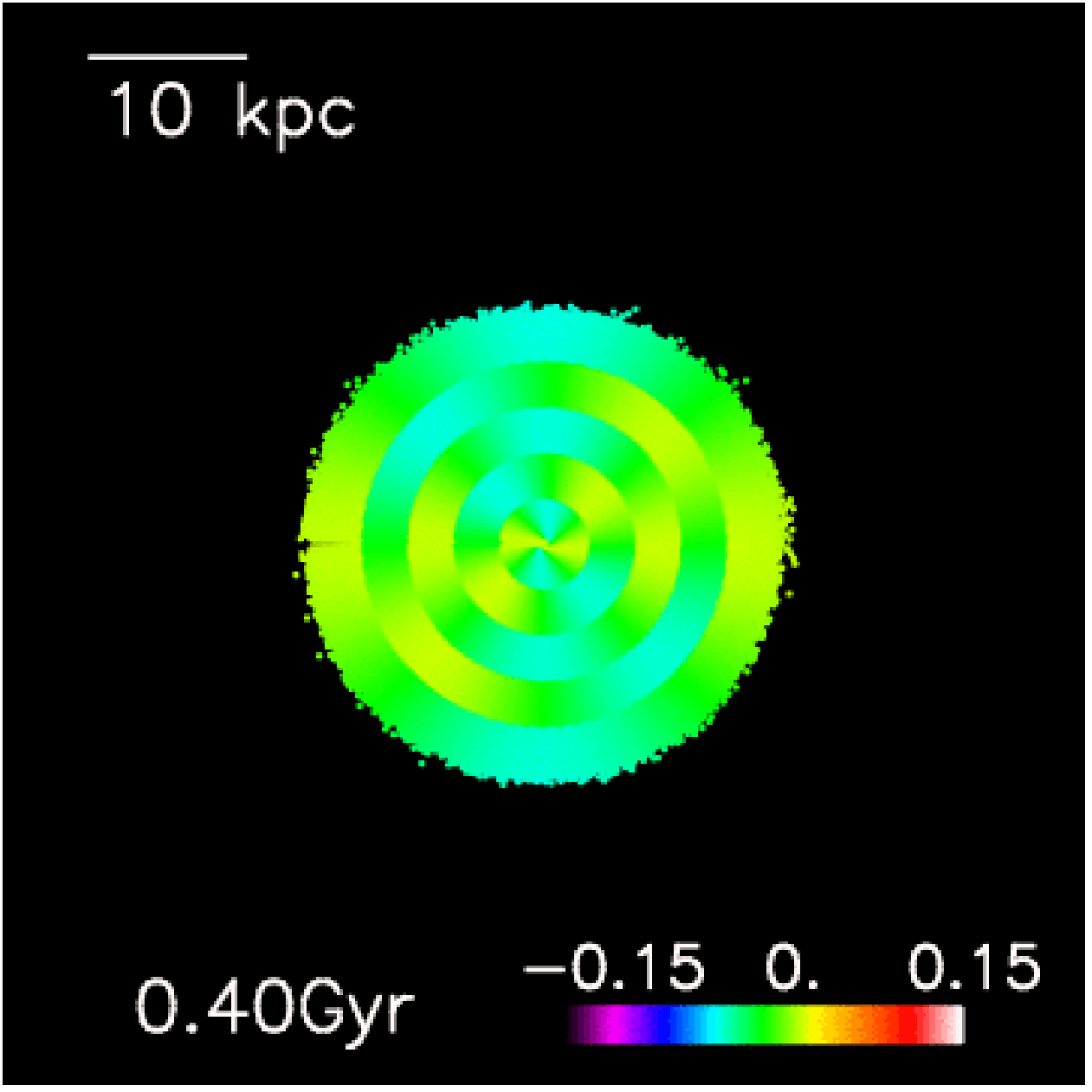}
\hspace{-0.2cm}
\includegraphics[width=3.cm,angle=270]{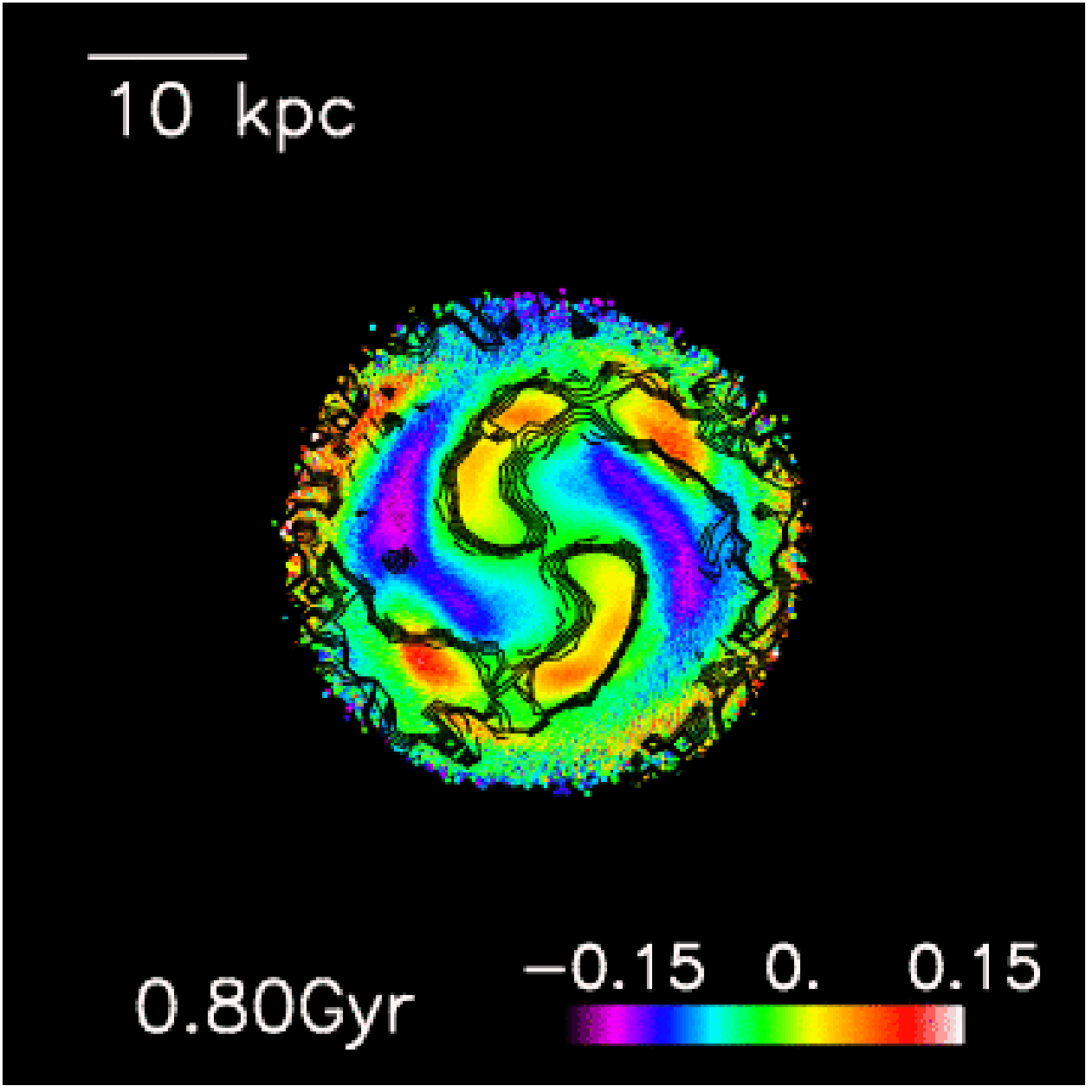}
\hspace{-0.2cm}
\includegraphics[width=3.cm,angle=270]{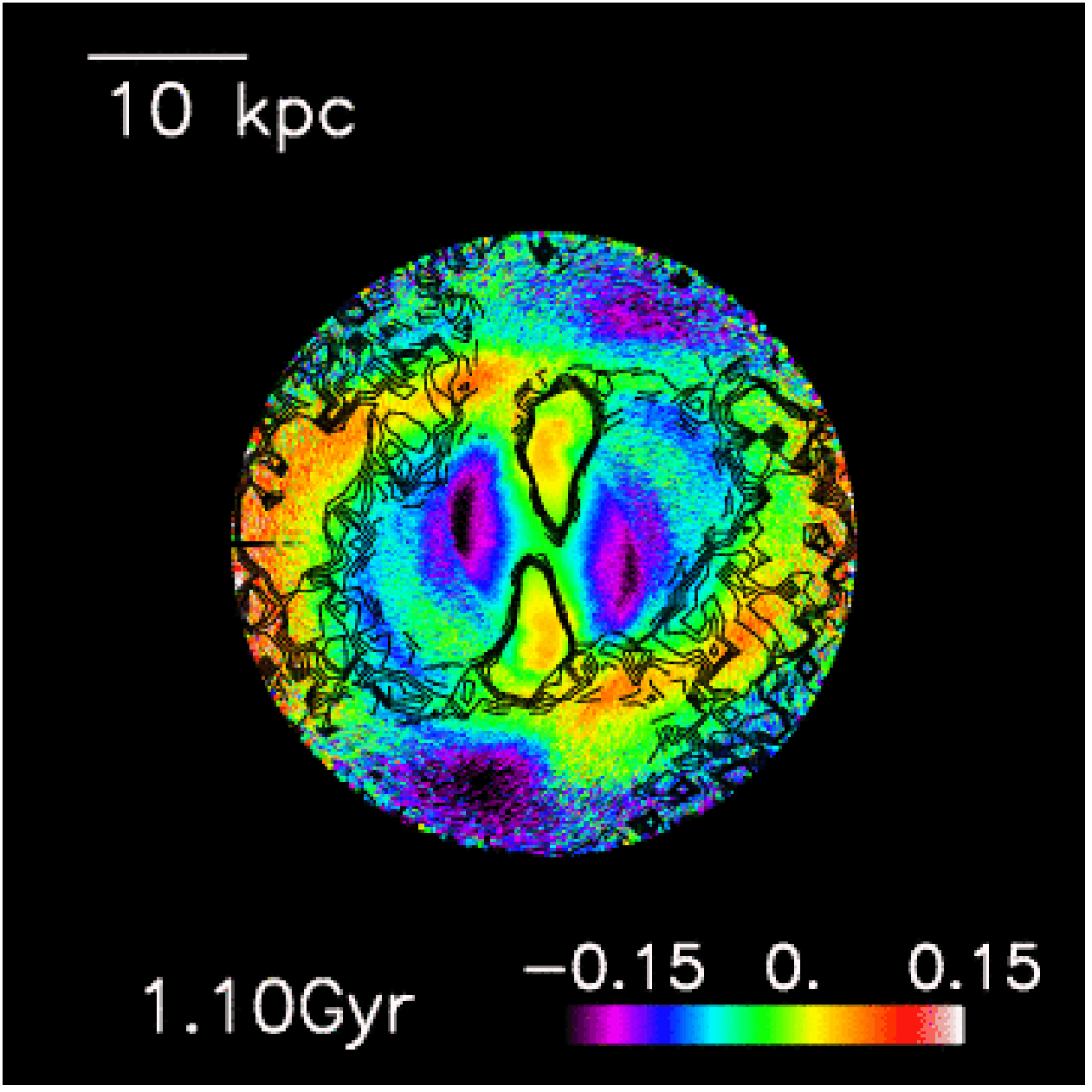}

\includegraphics[width=3.cm,angle=270]{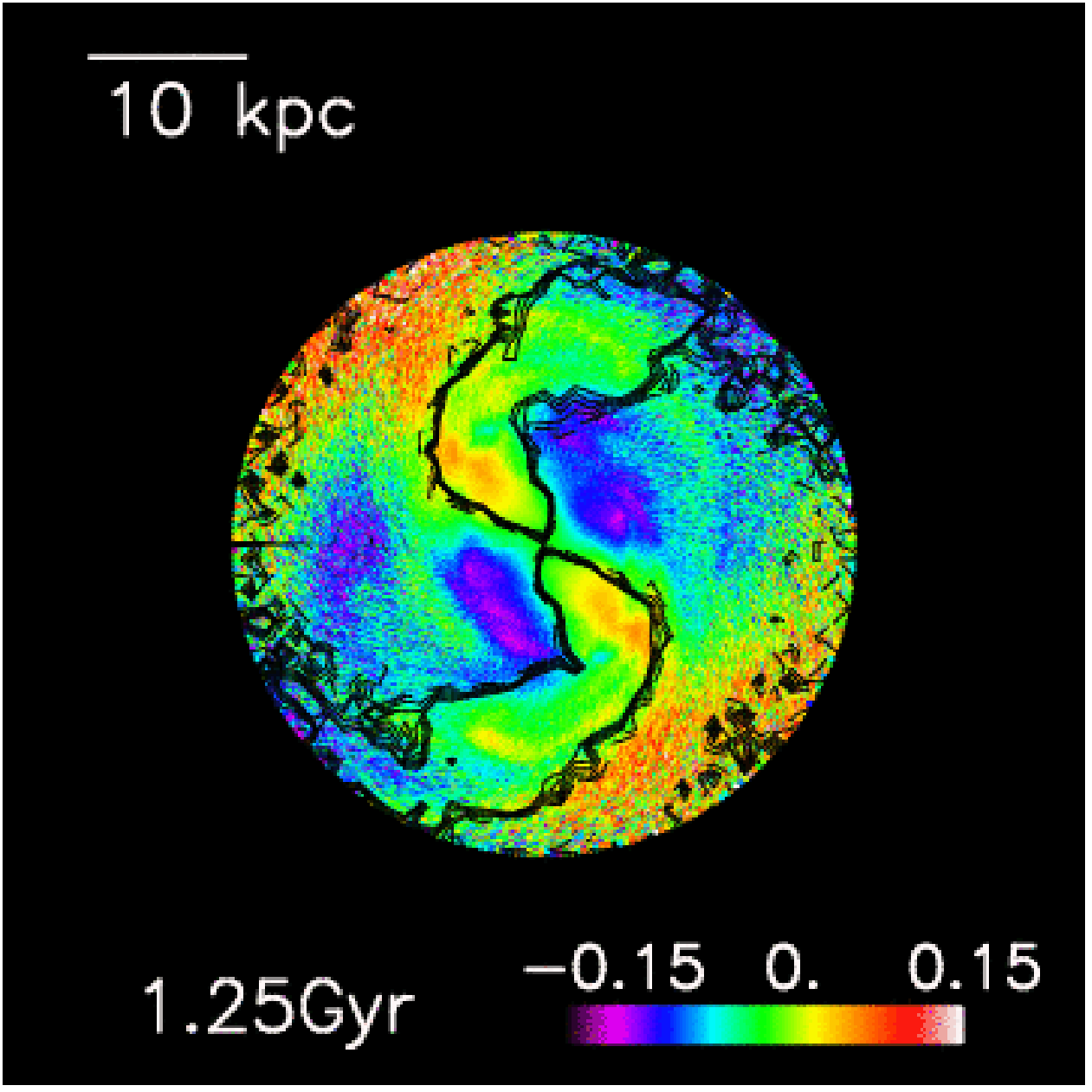}
\hspace{-0.2cm}
\includegraphics[width=3.cm,angle=270]{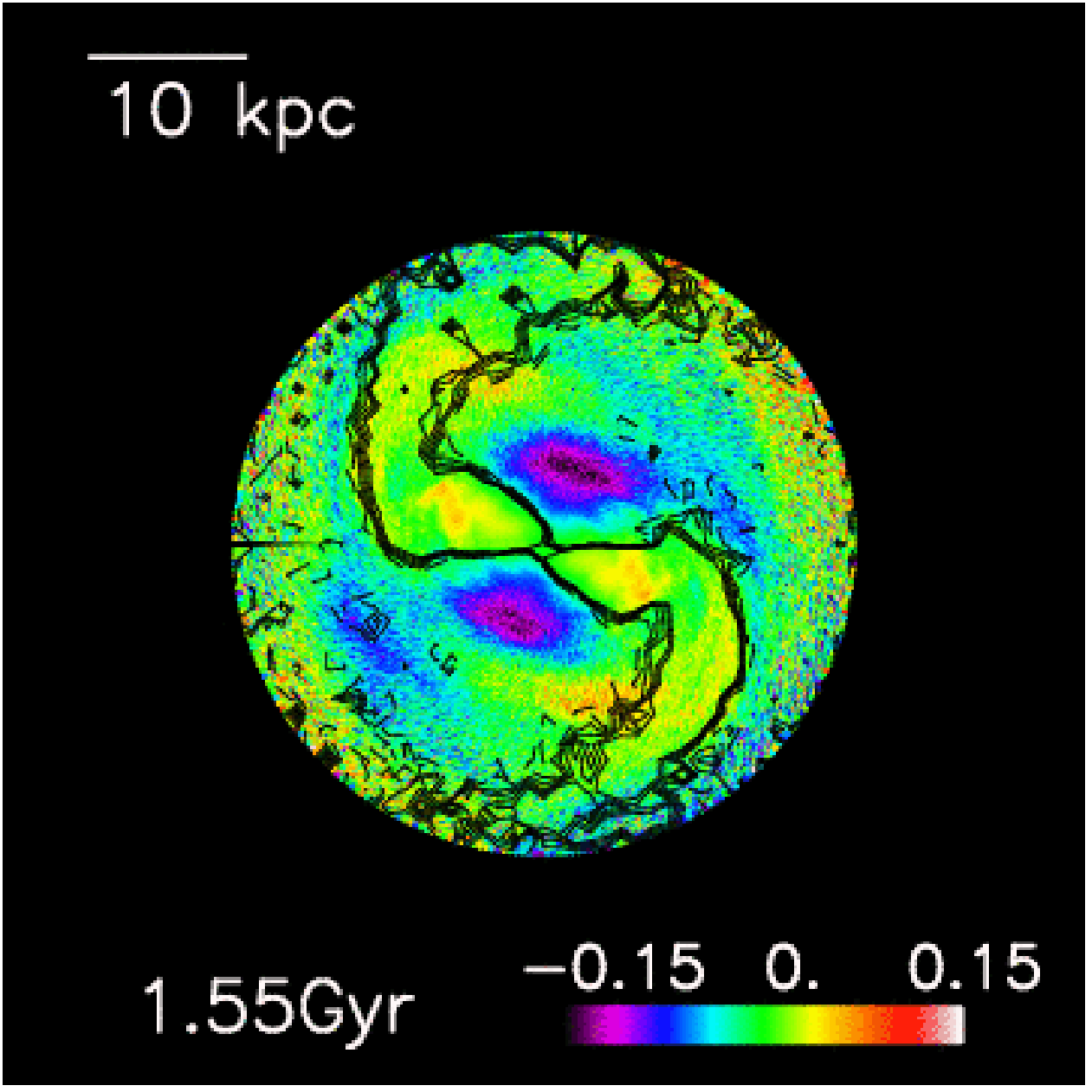}
\hspace{-0.2cm}
\includegraphics[width=3.cm,angle=270]{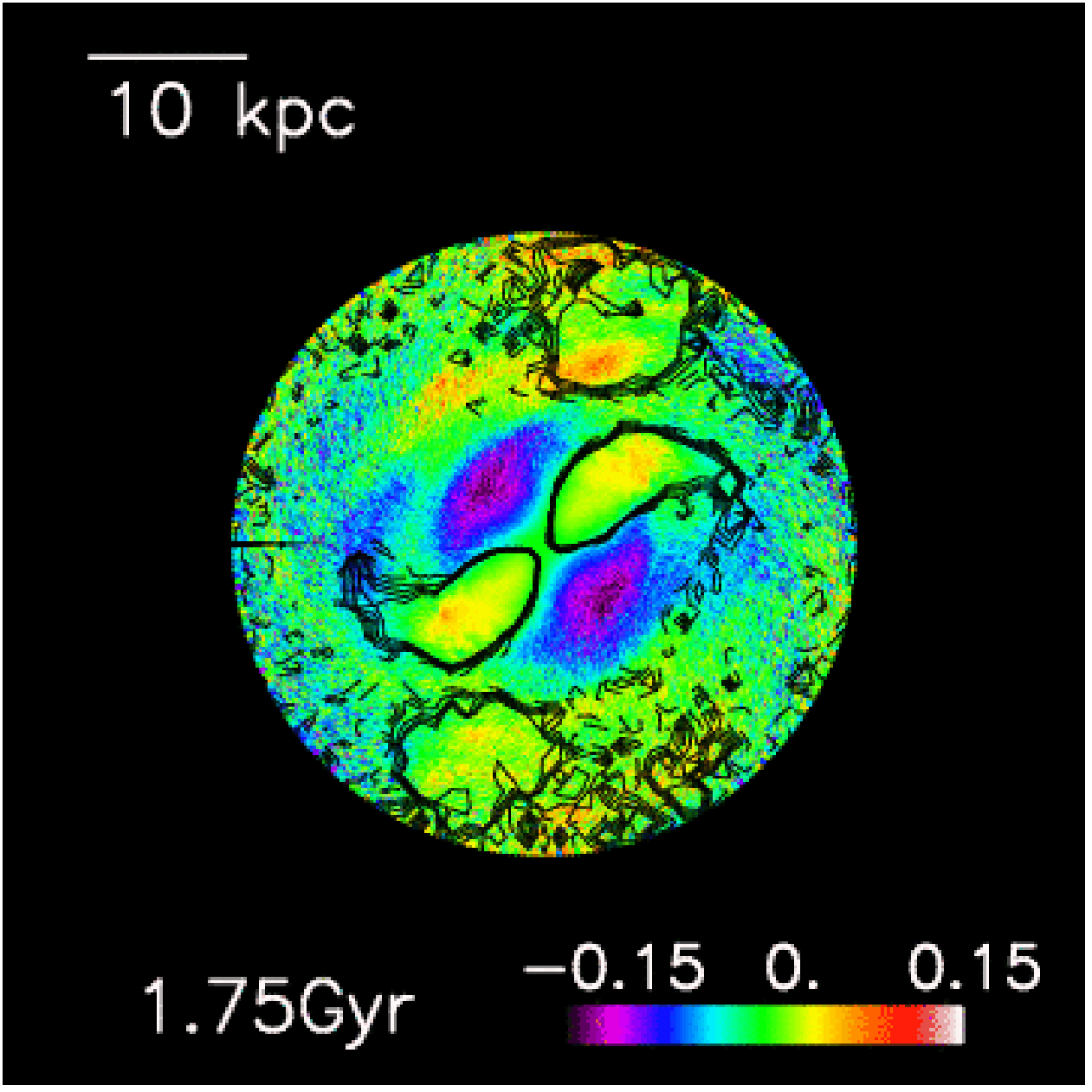}

\includegraphics[width=3.cm,angle=270]{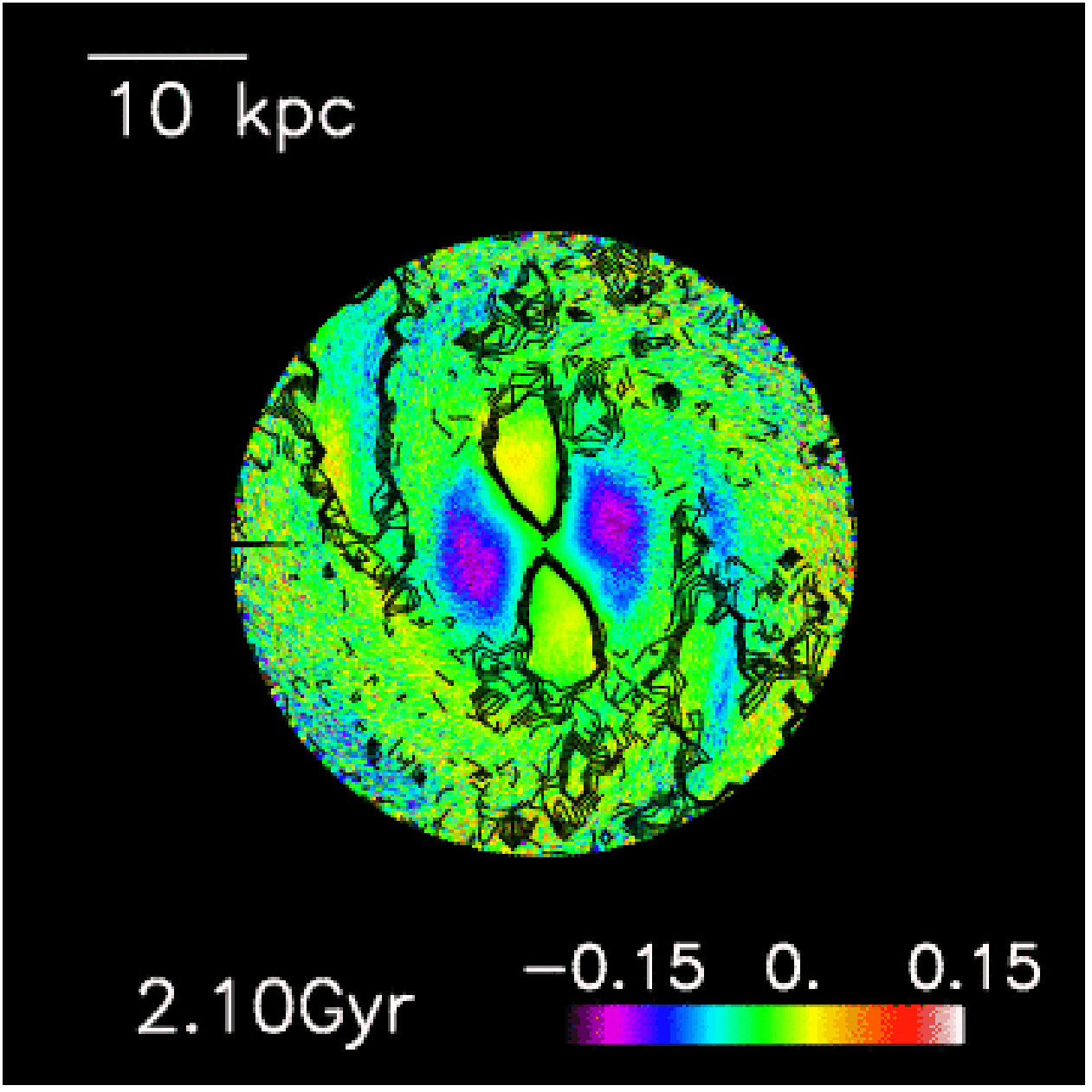}
\hspace{-0.2cm}
\includegraphics[width=3.cm,angle=270]{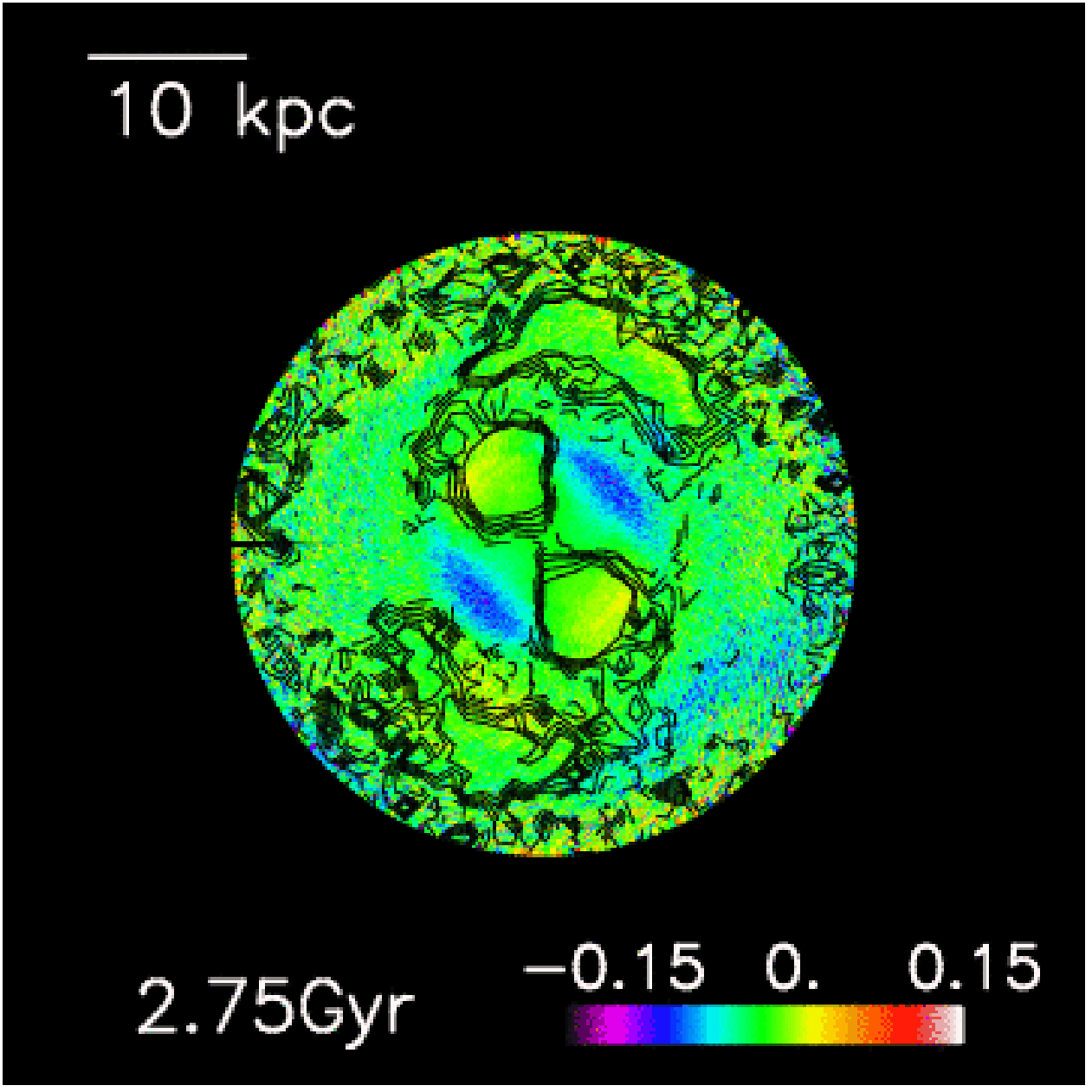}
\hspace{-0.2cm}
\includegraphics[width=3.cm,angle=270]{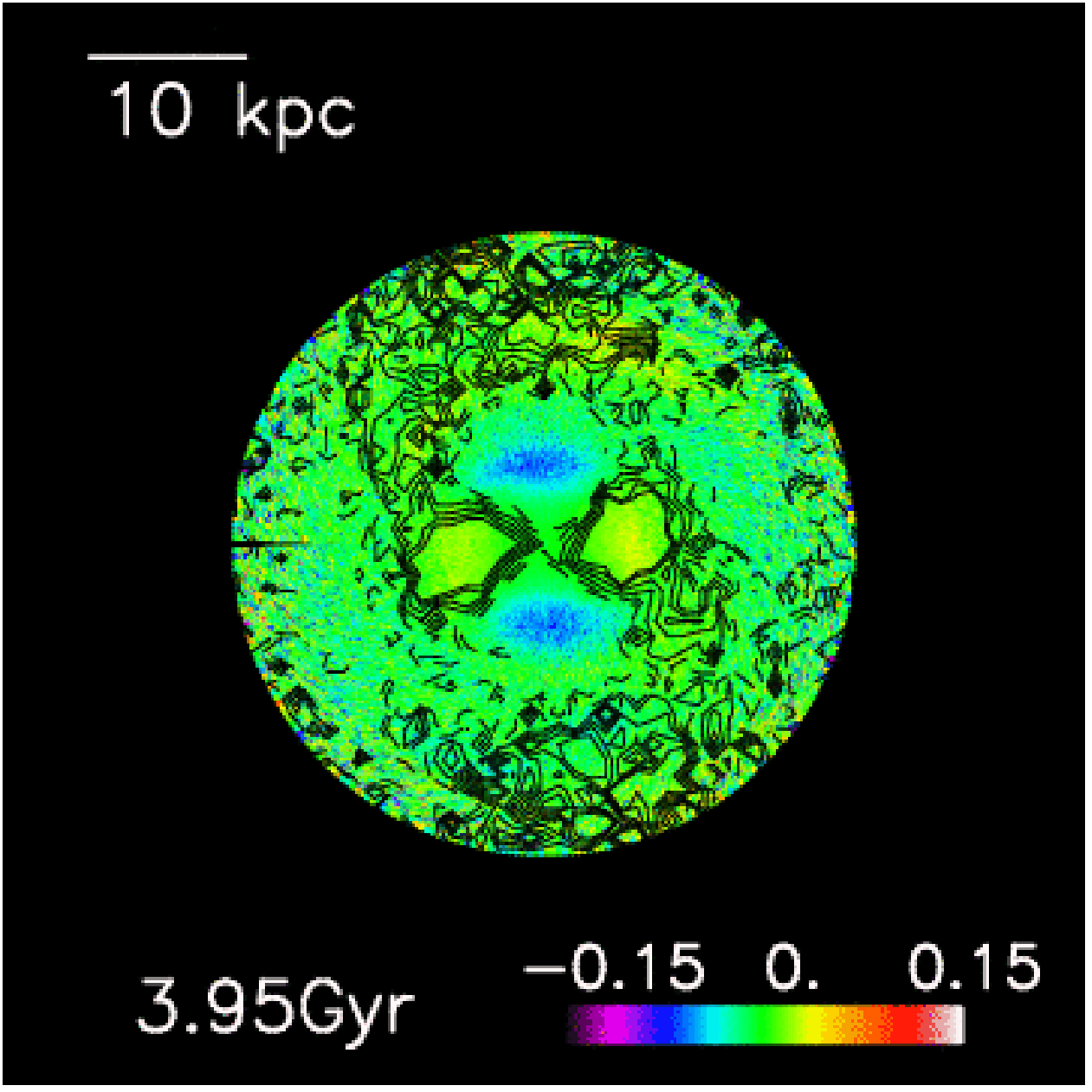}

\caption{Maps of azimuthal metallicity variations, $\delta_{\rm{[Fe/H]}}$ (dex), at different times for a stellar disk with a patchy metallicity distribution at t=0.4~Gyr (see text). Black contours correspond to equally spaced isosurfaces of the differential stellar density, $\Sigma_{diff}$. See Fig.~\ref{initvar} and Fig.\ref{var1} for comparison. }
\label{initvar2}
\end{figure}

Our initial conditions are compatible with a galaxy without prominent stellar asymmetries in its disk which experiences a phase of bar formation and strong bar activity, before the bar strength fades again. These models thus capture a possible recurrent phase in the evolution of galaxy disks (see for ex Bournaud \& Combes 2002).
About the choice of adopting initially a null azimuthal variation in the metallicity distribution, this is in agreement with the fact that at the end of the simulation, once the bar activity diminishes, azimuthal variations fade also, with typical values $\delta\sim0.4 \Delta$, $\Delta$ being the metallicity gradient (see Fig.\ref{var1} and \ref{vargrad}). 
Our choice is thus compatible with our findings: the azimuthal variations fade, as the stellar distribution tends to axisymmetry.
It can be noted also that the amplitude of azimuthal variations we find at the end of the simulation in the disk, for an initial $\Delta=-0.07$dex/kpc is of the order of 2 $\times$ 0.03 dex\footnote{The multiplication by a factor 2 is motivated by the fact that $\delta$ corresponds to half of the total variation}= 0.06 dex (see Fig.\ref{vargrad}). This value is lower than the dispersion induced in the system by blurring alone (see Fig. 1), thus ensuring  that the choice of our initial conditions is realistic (any azimuthal variation of this order is included in the metallicity dispersion at the pre-migration phase).
To further prove that the choice of initializing the stellar disk with an axisymmetric metallicity distribution does not change our conclusions, we have post-processed our simulations, assigning, at any radius, an initial azimuthal variation in metallicity $\delta \propto cos(2\theta)$, i.e. $z_m(R,\theta)=z_010^{-0.07R}10^\delta$. The resulting metallicity variation maps are given in Fig.~\ref{initvar}, for the case of $\delta=0.03cos(2\theta)$, and show that even adopting initial azimuthal variations similar to those we get at the end of the simulation ($\delta\sim0.4\Delta$), we still recover the same trends found in Fig.\ref{var1}. The same results are found if the stellar disk has initially a more patchy metallicity distribution (Fig.~\ref{initvar2}). Note that, for producing these maps, we have assigned non-zero azimuthal variations to the stellar disk at time t=0.4 Gyr. Indeed, any metallicity variation assigned at earlier times is completely washed out by shear before migration starts. Thus the amplitudes found in Fig.~\ref{initvar} and \ref{initvar2} can be taken as upper limits.

\end{appendix}


\begin{thebibliography}{}

\bibitem[Athanassoula \& Misiriotis(2002)]{atha02}Athanassoula, E. \& Misiriotis A. 2002, MNRAS, 330, 35

\bibitem[Athanassoula(2008)]{atha08}Athanassoula, E. 2008, Formation and Evolution of Galaxy Bulges (IAU Symp. 245), ed. M. Bureau, E. Athanassoula, \& B. Barbuy (Cambridge Univ. Press), 93 

\bibitem[Bakos et al.(2011)]{bak11}Bakos, J., Trujillo, I., Azzollini, R., et al. 2011, MSAIS, 18, 113%Beckman, J. E. \& Pohlen, M. 

%\bibitem[Barnes \& Hut(1986)]{bar86}Barnes, J., \& Hut, P. 1986, Nature, 324, 446

\bibitem[Bekki \& Tsujimoto(2011)]{bekki11}Bekki, K. \& Tsujimoto, T. 2011, ApJ, 738, 4

\bibitem[Berentzen et al.(2007)]{ber07}Berentzen, I., Shlosman, I., Martinez-Valpuesta, I. et al. 2007, ApJ, 666, 189
 %\& Heller, C. H. 

\bibitem[Binney \& Tremaine(1987)]{BT87}Binney, J., \& Tremaine, S. 1987, Galactic Dynamics (Princeton Univ. Press)

\bibitem[Bird et al.(2012)]{bird12}Bird, J. C., Kazantzidis, S. \& Weinberg, D. H. 2012, MNRAS, 420, 913

\bibitem[Bournaud \& Combes(2002)]{bour02}Bournaud, F. \& Combes, F. 2002, A\&A, 392, 83

\bibitem[Bournaud et al.(2009)]{bour09} Bournaud, F., Elmegreen, B. G. \& Martig, M. 2009, ApJL, 707, 1

\bibitem[Bovy et al.(2012)]{bov12}Bovy, J., Rix, H.-W. \& Hogg, D. W. 2012,  ApJ, 751, 131

\bibitem[Brunetti et al.(2011)]{bru11}Brunetti, M., Chiappini, C. \& Pfenniger, D. 2011, A\&A, 534, 75

\bibitem[Chilingarian et al.(2010)]{chili10} Chilingarian, I. V., Di Matteo, P., Combes, F., Melchior, A.-L. \& Semelin, B. 2010, A\&A, 518, 61

\bibitem[Combes \& Sanders(1981)]{cosan81}Combes, F. \& Sanders, R.~H. 1981, A\&A, 96, 184

\bibitem[Combes et al.(1990)]{com90}Combes, F., Debbasch, F., Friedli, D., \& Pfenniger, D. 1990, A\&A, 233, 82

\bibitem[Davies et al.(2009)]{davies09}Davies, B., Origlia, L., Kudritzki, R-P., Figer, D. F., Rich, R. M. et al. 2009, ApJ, 696, 2014

\bibitem[Debattista et al.(2006)]{deb06}Debattista, V. P., Mayer, L., Carollo, C. M., et al. 2006, ApJ, 645, 209%Moore, B., Wadsley, J., \& Quinn, T. 

	
\bibitem[Elmegreen \& Hunter(2006)]{elm06}Elmegreen, B. G. \& Hunter, D. A. 2006, ApJ, 636, 712

\bibitem[Friedli et al.(1994)]{friedli94}Friedli, D., Benz, W. \& Kennicutt, R. 1994, ApJL, 430, 105

%\bibitem[Gingold \& Monaghan(1982)]{gin82}Gingold, R. A., \& Monaghan, J. J. 1982, JCoPh, 46, 429

\bibitem[Gadotti \& de Souza(2005)]{gad05}Gadotti, D.~A. \& de Souza, R.~E. 2005, ApJ, 629, 797

\bibitem[Grand et al.(2012)]{grand12}Grand, R. J. J., Kawata, D. \& Cropper, M. 2012, MNRAS, 421, 1529

\bibitem[Haywood(2008)]{hay08}Haywood, M. 2008, MNRAS, 388, 1175

\bibitem[Hernquist(1993)]{hern93}Hernquist, L. 1993, ApJS, 86, 389

\bibitem[Yoachim et al.(2012)]{yoa12}Yoachim, P., Ro$\breve{s}$kar, R. \& Debattista, V. P. 2012, ApJ, 752, 97

\bibitem[Lynden-Bell \& Kalnajs(1972)]{lynkal72}Lynden-Bell, D. \& Kalnajs, A.~J. 1972, MNRAS, 157, 1	
	
\bibitem[Loebman et al.(2011)]{loe11}Loebman, S. R., Ro$\breve{s}$kar, R., Debattista, V. P.  et al. 2011, ApJ,  737, 8

\bibitem[Luck et al.(2006)]{luck06}Luck, R.~E., Kovytuk, V. V., \& Andrievsky, S. M. 2006, AJ, 132, 902

\bibitem[Luck \& Lambert(2011)]{luck11}Luck, R.~E. \& Lambert, D.~L. 2011, AJ 142, 136

%\bibitem[Lucy(1977)]{lucy77}Lucy, L. B. 1977, AJ, 82, 1013
	
\bibitem[Maciel \& Koppen(1994)]{maciel94}Maciel, W. J. \& Koppen, J. 1994, A\&A, 282, 436
	
\bibitem[Martinez-Valpuesta et al.(2006)]{mar06}Martinez-Valpuesta, I., Shlosman, I. \& Helle, C. 2006, ApJ, 637, 214

	
\bibitem[Minchev \& Famaey(2010)]{min10}Minchev, I. \& Famaey, B., 2010, ApJ 722, 112 

\bibitem[Minchev et al.(2011)]{min11}Minchev, I., Famaey, B., Combes, F., et al. 2011, A\&A 527, 147%Di Matteo, P., Mouhcine, M. \& Wozniak, H. 

\bibitem[Minchev et al.(2012)]{min12}Minchev, I., Famaey, B., Quillen, A. C., et al. 2012, A\&A 548, 126 %arXiv-1203.2621%Di Matteo, P., Combes, F., Vlajic, M., Erwin, P. \& Bland-Hawthorn, J. 2012, A\&A submitted; arXiv-1203.2621

\bibitem[Qu et al.(2011)]{qu11}Qu, Y., Di Matteo, P., Lehnert, M. D. \& van Driel, W. 2011, A\&A, 530, 10

\bibitem[Quillen et al.(2009)]{min09}	Quillen, A. C., Minchev, I., Bland-Hawthorn, J. \& Haywood, M. 2009, MNRAS, 397, 1599

\bibitem[Radburn-Smith et al.(2012)]{rad12}Radburn-Smith, D. J.,  Ro$\breve{s}$kar, R.,  Debattista, V. P., et al. 2012, ApJ, 753, 138%Dalcanton, J. J., Streich, D. 
 
\bibitem[Ro$\breve{\textrm{s}}$kar et al.(2008a)]{ros08}Ro$\breve{\textrm{s}}$kar, R.,  Debattista, V. P., Quinn, T. R., et al.  2008a, ApJ, 684, L79%Stinson, G. S., \& Wadsley, J.

\bibitem[Ro$\breve{\textrm{s}}$kar et al.(2008b)]{ros08b}Ro$\breve{\textrm{s}}$kar, R.,  Debattista, V. P., Stinson, G. S.,  et al. 2008b, 675, L65% Quinn, T. R.,  Kaufmann, T.,  \& Wadsley, J. 

\bibitem[Ro$\breve{\textrm{s}}$kar et al.(2012)]{rok12}Ro$\breve{\textrm{s}}$kar, R.,  Debattista, \& Loebman, S. 2012, MNRAS submitted; astro-ph/1211.1982


\bibitem[S$\acute{\textrm{a}}$nchez-Bl$\acute{\textrm{a}}$zquez et al.(2009)]{san09}S$\acute{\textrm{a}}$nchez-Bl$\acute{\textrm{a}}$zquez, P., Courty, S., Gibson, B.K., et al. 2009, MNRAS, 398, 591

\bibitem[S$\acute{\textrm{a}}$nchez-Bl$\acute{\textrm{a}}$zquez et al.(2011)]{psb11}S$\acute{\textrm{a}}$nchez-Bl$\acute{\textrm{a}}$zquez, P., Ocvirk, P., Gibson, B. K., et al. 2011, MNRAS, 415, 709%P$\acute{\textrm{e}}$rez, I. \& Peletier, R. F. 


\bibitem[Scarano \& L$\acute{\textrm{e}}$pine(2012)]{lep12}Scarano, S.~Jr. \& L$\acute{\textrm{e}}$pine, J.~R.~D. 2012, MNRAS, accepted; astro-ph/1209.5031

\bibitem[Sellwood \& Binney(2002)]{sel02}Sellwood,  J. A. \& Binney, J. J. 2002, MNRAS, 336, 785

\bibitem[Semelin \& Combes(2002)]{sem02}Semelin, B., \& Combes, F. 2002, A\&A, 388, 826

\bibitem[van Zee et al.(1998)]{vanzee98}van Zee, L., Salzer, J. J., Haynes, M. P., et al. 1998, AJ, 116, 2805% O'Donoghue, A. A., \& Balonek, T. J. 
	
\end{thebibliography}
\end{document}